\documentclass[fleqn,usenatbib]{mnras}

\usepackage{newtxtext,newtxmath}
\usepackage[T1]{fontenc}

\DeclareRobustCommand{\VAN}[3]{#2}
\let\VANthebibliography\thebibliography
\def\thebibliography{\DeclareRobustCommand{\VAN}[3]{##3}\VANthebibliography}

\usepackage{graphicx}	% Including figure files
\usepackage{amsmath}	% Advanced maths commands
\usepackage{caption}
\usepackage[utf8]{inputenc}
\usepackage{hyperref}
\usepackage{pdfpages}
\usepackage{xr-hyper}
\usepackage{textcomp}
\usepackage{subcaption}
\usepackage{tablefootnote}[2014/01/26]
\usepackage{threeparttable}
\usepackage{longtable}
\usepackage{multirow}

\usepackage{subcaption}
\title[Gaia RV standard stars with companion]{Analysis of Gaia radial-velocity standards: stability and new substellar companion candidates \thanks{Based on data obtained from the European Southern Observatory (ESO) Science Archive Facility}}

\author[A. Boulkaboul et al.]{A. Boulkaboul$^{1,2}$\thanks{E-mail: amina.boulkaboul@craag.edu.dz},
Y. Damerdji$^{1,3}$, T. Morel$^{3}$, Y. Frémat$^{4}$, C. Soubiran$^{5}$, E. Gosset$^{3}$, T.E. Abdelatif$^{1}$
\\
$^{1}$CRAAG - Centre de Recherche en Astronomie, Astrophysique et Géophysique, Route de  l’Observatoire, Bp 63 Bouzareah, DZ-16340, Algiers, Algeria\\
$^{2}$University of Science and Technology Houari Boumediene, Alia, DZ-16000, Algiers, Algeria\\
$^{3}$Space sciences, Technologies and Astrophysics Research (STAR) Institute, Universit\'e de Li\`ege, Quartier Agora, All\'ee du 6 Ao\^ut 19c, B\^at. B5c, B4000-Li\`ege, Belgium\\
$^{4}$Royal Observatory of Belgium, av. circulaire 3, B1180-Brussels, Belgium\\
$^{5}$Laboratoire d’astrophysique de Bordeaux, Univ. Bordeaux, CNRS, B18N, allée Geoffroy Saint-Hilaire, 33615 Pessac, France\\
}

\date{Accepted XXX. Received YYY; in original form ZZZ}

\pubyear{2020}

\begin{document}
\label{firstpage}
\pagerange{\pageref{firstpage}--\pageref{lastpage}}
\maketitle

\begin{abstract}
Our main aim is to test the non-variability of the radial velocity (RV) of a sample of 2351 standard stars used for wavelength calibration of the RVS instrument onboard Gaia. In this paper, we present the spectroscopic analysis of these stars with the determination of their physical parameters by matching observed and synthetic spectra. We estimate the offset between different instruments after determining the shift between measured and archived RVs since the instrument pipelines use various numerical masks. 
Through the confirmation of the stability of the target RVs, we find 68 stars with a long-term variation having an acceleration which exceeds $10 \,\rm{m\, s^{-1}yr^{-1}}$. This suggests a barycentric reflex motion caused by a companion. As activity phenomena may be the source of periodic and trend-like RV variations in stars with putative planetary companions, we analysed various activity indicators in order to check their correlations to the RV changes. Among the trend stars, 18 have a trend model scatter greater than $100 \,\rm{m\, s^{-1}}$ over a time span from 10 to 12 years. We also confirm that six stars with known substellar companions have a total model scatter, $3\sigma$,  exceeding the threshold set by Gaia, that is, $300 \,\rm{m\, s^{-1}}$. In addition, TYC8963-01543-1 an SB2 star has data scatter $\sigma = 176.6\, \rm{m\, s^{-1}}$. Four more other stars are revealed to be variable after combining data from different instruments.  
Despite the presence of low-amplitude changes, a very large fraction of our sample (98.8\%) appears suitable as RV calibrators for Gaia RVS.
\end{abstract}


%%%%%%%%%%%%%%%%%%%%%%%%%%%%%%%%%%%%%%%%%%%%%%%%%%

%%%%%%%%%%%%%%%%% BODY OF PAPER %%%%%%%%%%%%%%%%%%

\section{Introduction}

According to \citet{raghavan2010survey}, the %multiplicity frequency 
 fraction of main-sequence, solar-type stars with stellar or brown dwarf companions is $46\pm2\%$. The orbital periods are generally long (distribution peaks at $\log P$ [d] = 5.03).  
 Since stellar multiplicity plays a major role in
  the formation and evolution of stars, several surveys have investigated the occurrence of multiplicity. As an illustration, single-lined
 (SB1) and double-lined (SB2) spectroscopic binaries 
 make up 33$\%$ of the multiple systems in the catalogue of \citet{raghavan2010survey}, while the SB1 proportion in the Gaia-European Southern Observatory (ESO) Survey is in the range 7-14$\%$ \citep{merle2020gaia}. 

In SB1's, the periodic RV variation of the primary/host star is induced by the gravitational influence of its orbiting companion. Unfortunately, stellar activity phenomena (spots, plages and activity cycles), which affect line profiles may produce a variation in RV that mimics or hides a Keplerian signal resulting from a substellar companion (\citealt{queloz2001no}, \citealt{dumusque2012earth}, \citealt{robertson2013halpha}, \citealt{santos2014harps}). Therefore, the identification of the source of the RV variations is difficult.

Binaries with low velocity amplitude (large mass ratio) help constraining binary formation models. However, the Ninth Catalog of Spectroscopic Binary Orbits (SB9)\footnote{\url{http://sb9.astro.ulb.ac.be}} \citep{pourbaix2004s} does not contain many binaries with primary velocity amplitude lower than 1 $\rm{km\, s^{-1}}$: around 0.4\% of the catalogue, with a mean mass function value of $6.3 \times 10^{-5} \, M_{\odot}$, 
according to the latest version (2021-03-02). Our analysis can provide them.

We look at well-studied and supposedly stable stars as a validation set to establish the detection thresholds to apply for the detection of spectroscopic multiple stars in our future studies. Indeed, the detection of RV variations depends on the stability of the instrument, the quality of the measured RVs, their derived uncertainties and time baseline. 
Therefore, our main aim in the present paper is to assess the stability of a sample of RV standard candidates used to calibrate the Radial Velocity Spectrometer \citep[RVS;][]{cropper2018gaia} onboard of Gaia \citep{prusti2016gaia}. To this aim, we systematically apply an efficient procedure that searches for a binary signature in the RVs, even if it is of low amplitude (i.e. in the substellar regime). 
The criteria to select the Gaia RV standard stars and the list of candidates are described in \citet[][hereafter CS13 and CS18,  respectively]{soubiran2013catalogue, soubiran2018gaia}. These stars should be stable over at least 300 days and not have any bright neighbours ($\Delta I < 4$ mag) within a circular region of $20^{''}$ radius to make sure the RVS spectrum is not contaminated. Their RV scatter ($3 \sigma$) should not exceed 300 $\rm{m\, s^{-1}}$ throughout the duration of the mission (originally planned for five years when it was selected by ESA, but the mission was extended). The selection of the most stable candidates was performed in advance of the mission, from a compilation of RV measurements from different spectrographs. It implied a necessary compromise between the number of standards needed for the RVS calibrations (several thousands), and the number of stars followed up from the ground over a sufficient time baseline to test their stability. As a consequence, the selected standard stars are only candidates, and some of them revealed variations with a remarkable trend during the Gaia observations (CS18). For the target stars, we thus track SB1 systems with long-term RV variations in the catalogue of RV standards for Gaia RVS to clean it and improve the calibration of the next releases. 

This paper is structured as follows. In Sect. \ref{targ}, we give an overview about our target stars; in Sect. \ref{method}, we outline the methodology adopted in the spectroscopic analysis; in Sects. \ref{fullSamp}  and \ref{orbitSol} we describe the overall results for the sample; in Sect. \ref{trend}, we give details about stars showing trends; in Sect. \ref{activi}, we focus on the long-term RV variations and their correlations with different activity indicators and cross-correlation function (CCF) parameters. Finally, we draw our main conclusions in Sect. \ref{conc}.

\section{The target stars}
\label{targ}

We initially considered the sample of 2351 stars from the catalogue of Gaia RV standard star candidates (CS18) with CAL1 quality (wavelength calibrators for DR2 and DR3). The CAL1 stars were chosen to have at least two ground-based RV measurements spread over more than 300 days with a standard deviation of the mean less than 100\,$\rm{m\, s^{-1}}$. 

\begin{figure}
\begin{center}
\includegraphics[width=.5\columnwidth]{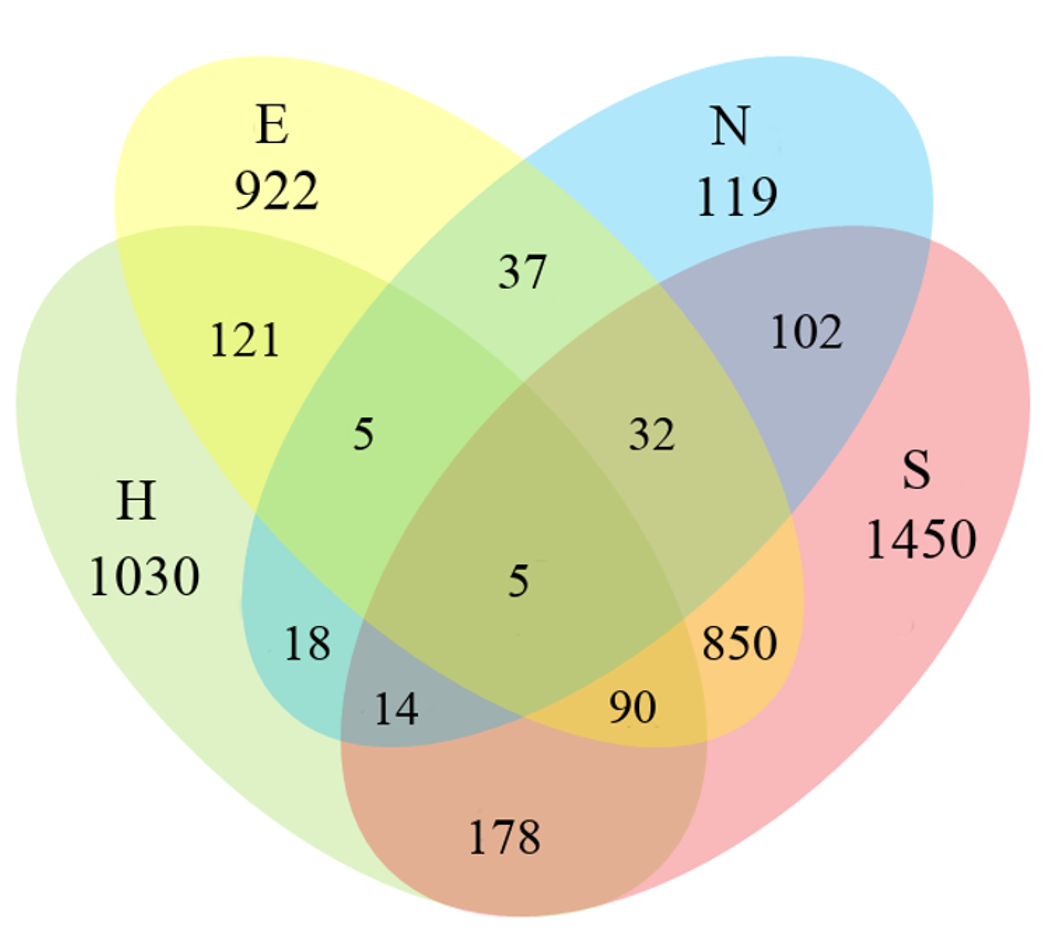}
\end{center}
\caption{Venn diagram showing the number of stars shared by different instruments: HARPS (H), SOPHIE (S), ELODIE (E), and NARVAL (N).}
\label{diagVenn}
\end{figure}

The HARPS, SOPHIE, ELODIE and NARVAL samples contain 1030, 1450, 922, and 119 stars, respectively. For 11 stars among the HARPS sample, CS18 used CORALIE spectra. Stars in common between different instruments are shown in Fig.~\ref{diagVenn}.
We provide an excerpt of the sample in Table~\ref{tab:tableTarget}, which gives the target coordinates ($\alpha$, $\delta$), visual apparent magnitude $V$, spectral type, observation duration, signal-to-noise ratio (SNR) range, the number of observations before and after outlier filtering (Sect. \ref{orbitSol}), as well as the instrument used. 
\begin{table*}
    \caption{List of targets:  Target Id, 
    %along with their 
    coordinates ($\alpha$, $\delta$) at the specific epoch, visual apparent magnitude, $V$, spectral type taken from CS18, SpT, total time span, $T$, minimum and maximum SNR values, total number of analysed spectra, $N_\mathrm{init}$, the final number of measurements after filtering, $N$, the number of additional spectra compared to CS18, $N_\mathrm{add}$}, and the instrument used.
    \label{tab:tableTarget}
    \begin{tabular}{|r|r|r|r|r|r|r|r|r|r|r|r|r|}
    \hline
    {HIP/TYC} &   {$\alpha$ [deg]} & {$\delta$ [deg]} & {Epoch}&{$V$ [mag]} & {SpT} & {$T$ [days]} & {min(SNR)} & {max(SNR)} & {$N_\mathrm{init}$} & {$N$} & {$N_\mathrm{add}$} & {Instrument}\\
\hline
47&     $ 0.135 $ & $  -56.836$   & J2015.5  &  10.75  &  K3V &   6154.04 &   23.5 &   44.7 &  10 & 10 & 3 & H \\
57&     $ 0.168  $ & $  -69.676 $   & J2000.0  &   8.24  &  K1V &   4766.05 &   42.4 &  143.0 &  38 &  37 & 3 & H \\
80&     $ 0.245 $ & $  -11.824$   & J2015.5  &   9.11  &  G2V &   4266.28 &   40.7 &  158.2 &  124 &  121 & 86 & H \\
142&     $ 0.454 $ & $   66.306$  &  J2015.5  &   7.32  &  K0 &   1473.00 &    43.3 &  133.0 &  32 &  32 &  1 &S \\
184&     $ 0.590 $ & $   11.006$  &  J2015.5  &   8.47  &  K0V &      0.00 &   59.3 &   59.3 &  1 &  1 & 0 & E \\
184&     $ 0.590 $ & $   11.006$  &  J2015.5  &   8.47  &  K0V &   4035.58 &    43.8 &   81.8 &  5 &  5 & 1 & S \\
348&     $ 1.091 $ & $   12.958$  &  J2015.5  &   8.64  &  G5 &   3273.99 &    40.4 &   60.7 &  4 &  4 & 0 & S \\
400&     $ 1.236 $ & $   23.270$  &  J2015.5  &   7.81  &  G9V &   5096.60 &   27.3 &  195.1 &  348 &  346 & 297 & S \\
413&     $ 1.262 $ & $  -36.015$   & J2015.5  &   7.74  &  G0V   &   5739.22 &  26.0 &  131.8 &  18 &  18 & 7 & H \\
436&     $ 1.322 $ & $  -67.835$   & J2015.5  &   8.49  &  K4.5V &   6229.99 &  32.4 &  127.4 &  82 &  80 & 69 & H \\
\multicolumn{10}{l}{...}\\
\hline
\end{tabular}
\begin{tablenotes}
      \footnotesize
      \item Note: This table is available in its entirety online at the CDS. A portion is shown here for guidance regarding its form and content.
    \end{tablenotes}
\end{table*}

The archive spectra we analysed are available in the ESO database archive for HARPS\footnote{\url{http://archive.eso.org/wdb/wdb/adp/phase3_spectral/form}} and in the relevant instrument website for SOPHIE\footnote{\url{http://atlas.obs-hp.fr/sophie/}}, ELODIE\footnote{\url{http://atlas.obs-hp.fr/elodie/index.html}} \citep{moultaka2004elodie}, and NARVAL\footnote{\url{http://polarbase.irap.omp.eu/}}. As recent HARPS, NARVAL, and SOPHIE observations are available, we benefit from more measurements than in CS18 and, thus, also from a longer time span. These additional spectra have been retrieved through queries to ESO, NARVAL and SOPHIE archives. The number of these additional spectra for each target is listed in Table~\ref{tab:tableTarget}. Taking these additional data into account increases the total number of spectra by $19.5\%$, $11.8\%$, $38.2\%$, and $74.8\%$ for NARVAL, ELODIE, SOPHIE, and HARPS respectively. The additional ELODIE spectra are those removed from CS13 and CS18, they either have low SNR value or do not have archival RVs. We filtered out those with low SNR when analysing the RVs.

HARPS spectra have a resolving power $R$ = $\lambda/\Delta\lambda$ = 115,000 covering a wavelength range of $\lambda\lambda$\, $3780$--$6910$\,\AA, while SOPHIE achieves 
$R =75,000$ and covers the wavelength range $\lambda\lambda$\, $3872$--$6943$\,\AA . ELODIE, on the other hand, covers the wavelength range $\lambda\lambda$\, $3850$--$6800$\,\AA \, with a spectral resolution of 42,000. Covering the optical range $\lambda\lambda$\, $3700$--$10000$\,\AA , NARVAL has $R$ $\sim$ 81,000.

\section{METHODS}
\label{method}
We developed a pipeline different from the instrument ones to derive the RVs based on a synthetic spectrum and initial stellar parameters rather than on numerical masks.
We simultaneously derive the RV and the line broadening (Vbroad: which makes no distinction between rotation and macroturbulence velocities) using the minimum distance algorithm based on the minimisation of the quadratic sum,  $\chi^2$, of the difference between the observed spectrum and synthetic spectra (theoretical template). We vary as free parameters: the effective temperature, $T_{\mathrm{eff}}$, surface gravity, $\log\,g$, total line broadening, Vbroad, metallicity, [Fe$\slash$H], and abundance ratio  of the $\alpha -$process elements$, [\alpha/\rm{Fe}]$. The minimum $\chi^2$ gives the template that best matches the observed spectrum. This procedure is similar to that adopted for the ground-based processing of the Gaia RVS spectra \citep{sartoretti2018gaia}.

The template is a synthetic spectrum $\mathrm{s(\lambda_s)}$ resampled to the observations, as indicated in Eq. \ref{resample} \citep{david2014multi}, where $\lambda_s$ is the rest frame wavelength scale and $G$ is Green's function convolved with a Gaussian kernel. In order to have the template and observed spectra fully overlapping in wavelength space after applying the Doppler shift, the wavelength range of the former must extend beyond the observed one \citep{david2014multi}. It is then convolved with the rotation profile u[Vbroad] given by \cite{gray2021observation}. We thus have

\begin{equation}
\label{resample}
T(\lambda, RV, \rm{Vbroad}) = u[\rm{Vbroad}]*G\left[\lambda, \lambda_s\left(1 + \frac{RV}{c}\right)\right]*s(\lambda_s) .
\end{equation}

We minimise the quadratic sum over each wavelength element
\begin{align}
\label{eqChi2}
\chi^2 = &\sum_i w_i(S_i - T'_i)^2& \\
\label{sb1}
T'_i =& P_n(\lambda_i) \: T(\lambda_i, \mathrm{RV}, \mathrm{Vbroad}) + b&
\end{align}
where $S_i$ is the observed spectrum, $w_i$ is the statistical weight, and $T'_i$ is the single-lined model given by Eq. \ref{sb1},  $P_n(\lambda)$ is the polynomial modelling the continuum as a function of wavelength, $\lambda$. $T$ is the normalised synthetic spectrum used. It is convolved with the instrument 
line-spread function (LSF) assumed to be Gaussian, then with the rotational broadening profile, and finally Doppler shifted. $b$ refers to the background flux. Even though macroturbulent velocity has an effect on the line shape distinct from that from rotation, we only considered the rotational profile. The Vbroad parameter, thus, includes both effects.

The grid of synthetic spectra, $T$ ($R$ $>$ 150,000), is extracted from the Pollux database \citep{palacios2010pollux}. We chose the MARCS \cite{gustafsson2008grid}-AMBRE \citep{de2012ambre} plane-parallel and spherical atmosphere models with a microturbulent velocity (1 $\rm{km \, s^{-1}}$ for the former grid and either 1 or 2 $\rm{km \, s^{-1}}$ for the latter) that is suitable to cool stars in the range 3500--8000 K. Because they are also adopted for the ground-based processing of the Gaia RVS data \citep{sartoretti2018gaia}, we preferred the MARCS-AMBRE synthetic spectra over the PHOENIX ones \citep{husser2013new}.

The astrophysical parameters (APs: $T_{\mathrm{eff}}$, $\log\,g$, [Fe$\slash$H]
and [$\alpha\slash$Fe]) of the template that best matches our target spectrum are derived by using a minimum distance algorithm. 
A first guess of the RV and Vbroad is derived by a least square fitting method in Fourier space to find an approximate solution for each spectrum.  
The solution is refined afterwards by Levenberg-Marquardt minimisation (\citealt{press1986numerical}, Sect.  15.5.2). More details about this method can be found in Sect. 7.5 of \citet{sartoretti2018gaia}.

The line broadening is considered a non linear parameter, which is required to be the same for all spectra of a given target. We initially fit each spectrum individually, allowing us to exclude spectra with an outlying $\chi^2$ (outside the range median$\pm 5\sigma$) from the subsequent global fit, in which we measure a common Vbroad for all spectra. Then, we fix Vbroad and re-measure the RVs to obtain non-correlated values. 

The RVs estimated using this method should be more homogeneous than archival pipeline RVs that are occasionally based on an unsuitable numerical correlation mask or on different masks for the same star. As an example, choosing either a K5 or a G2 mask induces a RV shift reaching up to 20 $\rm{m\, s^{-1}}$ \citep{anglada2012harps}. 

\label{FitSpec}
We fitted the observed spectra 
within the five spectral ranges: $\lambda\lambda$\,$4000$--$4300$\,\AA,  $\lambda\lambda$\,$4400$--$4800$\ \AA, $\lambda\lambda$\,$4900$--$5200$\ \AA, $\lambda\lambda$\,$5400$--$5800$\ \AA, and $\lambda\lambda$\,$6000$--$6200$\ \AA . These spectral domains present the advantage of not being contaminated with telluric lines. We also excluded the Balmer lines that would hinder the RV measurement, except the H$_{\delta}$ line because it is weak.

We looped over MARCS-AMBRE plane-parallel and spherical synthetic spectra with microturbulence velocity of 1 and 2 $\rm{km \, s^{-1}}$; for APs initially ranging from [$p - \rm{\Delta p}$] to [$p + \rm{\Delta p}$] with $p$ referring to the AP: either $T_{\mathrm{eff}}$, $\log\,g$ or [Fe$\slash$H] whose initial estimates are extracted from the most recent estimates in the PASTEL catalogue \citep{soubiran2016pastel}.
 The parameters are taken from Gaia DR2 \citep{brown2018gaia} for the 13 stars not included in PASTEL.
The parameter steps, $\rm{\Delta p}$, are: $\Delta T_{\mathrm{eff}} = 250$ K, $\Delta \log\,g = 0.5$ dex and $\Delta$[Fe$\slash$H] = 0.25 dex, while we consider all possible $[\alpha/\rm{Fe}]$ for every combination. If one of the determined parameters is on the interval edge, $p_0 =[p \pm \rm{\Delta p}]$, the template parameters for the second iteration lie in the interval $[p , p + 2 \rm{\Delta p}]$ or $[p - 2 \rm{\Delta p} , p]$.
This process is repeated until the parameter is at the centre of the box. We find the best template for each domain separately, then we choose the one which yields the minimum $\chi^2$ sum for all domains. We do not fit the domains simultaneously, as there is a shift reaching $100 \,\rm{m\, s^{-1}}$ between the RVs estimated for each of them (as detailed in the following section).

The polynomial degree of the continuum function in each interval is chosen using an $F$-test, which compares successive polynomial models fitting residuals with a degree starting 
from three and reaching up to ten to identify the polynomial of the lowest degree that yields an acceptable solution.

As the flux errors are not provided by the HARPS pipeline, we have opted for the unit weighting: $w_i = 1/ \sigma_i^2$, with $\sigma_i = 1$ and $w_i = 1$. For consistency, it was also the case for the spectra obtained with other spectrographs. The Poisson weighting ($\sigma_F = \sqrt{F}$, where $F$ is the flux expressed in photon counts) accounts for the loss in RV measurement precision as the broad and deep spectral lines broaden the $\chi^2$ function. 
The $\chi^2$ was rescaled for each spectrum in order to obtain normalised flux residuals ($S-T'$) with unit variance \citep{andrae2010and}. There is no need to rescale if the flux residuals are good enough (i.e. according to Anderson-Darling normality test; \citealt{anderson1952asymptotic}). Great care is taken to standardise input fluxes when applying unit weight scheme.
 
Our Vbroad is a measure of the line broadening arising from rotation and macroturbulence after accounting for the effect of microturbulence and instrumental resolution. Since Vbroad depends on APs, we adopted as uncertainty the standard deviation of measurements obtained for synthetic spectra with parameters [$p \pm \rm{\Delta p}$]. Namely, we compared Vbroad of the best-fit synthetic spectrum and that obtained for closely similar stellar parameters ($T_{\mathrm{eff}}$, $\log\,g$ and [Fe$\slash$H]).

\section{Analysis of the full sample}
\label{fullSamp}
\subsection{RV offset and Vbroad analysis}
We measured the RV and Vbroad of 2351 stars by fitting HARPS/SOPHIE/ELODIE/NARVAL spectra with MARCS-AMBRE synthetic spectra. 
HARPS, then SOPHIE if not available, spectra were always preferred to select the best synthetic spectrum used to estimate the two quantities.

\begin{figure}
\begin{center}
\includegraphics[width=\columnwidth]{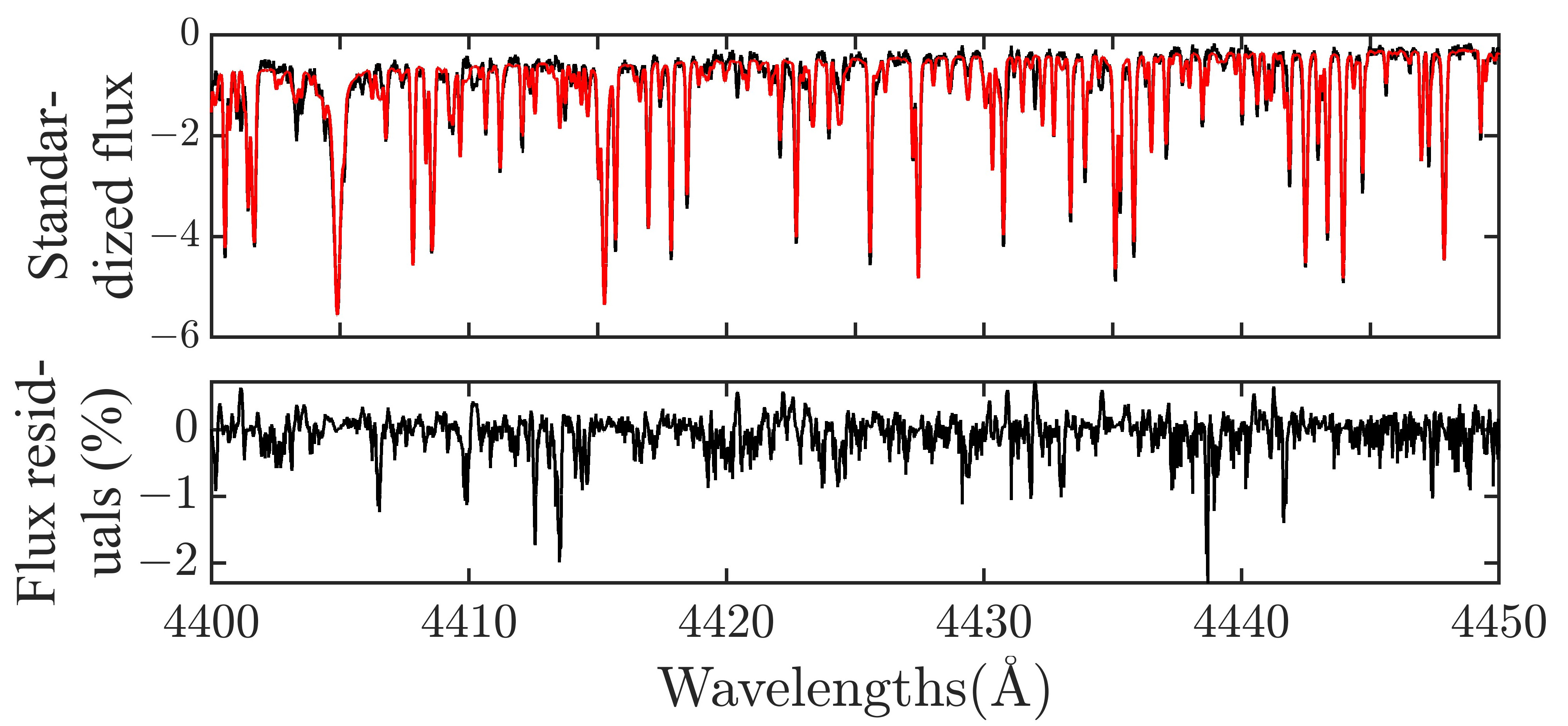}
\end{center}
\caption{Top panel: HIP21821 best fit with MARCS-AMBRE synthetic spectrum for observation BJD = 2455493.8 in the interval $\lambda\lambda$\,$4400$--$4450$\,\AA \,  with $RV = 9.0782 \pm 0.0057\, \rm{km\, s^{-1}}$ and Vbroad $ = 4.79 \pm 0.26\,  \rm{km\, s^{-1}}$. Observed spectrum in black and synthetic spectrum in red ($T_{\mathrm{eff}} = 6000$ K, $\log\,g = 4.0$ dex, [Fe$\slash$H] $= -0.5$ dex, and [$\alpha\slash$Fe] $= 0.2$ dex). The fit residuals divided by the standardised flux in that domain are shown in the bottom panel.}
\label{fig:21821Fit}
\end{figure}

 An example of our fitted spectrum along with the fit residuals are shown in Fig. \ref{fig:21821Fit} for HIP21821 in the range $\lambda\lambda$\,$4400$--$4450$\ \AA.
  It can be seen that the fit is globally good, except for a few weak lines. This could be due to template mismatch and the fact that the line list adopted to compute the synthetic spectrum is necessarily imperfect.
  
\begin{figure*}
\begin{center}
\includegraphics[width=\textwidth]{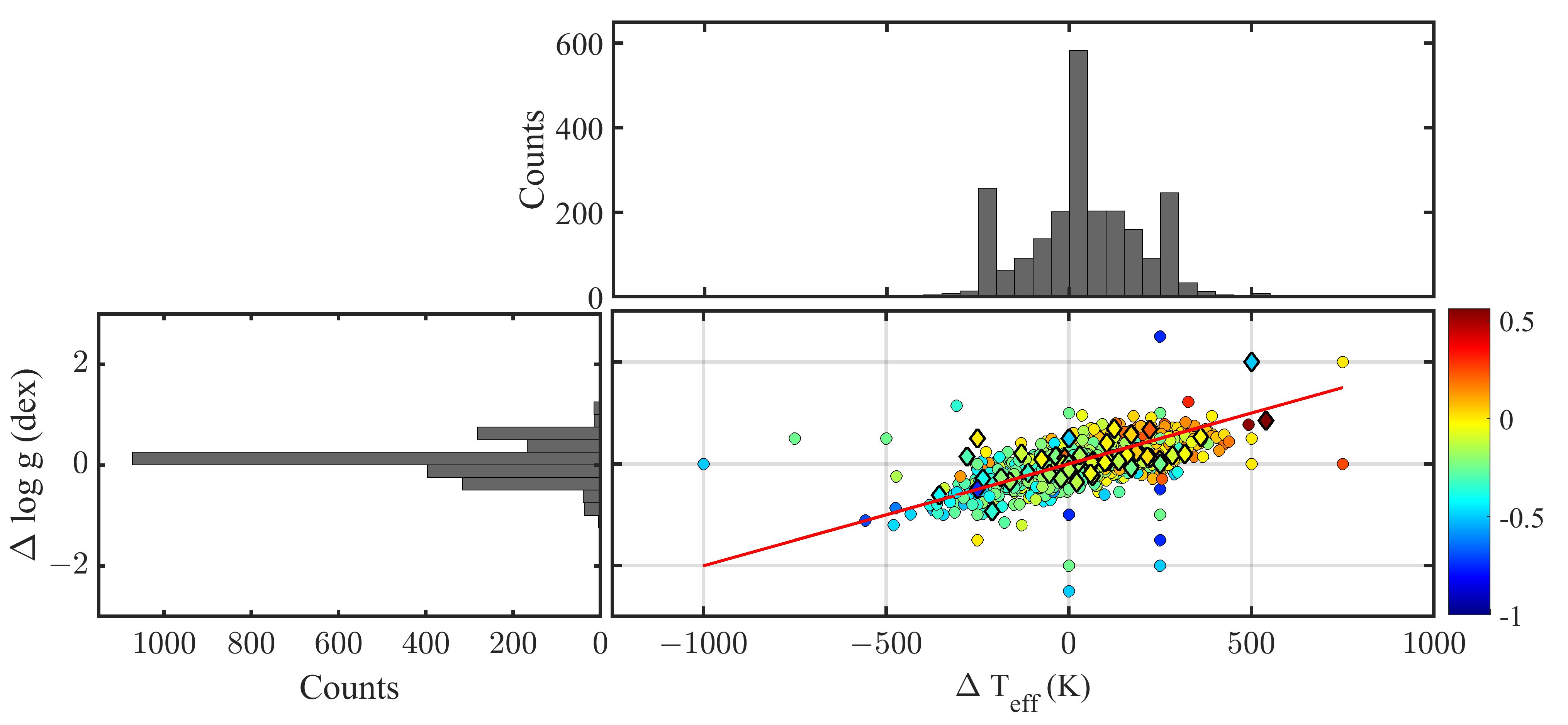}\end{center}
\caption{Top panel: distribution of the differences between the best template and PASTEL's $T_{\mathrm{eff}}$, $\Delta T_{\mathrm{eff}}$. Bottom left panel: $\Delta \log\,g$ distribution. Bottom right panel: $\Delta \log\,g$ as a function of $\Delta T_{\mathrm{eff}}$ where the trend stars (see next Section) are plotted as diamonds. The colourbar denotes $\Delta$[Fe$\slash$H]. The red line shows the relation $\Delta \log\,g = \Delta T_{\mathrm{eff}} / 500$.}
\label{fig:diffAPS_hist}
\end{figure*}

\begin{figure}
\begin{center}
\includegraphics[width=\columnwidth]{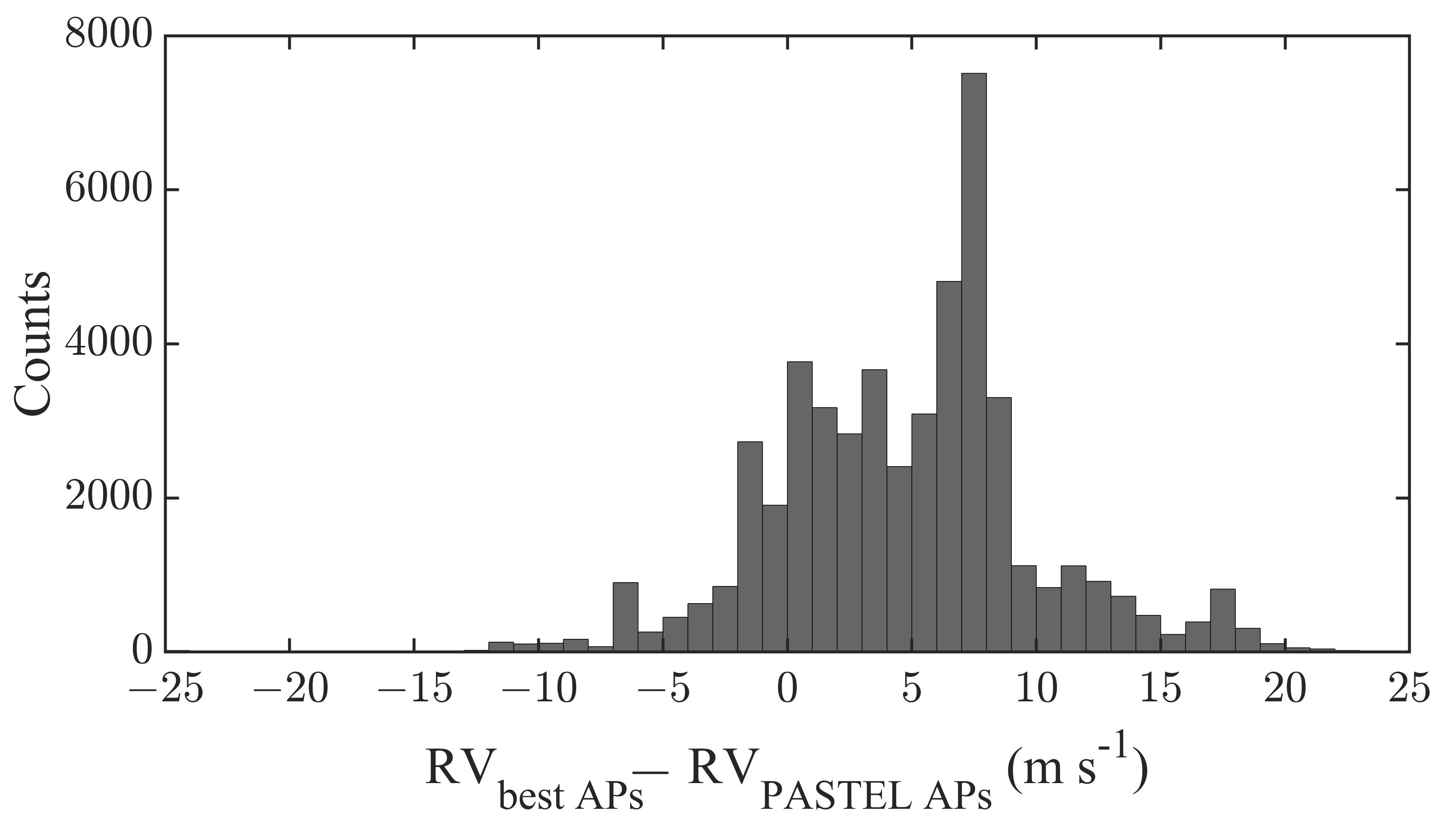}\end{center}
\caption{Distribution of the difference in RVs estimated from the best-fit and PASTEL templates with a median of $5.30 \pm 0.02 \, \rm{m\, s^{-1}}$.}
\label{fig:histDeltaRV}
\end{figure}

 The APs of the best synthetic spectra selected after the minimisation are listed for a portion of the sample in Table~\ref{tab:APS}. We computed $\Delta \rm{AP}$, which is the difference between the AP of the best template and that selected in PASTEL. We plot $\Delta \log\,g$ as a function of $\Delta T_{\mathrm{eff}}$ in the bottom right panel of Fig. \ref{fig:diffAPS_hist} where the colourbar denotes $\Delta$[Fe$\slash$H]. The trend stars (detailed in the next Section) are plotted as diamonds. The distribution of $\Delta T_{\mathrm{eff}}$ is shown in the top right panel, where $92.9\%$ of the stars have $\lvert\Delta T_{\mathrm{eff}}\rvert \le 250$ K while $90.1 \%$ of the stars have $\lvert\Delta \log\,g \rvert \le 0.5$ dex, as shown in the distribution in the bottom left panel. We have $86.3 \%$ of the targets with APs inside the box [$\pm 250$ K, $\pm 0.5$ dex]. The rate drops to $68.1\%$ if we include those with $\lvert \Delta$[Fe$\slash$H]$\rvert \le 0.25$ . We show in Fig. \ref{fig:histDeltaRV} the distribution of the difference in RVs estimated using the best template and that with APs closest to those in 
 %e template with closest APs to 
 PASTEL. The median is $5.30 \pm 0.02 \, \rm{m\, s^{-1}}$. The multiple modes present in the distribution could be related to whether all APs or just one of them differ between the two templates.  
 Even though the shift would not affect the orbital solutions in case of RV variations of high amplitude, it could induce a difference for lower-amplitude systems. This is expected as the shift between HARPS pre- and post-fiber-upgrade RVs is temperature dependent, as discussed below.

\begin{figure}
\begin{center}
\begin{subfigure}{.5\textwidth}
\includegraphics[width=\linewidth]{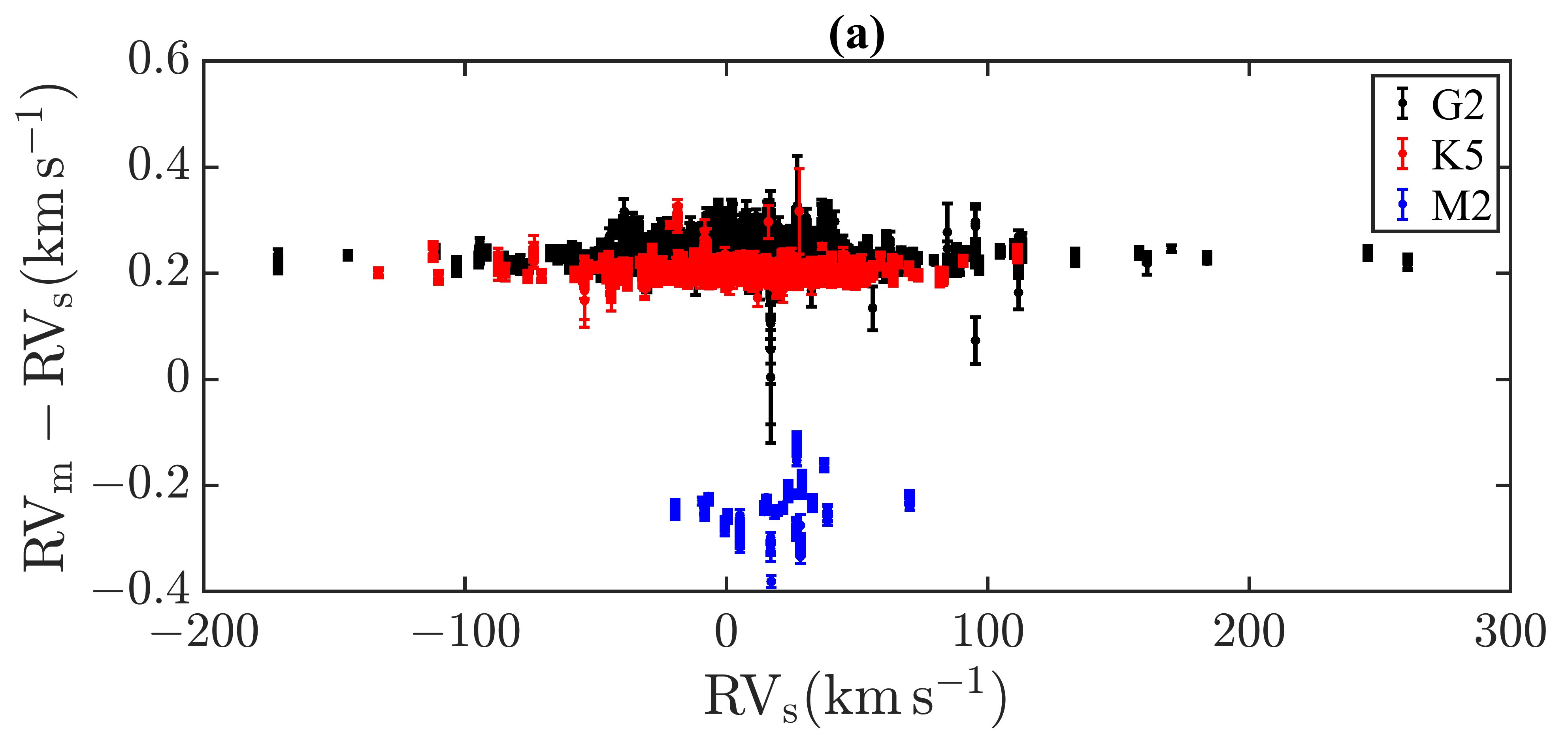}
\end{subfigure}
\begin{subfigure}{.5\textwidth}
\includegraphics[width=\linewidth]{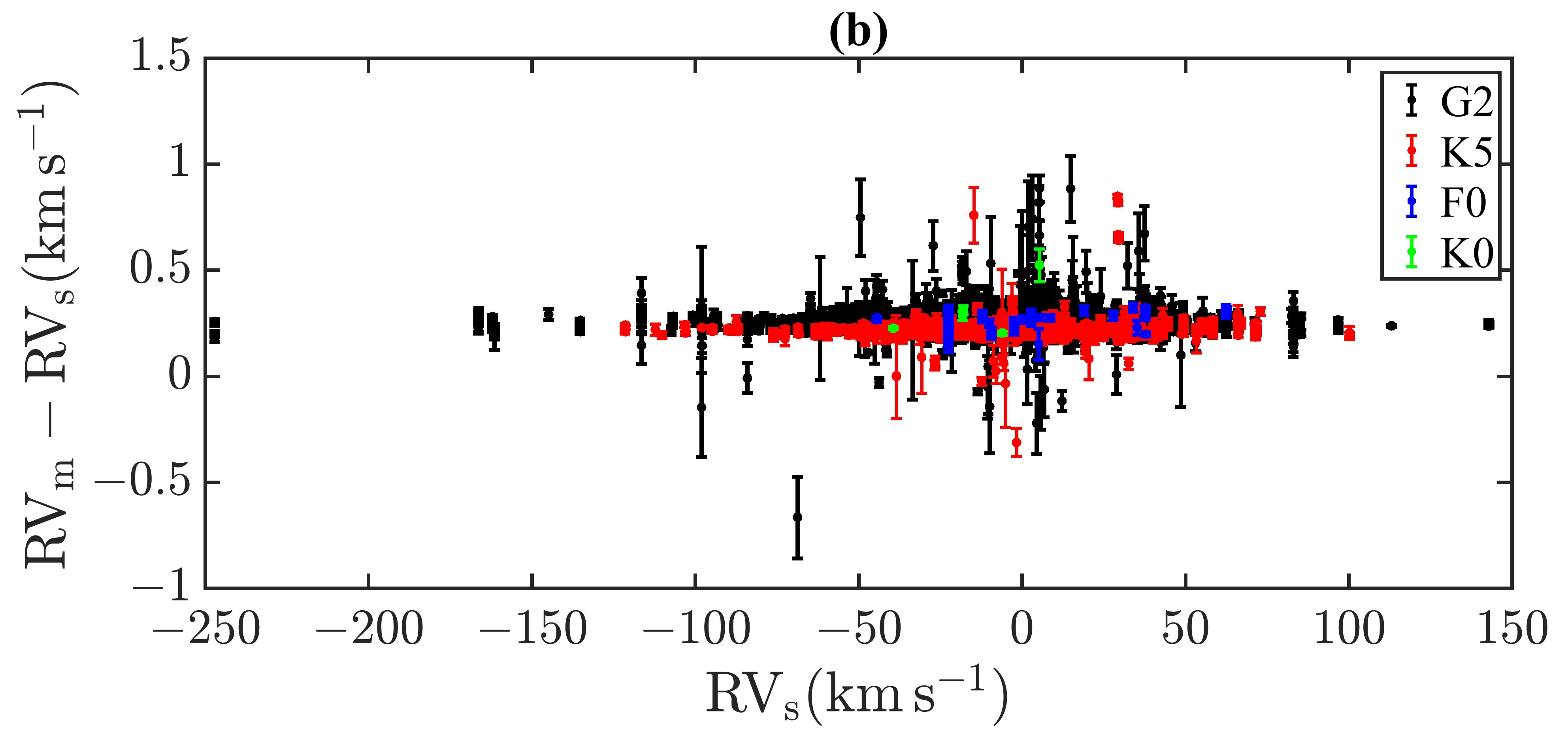}
\end{subfigure}
\begin{subfigure}{.5\textwidth}
\includegraphics[width=\linewidth]{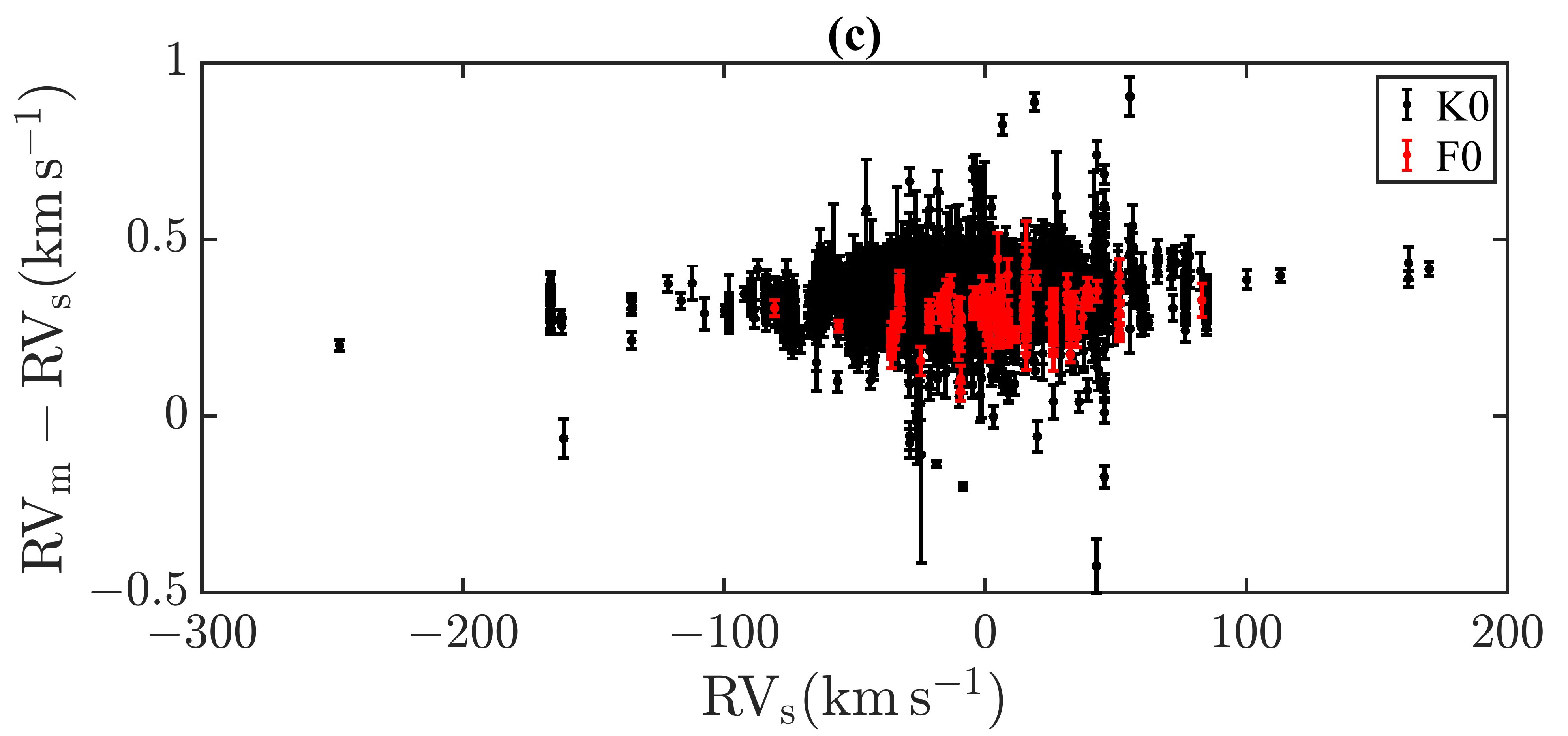}
\end{subfigure}
\begin{subfigure}{.5\textwidth}
\includegraphics[width=\linewidth]{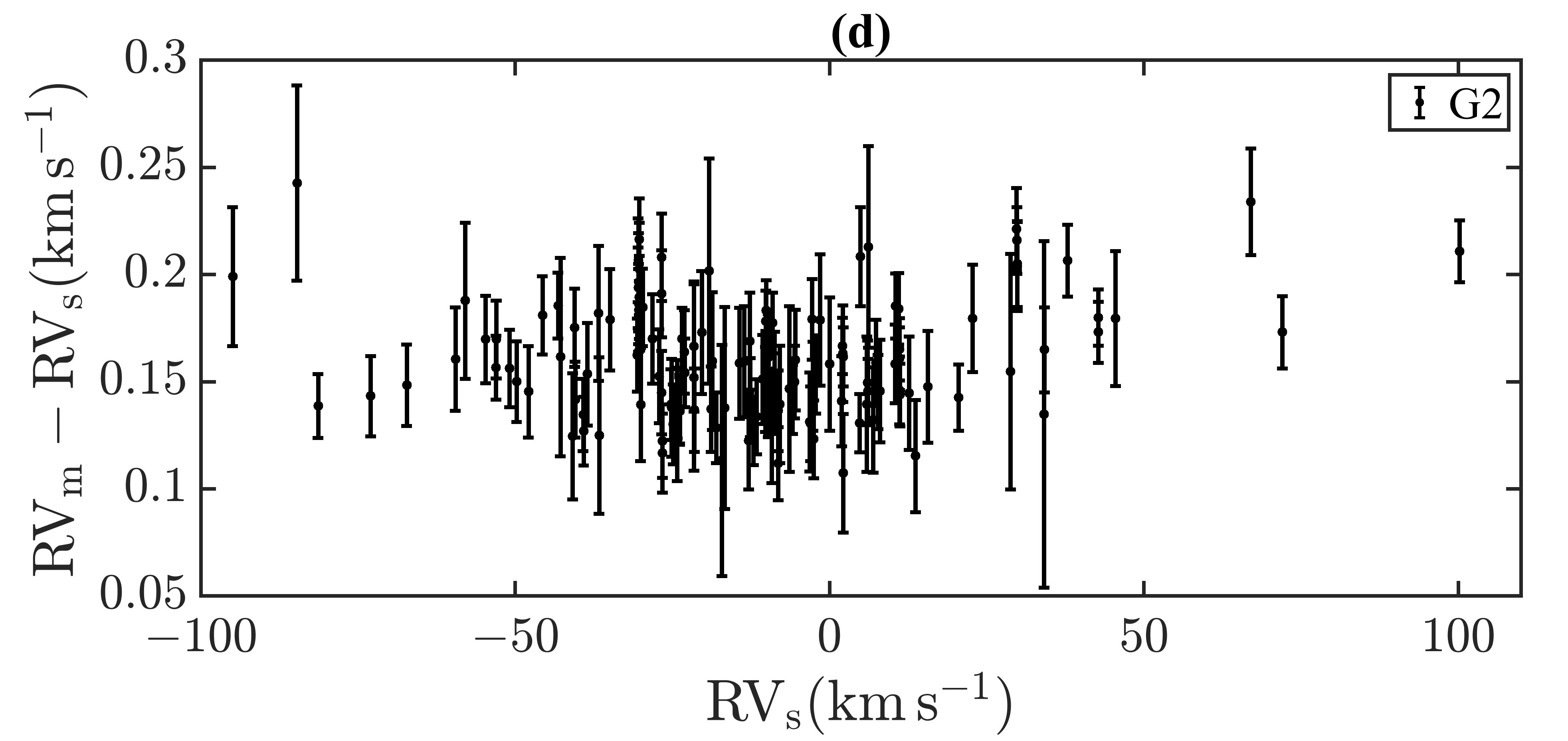}
\end{subfigure}
\end{center}
\caption{Difference between our measured RVs, $RV_m$, and those extracted from instrument pipeline, $RV_s$, as a function of $RV_s$. From top to bottom: (a) HARPS, (b) SOPHIE, (c) ELODIE, and (d) NARVAL. The shift is around $300\;  \rm{m\, s^{-1}}$ for ELODIE and $200\;  \rm{m\, s^{-1}}$ for the other three instruments. It is true for all masks except for the M2 mask (see upper panel) for which we argue that the RVs in the archives are overestimated.}
\label{compareESORV}
\end{figure}

We compare in Fig.~\ref{compareESORV} the difference between our measured RVs, $RV_m$, and those extracted from instrument archive, $RV_s$, as a function of $RV_s$. For HARPS (top panel), the difference between measured and archived RVs is $-241.70 \pm 2.09 \,\rm{m\, s^{-1}}$ for the M2 numerical mask, while it is $201.40 \pm 2.09 \,\rm{m\, s^{-1}}$ and $228.90 \pm 0.07 \,\rm{m\, s^{-1}}$ for the K5 and G2 masks, respectively. In the same way, we compare the RV measurements from SOPHIE spectra in Fig.~\ref{compareESORV} (b). The shift is close to the HARPS one with positive offsets of $217.8 \pm 0.3 \,\rm{m\, s^{-1}}$, $228 \pm 48\,\rm{m\, s^{-1}}$, $252.4 \pm 0.9 \,\rm{m\, s^{-1}}$, and $265 \pm 5$ for the K5, K0, G2 and F0 masks, respectively. ELODIE has the largest shifts: $336.4 \pm 0.8 \,\rm{m\, s^{-1}}$ and $ 282 \pm 4 \,\rm{m\, s^{-1}}$ for the K0 and F0 masks, respectively. Finally, the shift for NARVAL %instrument 
is $157 \pm 3 \,\rm{m\, s^{-1}}$ for the only mask used, G2.

Some archive HARPS RVs have been estimated using different masks for the same spectrum. For the same date, the median of the difference between RVs measured using either a M2 or a K5 mask is $439.1 \pm 2.0 \,\rm{m\, s^{-1}}$, which is very similar
to the difference between the shifts $RV_s - RV_m$ for the M2 and K5 masks: $443.1 \pm 2.1 \,\rm{m\, s^{-1}}$. Therefore, we conclude that our measurements are more consistent and that the archived RVs obtained from M2 masks are overestimated.

The average offset between our measurements and those of the archives is $240.4\,\rm{m\, s^{-1}}$ considering all spectrographs.
This offset is caused by the difference in instrument RV zero points and the method used. \cite{fremat2017test} found a similar shift of $220 \pm 30 \,\rm{m\, s^{-1}}$ between their UVES measurements and those found in the CS13 catalogue. We list in Table~\ref{tab:APS} the mean of our measured RVs, $<\rm{RV_m}>$, and the median given by CS18,
$<\rm{RV_{CS18}}>$. 

\begin{figure*}
\begin{center}
\begin{subfigure}{.49\textwidth}
\includegraphics[width=\linewidth]{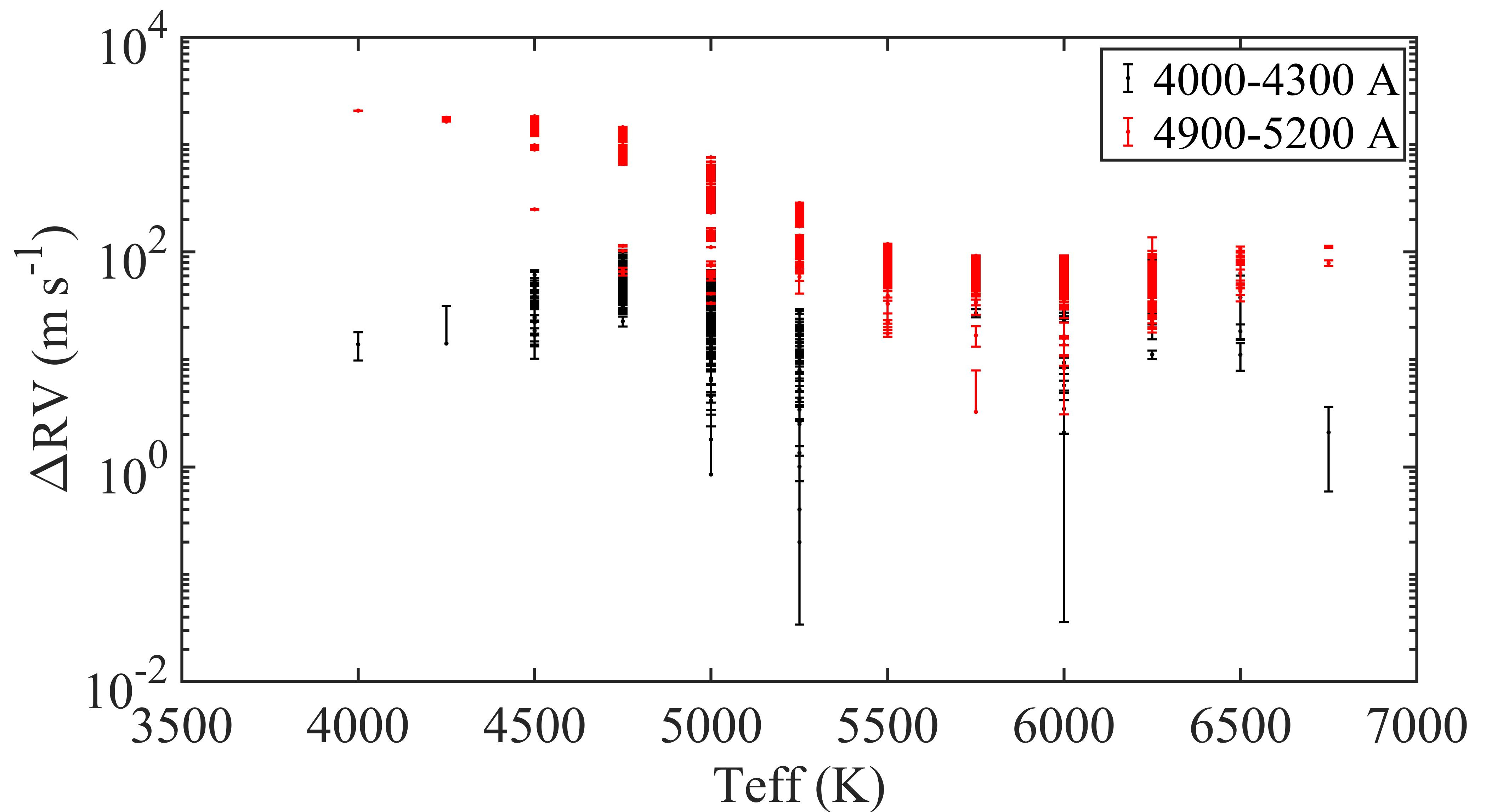}
\end{subfigure}
\begin{subfigure}{.49\textwidth}
\includegraphics[width=\linewidth]{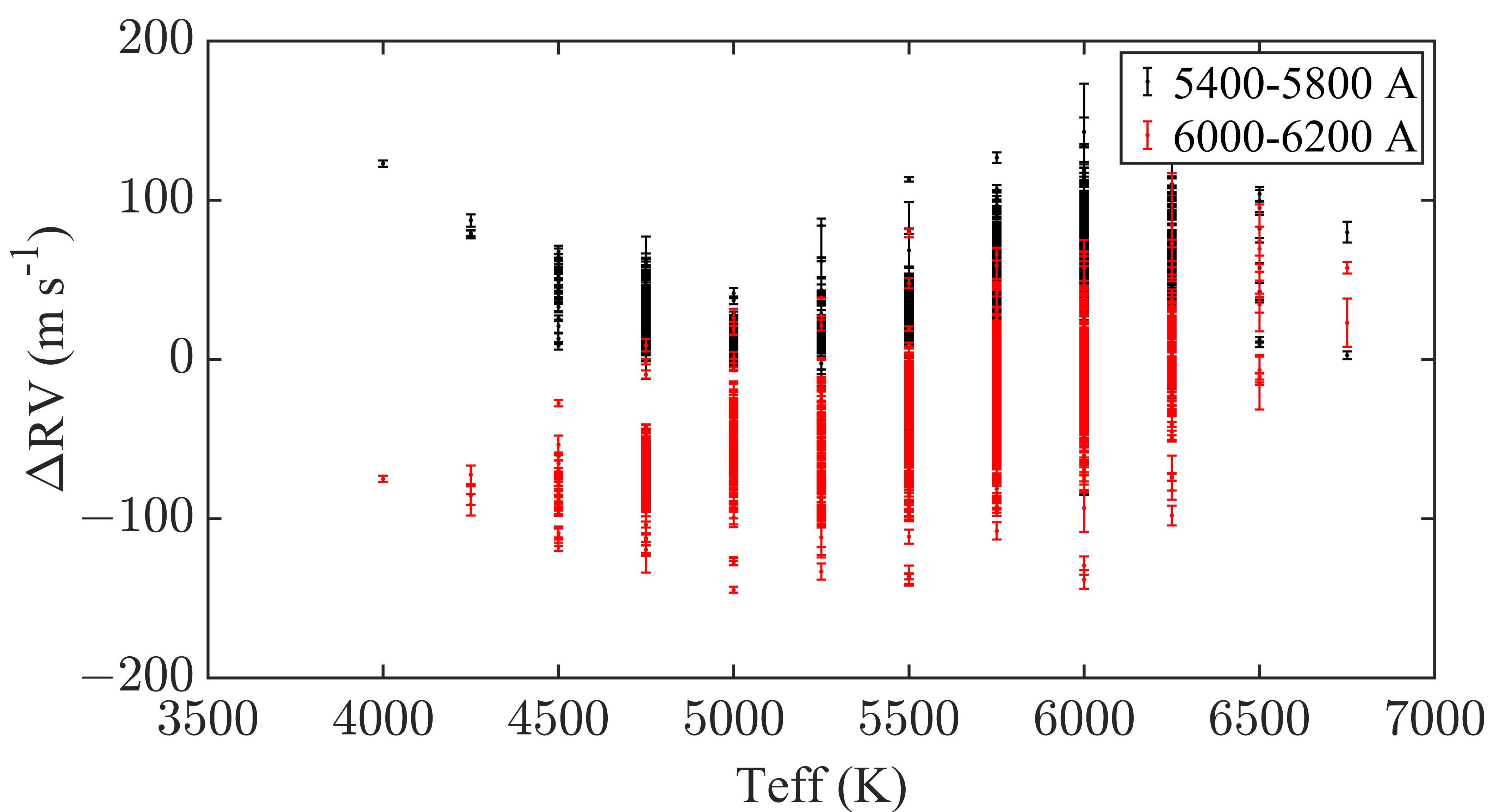}
\end{subfigure}
\end{center}
\caption{Variation of the RV offset of HARPS targets with respect to that for the wavelength domain $\lambda\lambda$\,$4400$--$4800$\ \AA \, for the four other spectral regions: $\lambda\lambda$\,$4000$--$4300$\,\AA \, and $\lambda\lambda$\,$4900$--$5200$\ \AA \, (left), and $\lambda\lambda$\,$5400$--$5800$\ \AA \, and $\lambda\lambda$\,$6000$--$6200$\ \AA \, (right). The data are shown as a function of $T_\mathrm{eff}$.}
\label{shiftDomain}
\end{figure*}

The mean shift between the RVs obtained for the reference domain $\lambda\lambda$\,$4400$--$4800$\ \AA \, and the four others is around 40 $\rm{m\, s^{-1}}$, except for $\lambda\lambda$\,$4900$--$5200$ \AA \, where it is 200 $\rm{m\, s^{-1}}$ and can occasionally reach up to 2 $\rm{km\, s^{-1}}$. We plot the shift for the four domains as a function of $T_{\mathrm{eff}}$
in Fig.~\ref{shiftDomain}. We attribute the temperature and wavelength dependencies of the shift to template mismatches, as well as to the instrumental calibration. 

\begin{figure}
\begin{center}
\includegraphics[width=\columnwidth]{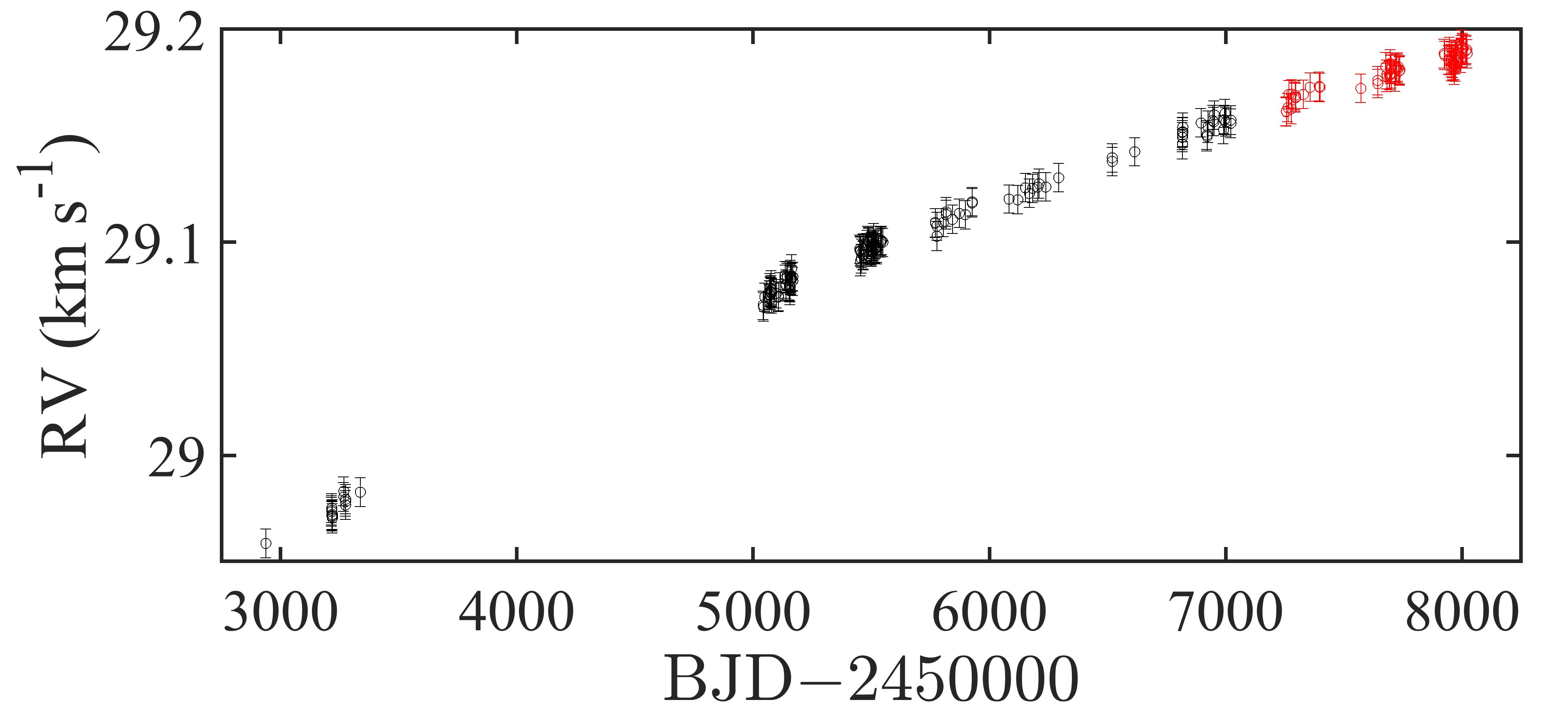}\end{center}
\caption{RV variation for HIP 1444 obtained from fitting the interval $\lambda\lambda$\,$4400-4800$\ \AA. HARPS fiber post-upgrade measurements are plotted in red. The RVs exhibit a trend variation with an amplitude of $233.6 \; \rm{m\, s^{-1}}$.}
\label{RVHIP1444}
\end{figure}

Furthermore, the HARPS fibre has been upgraded in June 2015, which caused an offset in the RV measurements compared to pre-upgrade observations. According to \cite{curto2015harps}, the RVs are correlated with the width of the spectral lines and the shift depends on the spectral type. As we have noticed that the recovery rate of known planetary periods depends on the shift, we decided to fit the shift as a function of $T_{\mathrm{eff}}$
 by maximising the number of known periods recovered (see next section). We obtain: $RV_{\rm{postUpgradeCorrected}}  \,= RV_{\rm{postUpgrade}} - 2.0\times 10^{-3}T_{\rm{eff}}(\rm{Template}) \, (\rm{K}) + 7.3 \,\rm{m\, s^{-1}}$. 
Figure~\ref{RVHIP1444} shows an example of the corrected RV time series 
in the range $\lambda\lambda$\,$4400$--$4800$\ \AA, where the HARPS post-upgrade measurements are plotted in red.

\begin{table*}
    \caption{Stellar parameters of the best-fit synthetic spectra ($T_{\mathrm{eff}}$, $\log\,g$, [Fe$\slash$H] and [$\alpha\slash$Fe]), along with Vbroad, our mean measured RV, $<\rm{RV_m}>$, and the median RV from CS18, $<\rm{RV_{CS18}}>$.
    A Vbroad value of 1.91 $\rm{km\, s^{-1}}$ is the lowest limit that can be measured by our algorithm.}
    \label{tab:APS}
    \begin{tabular}{|r|r|r|r|r|r|r|r|r|}
    \hline
    {HIP/TYC} & {$T_{\mathrm{eff}} [K]$} & {$\log\,g$ [dex]} & {[Fe$\slash$H] [dex]} & {[$\alpha\slash$Fe] [dex]} & {Vbroad [$\rm{km\, s^{-1}}$]} & {$<\rm{RV_m}>$ [$\rm{km\, s^{-1}}$]} & {$<\rm{RV_{CS18}}>$ [$\rm{km\, s^{-1}}$]} & {Instrument}\\
\hline
47 & $ 4750 $ & $ 5.0 $ & $ -0.25 $ & $ 0.1 $ & $ <1.91 \pm 1.28 $  &  $12.022 \pm 0.003 $ &  $11.752 \pm 0.001$ & H
\\
57 & $ 5500 $ & $ 5.0 $ & $ 0.00 $ & $ 0.2 $ & $ 3.52 \pm 0.50 $ & $ 37.137 \pm 0.001$ & $36.851\pm 0.0004$ & H
\\
80 & $ 5750 $ & $ 4.0 $ & $ -0.75 $ & $ 0.3 $ & $ 3.63 \pm 0.45 $&$-11.063\pm 0.001$&$-11.365\pm 0.001$ & H
\\
142 & $ 5750$ & $ 4.0 $ & $ 0.25 $ & $ -0.2 $ & $ 4.77 \pm 0.42 $ &  $-21.458\pm 0.002$ & $-21.721\pm 0.002$ & S
\\
184 & $ 5250$ & $ 4.5$ & $ 0.00$ & $ 0.0 $ & $ 1.98 \pm 1.11 $ & $-17.245\pm 0.005$ &$-17.512\pm 0.002$ & S
\\
184 & $ 5250$ & $ 4.5$ & $ 0.00$ & $ 0.0 $ & $ 3.00 \pm 1.11 $ & $-17.217\pm 0.028$ &$-17.512\pm 0.002$ & E
\\
348 & $ 6000$ & $ 4.5$ & $-0.25$ & $0.1$ & $3.52 \pm 0.67 $ & $18.877\pm 0.005$ & $ 18.587\pm 0.003$ & S
\\
400 & $ 5250$ & $ 4.5$ & $-0.50$ & $0.2$ & $<1.91 \pm 1.32$ & $7.8440\pm 0.0003$ &$  7.577\pm 0.001$ & S
\\
413 & $ 6250 $ & $ 4.5 $ & $ -0.25 $ & $ 0.1 $ & $ 5.01 \pm 0.47 $&$4.829 \pm 0.002$&$4.525 \pm 0.001$ & H
\\
436 & $ 4750 $ & $ 5.0 $ & $ -0.50 $ & $ 0.2 $ & $ <1.91 \pm 0.44 $&$40.516\pm0.001$&$40.228\pm0.0004$ & H
\\
\multicolumn{8}{l}{...}\\
\hline
    \end{tabular}
\begin{tablenotes}
      \footnotesize
      \item Note: This table is available in its entirety online at the CDS. A portion is shown here for guidance regarding its form and content.
    \end{tablenotes}
\end{table*}

\begin{figure}
\begin{center}
\includegraphics[width=\columnwidth]{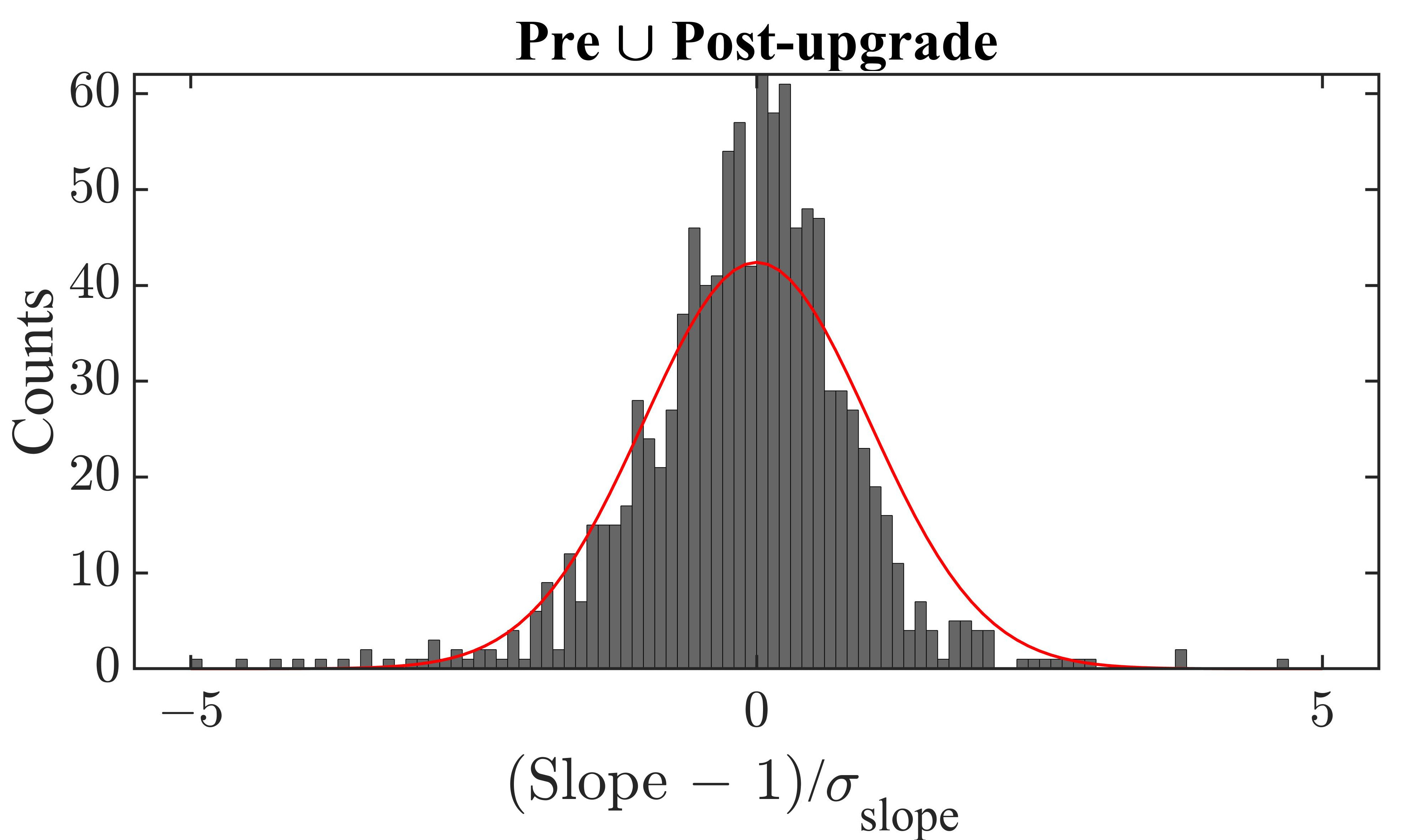}\end{center}
\caption{Distribution of the normalised and zero centred orthogonal regression slope between our measured and T20 HARPS RVs (Pre- $\cup$ Post-upgrade). The normal distribution is overplotted in red.}
\label{fig:histTrifo}
\end{figure}

\begin{figure}
\begin{center}
\begin{subfigure}{.5\textwidth}
\includegraphics[width=\columnwidth]{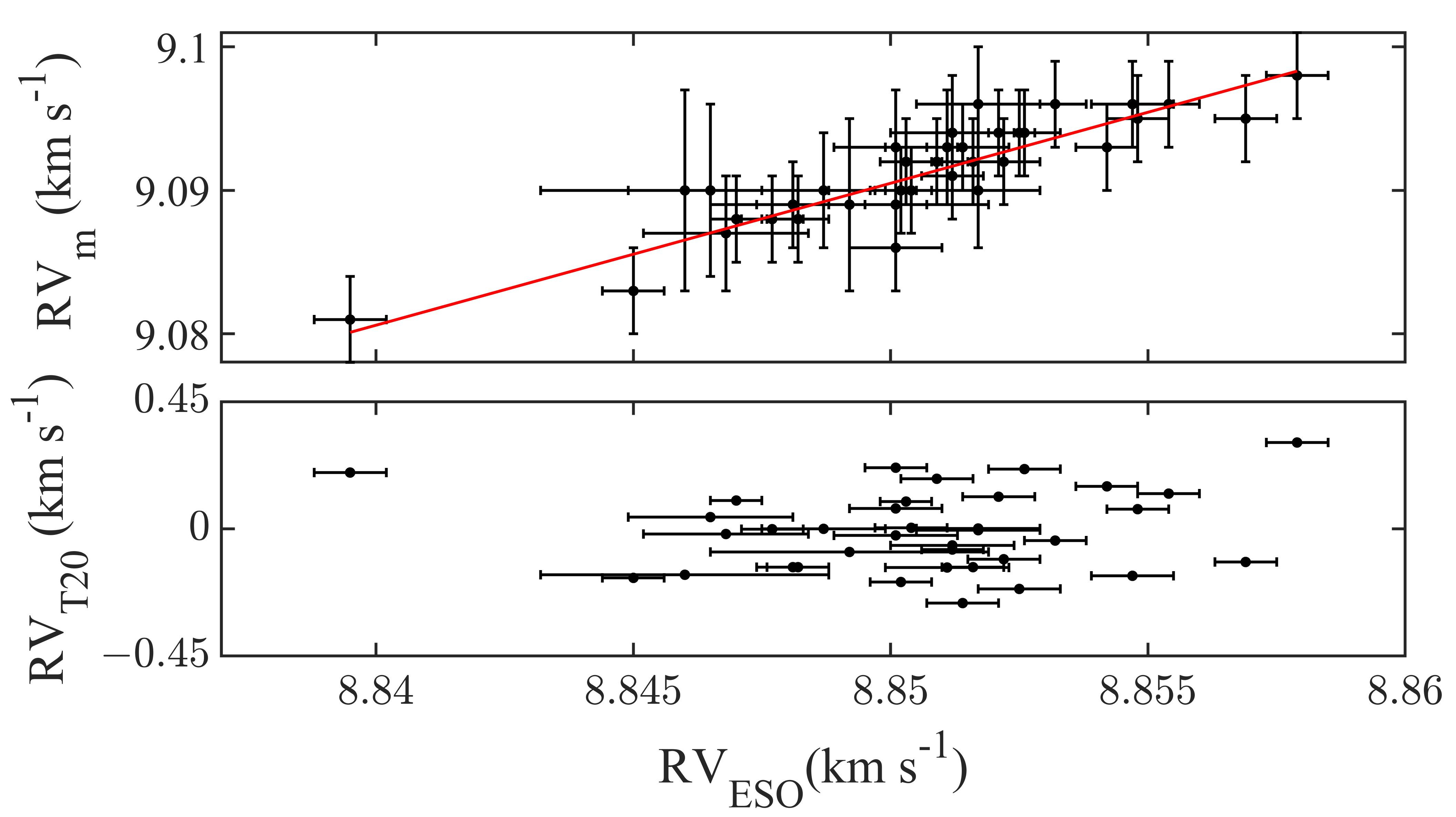}
\end{subfigure}
\end{center}
\caption{Top panel: HIP21821 measured RVs as a function of those derived by HARPS pipeline with a slope of $0.99 \pm 0.16$. Bottom panel: T20 RVs as a function of HARPS pipeline RVs with a slope of $317 \pm 113$.}
\label{fig:21821Eso_T}
\end{figure}

Applying the public SpEctrum Radial Velocity AnaLyser (SERVAL) pipeline to HARPS spectra, \cite{trifonov2020public} (hereafter T20) computed 
 RVs based on the $\chi^2$ minimisation where the template is created by shifting and co-adding all individual spectra of the target. Thus, they derived differential (relative) RV values. We fit our measured RVs to theirs using orthogonal regression \citep{krystek2007weighted} and list the slope and $y$-intercept of each target in Table \ref{tab:fitT20} for pre- and post-upgrade measurements, respectively. %\ref{tab:fitT20}. 
  These coefficients are estimated only for targets with more than three measurements. We show in Fig. \ref{fig:histTrifo} the distribution of the estimated slopes with overplotted normal distribution $\mathcal{N}$(0,1). The asymmetry is mainly due to the small uncertainties of T20 compared to ours. In addition, there are ten stars with $(Slope - 1)/\sigma_{\rm{slope}} \leqslant -5$. The lowest value is for HIP21821: its fitted spectrum is shown in Fig.  \ref{fig:21821Fit} for the domain $\lambda\lambda$\,$4400$--$4450$\ \AA. We compare our measured RVs, those derived by HARPS pipeline, and those of T20 in Fig. \ref{fig:21821Eso_T}. We find a linear relation between our RVs and those from  HARPS pipeline with a slope of $0.99 \pm 0.16$. The deviation from the 1:1 relation with respect to T20 RVs could either be due to the nightly zero-point corrections they applied to their RVs or for not taking into account the difference in RVs between different domains.

 Even though our measurements are less precise than T20's as we used portions of the spectrum, the difference in RVs between these portions is taken into consideration. In addition, these RVs are measured as absolute values with a shift around $200 \, \rm{m\, s^{-1}}$ compared to pipeline's RVs and show more homogeneity. On the other hand, T20 velocities are template free and therefore do not suffer from template mismatch problems.

\begin{table*}
    \caption{Orthogonal regression coefficients of the fit between our measured and T20 RVs.}
    \label{tab:fitT20}
    \begin{tabular}{|r|r|r|r|r|}
    \hline
    {HIP/TYC} & {Slope pre-upgrade} & {Y-intercept pre-upgrade [$\rm{km\, s^{-1}}$]} & {Slope post-upgrade} & {Y-intercept post-upgrade [$\rm{km\, s^{-1}}$]} \\
    \hline
47&$1.1 \pm 1.1$&$12.022\pm 0.003$& ... & ...\\
57&$ 0.925 \pm 0.085$&$37.138\pm 0.001$& ... & ...\\
80&$ 1.57 \pm 0.29$&$-11.063\pm 0.001$& ... & ...\\
413&$ 1.68 \pm 0.50$&$4.829\pm 0.002$& ... & ...\\
436&$ 1.08 \pm 0.71$&$40.518\pm 0.001$&$ 1.0 \pm 1.1$&$ 40.576\pm 0.046$\\
459&$ 1.27 \pm 0.17$&$7.126\pm 0.002$& ... & ...\\
490&$ 0.881 \pm 0.084$&$2.419\pm 0.004$&$ 0.87 \pm 0.12$&$2.481\pm 0.005$\\
569&$ 0.42 \pm 0.48$&$-27.367\pm 0.001$&$ 0.6 \pm 1.8$&$ -27.345\pm 0.002$\\
616&$ 0.3 \pm 1.1$&$-42.777\pm 0.001$& ... & ...\\
726&$ 0.95 \pm 0.10$&$-19.090\pm 0.001$&$ 1.01 \pm 0.24$&$ -19.092\pm 0.002$\\
\multicolumn{5}{l}{...}\\
\hline
\end{tabular}
\begin{tablenotes}
      \footnotesize
      \item Note: This table is available in its entirety online at the CDS. A portion is shown here for guidance regarding its form and content.
    \end{tablenotes}
\end{table*}

\begin{figure}
\begin{center}
\includegraphics[width=\columnwidth]{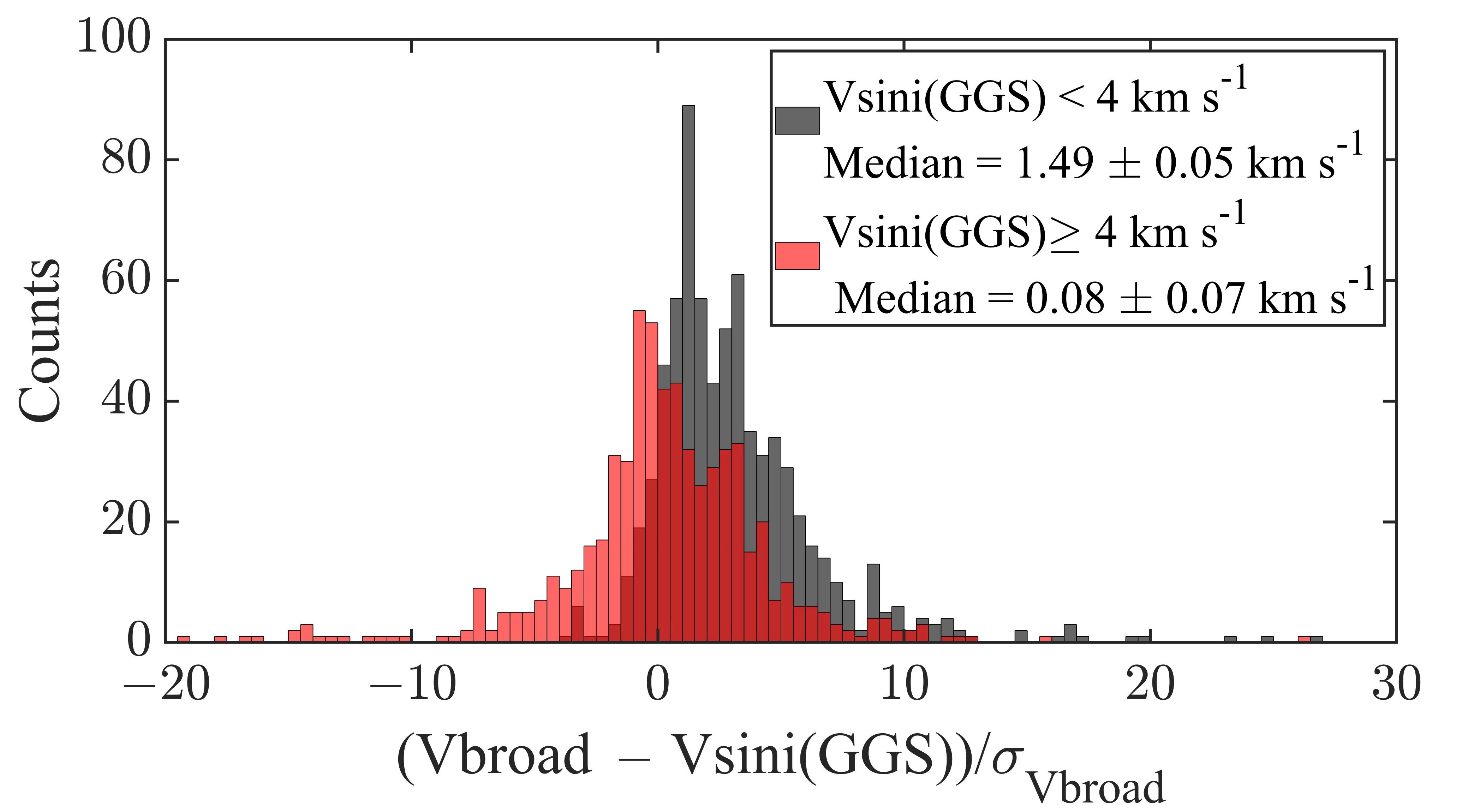}
\end{center}
\caption{Distribution of the Vbroad residuals, 
[$\rm{Vbroad}$-$\rm{V \sin i}$ (GGS)]/$\sigma_{\rm{Vbroad}}$
for 1354 stars. There are two histograms: for $\rm{V \sin i}$(GGS) $\geqslant 4 \ \rm{km \ s^{-1}}$ (red) and $\rm{V \sin i}$(GGS) $< 4 \ \rm{km \ s^{-1}}$ (black). Our measured Vbroad are overestimated with respect to GGS's for low $\rm{V \sin i}$ values, as revealed by the asymmetry.}
\label{vsiniGleb}
\end{figure}

We plot in Fig.~\ref{vsiniGleb} the histogram of the normalised residuals of Vbroad for 1354 stars obtained by comparing
our measurements and $\rm{V \sin i}$'s taken from the literature \citep{glebocki2005vizier} (hereafter GGS). The differences were divided by our uncertainties as GGS's catalogue does not provide uncertainties for all estimations. We overplot the distribution for slow and fast rotators: $\rm{V \sin i}$(GGS) below and above 4 \ $\rm{km \ s^{-1}}$. 
The precision of the measurements is expected to decrease when the true value is of the same order as or lower than the instrumental resolution ($\sim 2.6\,\rm{km \ s^{-1}}$). Therefore, the histogram exhibits a clear asymmetry at low $\rm{V \sin i}$ values
indicating that our Vbroad might be overestimated in this case.  On the other hand, we note that a large fraction of the measurements found in GGS are taken from \cite{nordstrom1997radial}, where they have been rounded off differently depending on $\rm{V \sin i}$. Some of the structures in the right wing of both distributions were found to be linked to this rounding.

\subsection{RV zero point between instruments}
As there are stars observed with several instruments (Fig.~\ref{diagVenn}), each of those having their own reduction pipeline and method for deriving the RVs, we must determine their relative RV offset. For all instruments, we adopted the HARPS RV reference frame. In Table \ref{tab:RV zero}, we compare the offset between
different instruments with respect to CS13's estimates. For all masks of a given instrument, we first estimate the median shift between measured and archived RVs. The offset is the difference between the shifts for the two instruments considered. The offset for HARPS vs SOPHIE is equal  within uncertainties to the one estimated by CS13, but it is much more precise. 
The agreement for NARVAL is less satisfactory because of the different wavelength calibration methods, although it remains within $3 \sigma$. We follow CS13 in fitting the offset between ELODIE and other instruments as a function of the $B - V$ colour index. The shift between RVs extracted from the instrument pipeline and our measurements as a function of $B - V$ for SOPHIE, ELODIE and HARPS is given by: $( 88.3 \pm 0.6 )\;(B-V) - ( 304.7 \pm 0.4 )  \rm{\; m\, s^{-1}}$, $( -175.1 \pm 2.7 )\;(B-V) - ( 213.4 \pm 1.8 ) \rm{\; m\, s^{-1}}$ and $( 66.7 \pm 0.2 )\;(B-V) - ( 271.3 \pm 0.1 ) \rm{\; m\, s^{-1}}$, respectively. The offset between ELODIE and SOPHIE as a function of the colour index is $ ( -263.4 \pm 2.7 )\;(B-V) + ( 91.4 \pm 1.8 ) \rm{\; m\, s^{-1}}$. The slope is close to the one found by CS13: $-259 \rm{\; m\, s^{-1}}$. The offset between HARPS and ELODIE as a function of $B - V$ is thus: $( 241.8 \pm 2.7 )\;(B-V) - ( 57.9 \pm 1.8 ) \rm{\; m\, s^{-1}}$. 

The offset between instruments has also been carefully taken into account in CS18, when all the measurements were put into the SOPHIE system. Since the shift between our measured and archived RVs is coherent for different instruments, we adopted the RVs obtained from $\lambda\lambda$\,$4400$--$4800$\ \AA \, as our reference in the following.
 
\begin{table}
    \caption{RV offsets between instruments (HARPS, SOPHIE and NARVAL) for all stars compared to those estimated by CS13.}
    \label{tab:RV zero}
    \centering
    \begin{tabular}{|l|l|l|}
    \hline
 & \multicolumn{2}{c}{Offset [$\rm{m\, s^{-1}}$]}\\
 {Instruments} & This work & CS13\\
     \hline
     HARPS--SOPHIE & $20.4 \pm 0.3$ & $17 \pm 5$ \\
     HARPS--NARVAL & $68.10 \pm 2.7$ & $43 \pm 7.96$ \\\hline
\end{tabular}
\end{table}

\section{Orbital solutions}
\label{orbitSol}
In order to confirm the stability of the targets with at least ten observations, we searched for high level of RV variability caused by possible companions.
 In the search for orbital parameters, we analysed the flux residuals ($S - T'$) of each fit and determined their standard deviation and normalised bias ($\rm{bias} = \rm{median}/\rm{\sigma_{median}}$). We filtered out the RVs based on these two metrics using a median absolute deviation filter. The RVs are the weighted mean value for the five wavelength ranges, where we considered the domain $\lambda\lambda$\,$4400$--$4800$\ \AA \, as the reference one and corrected the RVs from the other domains by subtracting the median of the shift. This procedure allows us to exclude outliers. For specific dates, we also omitted RV measurements that have a large drift in the pipeline wavelength calibration.

We applied the \citet*{heck1985period} periodogram (hereafter HMM), as revisited in \cite{zechmeister2009generalised}, to the RV dataset in order to perform a period search. To assign a false alarm probability (FAP) to peaks in the periodogram, we constructed an empirical cumulative probability distribution function (ECDF) from the highest peaks of $10^6$ realisations of white noise periodograms with the same time sampling. At first, we fitted the data by a trend with a maximum polynomial degree equal to three. We considered as significant peaks in the frequency periodograms, those that are above a probability threshold of $95 \%$. These are then used as an input in a Zechmeister \& K{\"u}rster Keplerian periodogram to find a first guess of the other orbital parameters. The trend is taken into account (subtracted) if the period of the highest peak in the periodogram exceeds the total time span. The period search process is repeated until there are no peak exceeding the probability threshold in the HMM periodogram. It is an iterative procedure, where the best solution is successively subtracted to search for significant periodic signals in the residuals, similarly to the method described in \cite{2001MNRAS.327..435G}. A global fit taking into account all the solutions derived during the previous steps is applied and RV residuals are then reanalysed in the next iteration. Then, we refined the solutions of all peaks above 
 the significance level simultaneously. In the case of a singular algebraic system, the fit was constrained to a circular orbit ($e$ and $\omega$ fixed to zero). The uncertainties in the orbital parameters are obtained from the variance-covariance matrix (\citealt{press1986numerical}, Sect.  15.4.1). In order to avoid the one-day period alias, we searched for periods longer than 2 days. The codes, developed in MATLAB \citep{MATLAB:2022} and Java Open JDK, are a generalisation to multiperiodic fit of those to produce the Gaia DR3 SB1 catalogue \citep{DR3-DPACP-178}. 

 We note that our approach includes some approximations. We considered all periodogram's peaks that are higher than FAP even though the test is only suitable for the highest one. In addition, a fitted curve subtraction introduces some correlations in the residuals that we have neglected by applying the same FAP to all the levels of time-series.
 
A total of 134 targets already have an orbital solution in the literature (190 systems). The known-period success rate with PASTEL template APs ($71.6\,\%$) is lower than with our pipeline best template APs ($72.6\,\%$) where the HARPS pre-/post-fiber-upgrade measurements are treated separately. 

Among the stars with an orbital solution from HARPS, six have an RV amplitude close to or exceeding the Gaia RV variability limit (\textsl{i.e.} $300 \, \rm{m\, s^{-1}}$): HIP1692, HIP26394, HIP36941, HIP62534, HIP71001, and HIP113834. However, only HIP26394, HIP71001, and HIP113834 exceed the stability level ($3\,\sigma_{RV} = 300 \, \rm{m\, s^{-1}}$) as defined in CS13. Among SOPHIE stars, HIP30057 has a scatter exceeding the CS13 stability level ($\sigma_{\rm{CS13}} = 106.7 \rm{m\, s^{-1}}$).

Estimating the total standard deviation, $\sigma_{\mathrm{Model}}$, of the orbital model (including all orbital solutions and trends), the following stars from the HARPS sample have $\sigma_{\mathrm{Model}} \ge 100 \ \rm{m\, s^{-1}}$: HIP5806 ($117.6 \, \rm{m\, s^{-1}}$), HIP26394 ($105.4 \, \rm{m\, s^{-1}}$), HIP62534 ($113.9 \, \rm{m\, s^{-1}}$), HIP71001 ($4694.1 \, \rm{m\, s^{-1}}$), HIP108095 ($159.6 \, \rm{m\, s^{-1}}$), HIP113834 ($164.7 \, \rm{m\, s^{-1}}$). For SOPHIE, the stars are HIP115714 ($117.7 \, \rm{m\, s^{-1}}$) and TYC3239-00992-1 ($105.7 \, \rm{m\, s^{-1}}$). The scatter of the model for HIP30057 ($96.6 \, \rm{m\, s^{-1}}$) does not exceed the threshold. The orbital solutions of these Keplerian signals are listed in Table \ref{table:orbit} where we were able to recover the known periods for five stars. 

\begin{figure}
\includegraphics[width=\columnwidth]{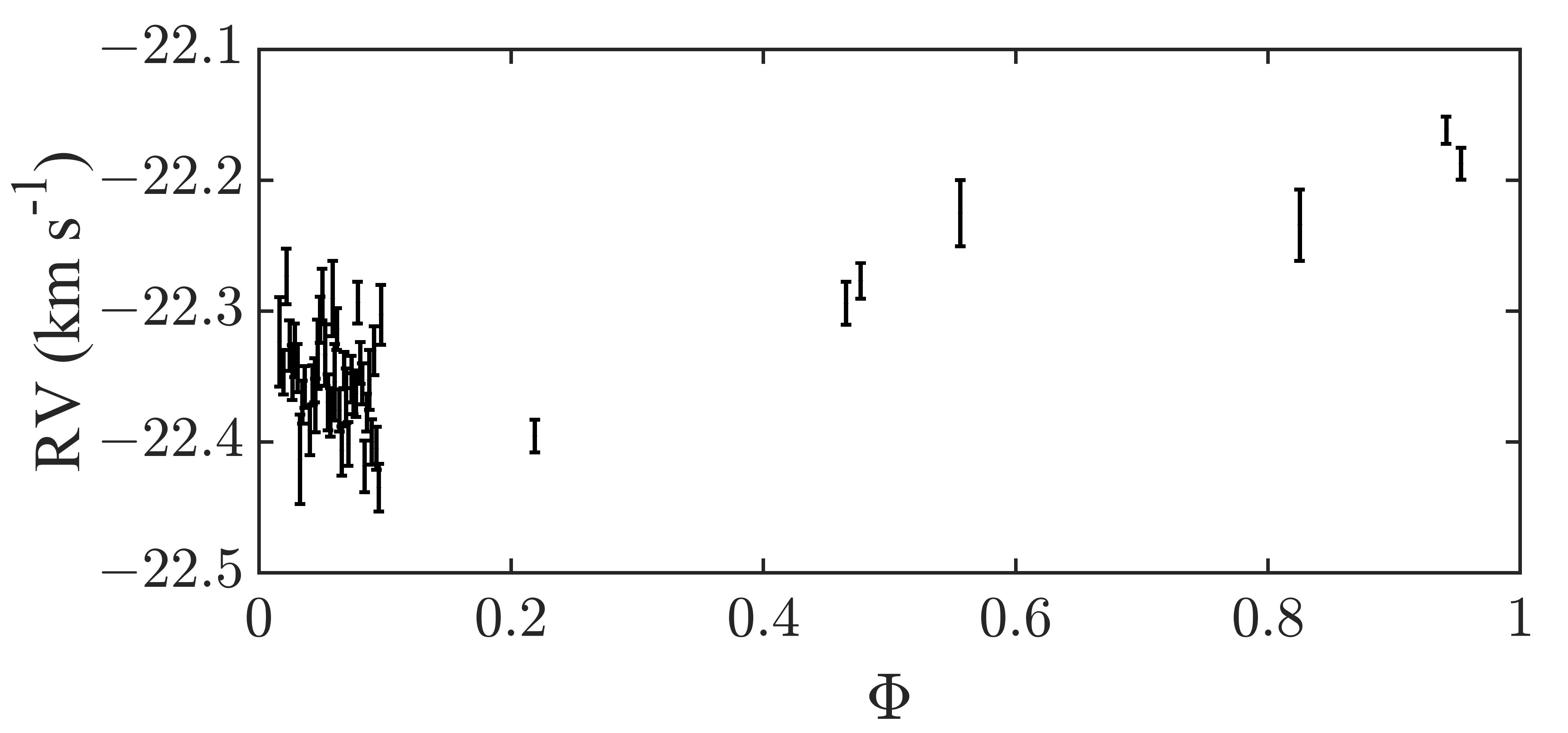}
\caption{TYC3239-00992-1 phase-folded RV curve with the known period of $3.852985$ d. The inhomogeneous phase sampling prevented us to derive this period.}
\label{fig:TYC992}
\end{figure}

However, the estimated period of TYC3239-00992-1 does not match within $5 \sigma$  the known transit period of $3.852985 \pm 0.000005$ d \citep{noyes2008hat}. We plot the phase-folded RV curve assuming this period in Fig. \ref{fig:TYC992}. We were not able to retrieve this period because of the bad sampling and missing data points.  

\begin{figure}
\includegraphics[width=\columnwidth]{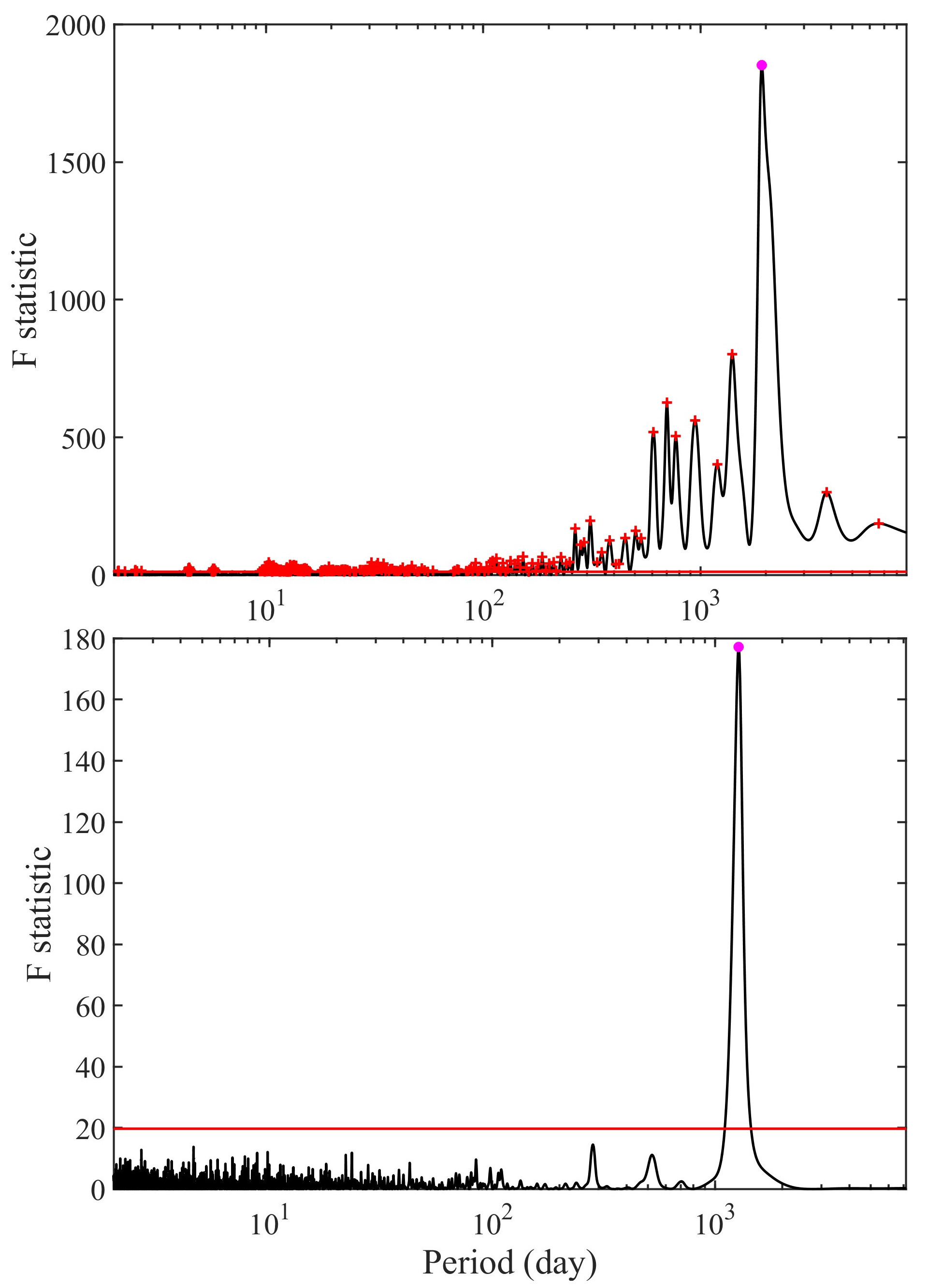}
\caption{Periodograms for HIP26394 (top) and HIP113834 (bottom). The red horizontal lines denote the adopted confidence threshold of 95\%. Peaks exceeding that threshold are marked by red crosses, while the highest peak is marked in magenta.}
\label{periodogram}
\end{figure}

\begin{figure}
\includegraphics[width=\columnwidth]{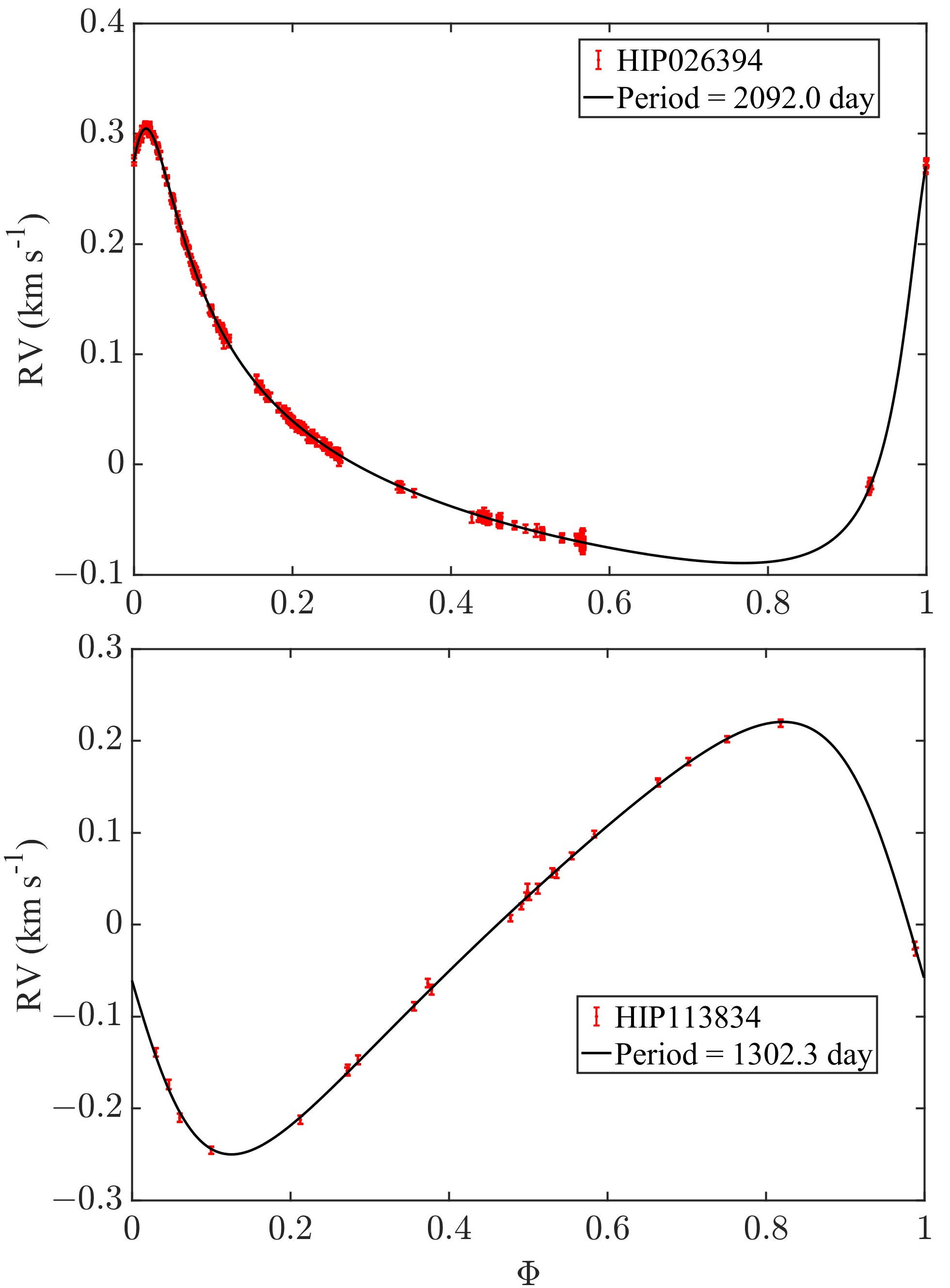}
\caption{Orbital solutions with semi-amplitude RV exceeding Gaia threshold for HIP26394 (top) and HIP113834 (bottom) with the periods of $2092.0 \pm 1.7$ d and $1302.3 \pm 1.5$ d respectively.}
\label{fig:orbit}
\end{figure}

We show HIP26394 and HIP113834 periodograms and orbital solutions in Figs. \ref{periodogram} and \ref{fig:orbit} respectively. The $F$ statistic denotes the ratio of two independent $\chi^2$ values ($F$ statistic of the pure sine model with respect to the variance), defined as $z$ variable from Eq. 23 of \cite{zechmeister2009generalised}.

In addition to the known period of HIP5806 we find two new ones which need to be confirmed. The short period of HIP108095 has not been detected previously and thus it could indicate the presence of a new substellar companion in addition to a second degree trend caused by a more massive companion. For these two stars, $\sigma_{\rm{Model}}$ is dominated by the trend scatter. The periodograms and phase-folded RV curves for these newly discovered periods are shown in Figs. \ref{fig:periodogramNew} and \ref{fig:orbitNew}, respectively. The long period of HIP71001 has a high uncertainty and its RV semi-amplitude is lower than its uncertainty thus we are cautious about this solution.

\begin{table*}
    \caption{Orbital solutions of stars with RV model scatter exceeding Gaia threshold. The solutions are computed with respect to (2450000 + 100 ${E}$) BJD epochs, where ${E}$, the reduced reference epoch, is given in column 2. Periods in bold are those recovered. $N$ denotes the number of data points for each target.}
    \label{table:orbit}
    \begin{tabular}{|l|r|r|r|r|r|r|r|r|r|}
    \hline
    {HIP/TYC} & {$E$} & {$P$ [d]} & {$K$ [$\rm{m\, s^{-1}}$]} & {$e$} & {$\omega$ [deg]} & {$T_{\rm{p}}$ [d]} & {$M_{\rm{min}} [M_{\rm{Jup}}]$} & {Instrument} & {$N$}
     \\
\hline
\, \, 5806$^a$  & 53  & $ \mathbf{1238 \pm 14} $ & $ 57 \pm 457 $ & $ 0.85  \pm 0.94 $ & $  -18  \pm  73 $ & $  1184  \pm  55 $ & $ 1.572 \pm 13.395$ & H & 131\\ 
& & $ 806 \pm 18 $ & $ 6.1  \pm 1.1  $ & $ 0.785 \pm 0.063 $ & $  135  \pm  14 $ & $  516.8  \pm  7.7  $ & $ 0.172 \pm 0.038$ \\ 
& & $ 51.154 \pm 0.047 $ & $ 2.97 \pm 0.39 $ & $ 0.21 \pm 0.13 $ & $  -54  \pm  38 $ & $  19.6  \pm  5.0  $ & $ 0.053 \pm 0.007$ \\ 
\, 26394$^b$ & 77 & $ \mathbf{2092.0 \pm 1.7} $ & $ 197.02 \pm 0.66 $ & $ 0.645 \pm 0.003 $ & $  -32.00  \pm  0.43 $ & $  686.35   \pm  0.57 $ & $ 9.222 \pm 0.207$ & H & 487\\
\, 62534$^c$ & 61 & $ \mathbf{1046.2 \pm 1.8} $ & $ 142.9 \pm 2.5 $ & $ 0.221 \pm 0.008 $ & $  77.3   \pm  1.8 $ & $  907.8  \pm  4.5 $ & $6.033 \pm 0.210$ & H & 66\\ 
\, 71001 & 51 & $5106 \pm 1222$ & $723 \pm 1044$ & $0.080 \pm 0.026$ &$99.3 \pm 4.7$ & $2878 \pm 659$ & $ 53 \pm 76$ & H & 29\\
108095 & 51 & $ 8.53127 \pm 0.00024 $ & $ 33.6 \pm 2.8 $ & $ 0.524 \pm 0.057 $  & $  -46.1  \pm  5.8 $ & $  1.287 \pm  0.079 $ & $ 0.261 \pm 0.025$ & H & 34\\
113834$^d$ & 55 & $ \mathbf{1302.3  \pm 1.5} $ & $ 235.3  \pm 2.2 $ & $ 0.318 \pm 0.008 $ & $  101.3   \pm  1.1  $ & $  539.5  \pm  3.6  $ & $ 11.596 \pm 0.247$ & H & 28\\ 
115714$^e$ & 75 & $ \mathbf{218.473 \pm 0.074} $ & $ 106.6 \pm 1.4 $ & $ 0.429 \pm 0.010 $ & $  -136.7 \pm  1.4 $ & $  214.31 \pm  0.70 $ & $ 2.867 \pm 0.290$ & S & 99\\ 
3239-00992-1 & 54 & $ 3.09410 \pm 0.00021 $ & $153 \pm 58$ & $0.23 \pm 0.22$ & $-63 \pm 26$ & $0.91 \pm 0.18$ & $ 1.208 \pm 0.463$ & S & 49\\
\hline
\end{tabular}
\begin{tablenotes}\footnotesize
\item References for literature values.
\item a: \cite{wittenmyer2019truly}
\item b: \cite{venner2021true}
\item c: \cite{stassun2017accurate}
\item d: \cite{moutou2011harps}
\item e: \cite{hebrard2016sophie}
\end{tablenotes}
\end{table*}

\begin{figure}
\includegraphics[width=\columnwidth]{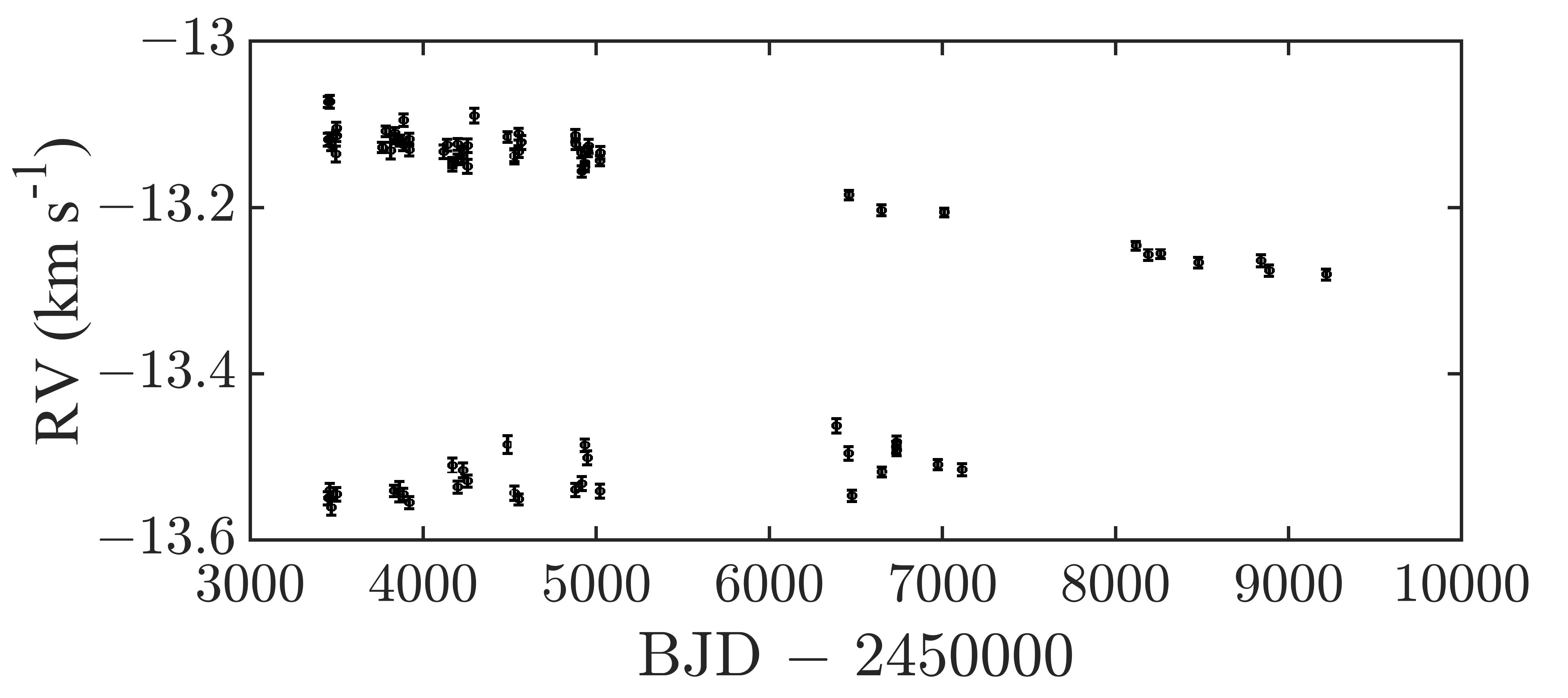}
\caption{RV variation of TYC8963-01543-1 similar to a trend SB2 where we only measured one RV component.}
\label{fig:TYC1534}
\end{figure}

Additionally, TYC8963-01543-1 does not have a significant period nor trend, but it has $\sigma_{\rm{CS13}} = 176.6\, \rm{m\, s^{-1}}$. Its RV curve (Fig. \ref{fig:TYC1534}) suggests the star to be an SB2. We will analyse it in a future work.

\section{Analysis of stars with RV trend }
\label{trend}
Some of the stars with long periods exceeding the observation duration have RV variations with an incomplete orbit or exhibit a polynomial trend. We list those with a mean acceleration (RV amplitude, $A$, divided by the time span, $T$) greater than $10 \,\rm{m\, s^{-1}yr^{-1}}$, along with their mean acceleration and trend degree in columns 3 and 4 of Table~\ref{tabe:TrendAct} for HARPS targets. There are four stars with previously detected Keplerian signal(s) for which we find an additional signal manifesting as a trend with an acceleration exceeding our adopted threshold:  
 HIP5806, HIP10278, HIP60689, and HIP89354. These trends indicate the presence of an additional companion with a long period. However, this is not necessarily always the case as such variations may have other origins (see Sect.~\ref{activi}).

There are stars with a very high acceleration value ($\frac{A}{T} > 25 \,\rm{m\, s^{-1}yr^{-1}}$) that could exceed the variability threshold that was set by the Gaia mission (300 $\rm{m\, s^{-1}}$ in ten years): HIP10278, HIP14596, HIP37233, HIP60689, HIP71001, HIP81533, HIP85158, HIP95849, HIP99385, HIP108095, HIP110156, HIP116374, TYC6792-01746-1, and TYC8681-00611-1. Adopting the constancy criterion of CS13, the following stars are variable: HIP12350, HIP14596, HIP37233, HIP71001, HIP99385, HIP108095, HIP110156, HIP116374, and TYC8681-00611-1; they should be reconsidered in the upcoming Gaia data release.

\begin{table*}
%\tiny
    \caption{Data for HARPS targets showing a significant RV trend ($\frac{A}{T} > 10\,\rm{m\, s^{-1}yr^{-1}}$): RV amplitude, acceleration value, order of polynomial fit, cross-correlation coefficients, $\rho$, and significance levels, $p$, for the RV vs $BIS$ and RV vs $S$-index variations, and chromospheric activity indicators. Stars with an asterisk have a Keplerian solution in addition to the trend, while double asterisks denote stars with confirmed companions. Blanks in the log$R'_{\rm{HK}}$ column indicate stars that are not present in Boro Saikia's catalogue.}
    \label{tabe:TrendAct}
    \begin{tabular}{|l|r|r|c|r|r|r|r|r|r|}
    \hline
     {HIP/TYC} & {$A$ [$\rm{m\, s^{-1}}$]} & {$\frac{A}{T} [\rm{m\, s^{-1}yr^{-1}}]$} & {Degree} & {$\rho_{\rm{BIS}}$} & {$\rm{pValue_{BIS}}$} & {$\rho_{S}$} & {$\rm{pValue_{S}}$}  & {log$R'_{\rm{HK}}$} & {RV jitter [$\rm{m\, s^{-1}}$]}\\
\hline
\, \, \, 948 & $ 186.4$ & $ 11.7 $ & $ 1$ & $ 0.110 $ & $ 0.594$ & $ -0.315 $ & $ 0.117 $& $ -5.171 $& $ 7.411 $\\
\, \, 1444* & $ 229.2$ & $ 16.5 $ & $ 2$ & $ -0.017 $ & $ 0.812$ & $ -0.373 $ & $ 0.000 $& $ -4.881 $& $ 9.185 $\\
\, \, 5806** & $ 160.7$ & $ 13.2 $ & $ 2$ & $ -0.221 $ & $ 0.010$ & $ -0.120 $ & $ 0.169 $& $ -4.724 $& $ 14.810$\\
\, 10278** & $ 223.3$ & $ 38.2 $ & $ 1$ & $ 0.353 $ & $ 0.066$ & $ 0.603 $ & $ 0.001 $ & $ ...$ & $ ...$\\ 
\, 12350* & $ 347.1$ & $ 21.3 $ & $ 2$ & $ 0.095 $ & $ 0.531$ & $ 0.008 $ & $ 0.960 $& $ -4.937 $& $ 8.551 $\\
\, 14596 & $ 434.1$ & $ 27.4 $ & $ 2$ & $ -0.074 $ & $ 0.524$ & $ 0.167 $ & $ 0.156 $& $ -4.819 $& $ 9.938 $\\
\, 19007 & $ 160.0$ & $ 17.0 $ & $ 2$ & $ 0.248 $ & $ 0.254$ & $ 0.163 $ & $ 0.457 $& $ -5.200 $& $ 6.134 $\\
\, 35305 & $ 90.0$ & $ 23.0 $ & $ 1$ & $ -0.124 $ & $ 0.465$ & $ -0.115 $ & $ 0.506 $ & $ ...$ & $ ...$\\ 
\, 37233 & $ 662.8$ & $ 41.3 $ & $ 3$ & $ 0.056 $ & $ 0.671$ & $ 0.363 $ & $ 0.005 $ & $ ...$ & $ ...$\\ 
\, 43449 & $ 35.6$ & $ 17.0 $ & $ 1$ & $ -0.145 $ & $ 0.622$ & $ 0.400 $ & $ 0.156 $ & $ ...$ & $ ...$\\ 
\, 45749 & $ 24.7$ & $ 17.2 $ & $ 1$ & $ 0.617 $ & $ 0.004$ & $ 0.682 $ & $ 0.001 $& $ -4.710 $& $ 8.506 $\\
\, 54102* & $ 306.6$ & $ 23.6 $ & $ 3$ & $ -0.280 $ & $ 0.088$ & $ -0.482 $ & $ 0.002 $ & $ ...$ & $ ...$\\ 
\, 60689** & $ 7.5$ & $ 49.1 $ & $ 1$ & $ 0.158 $ & $ 0.225$ & $ 0.540 $ & $ 0.004 $ & $ ...$ & $ ...$\\ 
\, 61743 & $ 150.3$ & $ 10.0 $ & $ 3$ & $ -0.041 $ & $ 0.824$ & $ -0.011 $ & $ 0.955 $ & $ ...$ & $ ...$\\ 
\, 70610 & $ 208.0$ & $ 14.8 $ & $ 1$ & $ 0.180 $ & $ 0.576$ & $ 0.489 $ & $ 0.127 $& $ -5.023 $& $ 7.745 $\\
\, 71001* & $ 3646.1$ & $ 246.2 $ & $ 3$ & $ -0.167 $ & $ 0.386$ & $ 0.230 $ & $ 0.239 $ & $ ...$ & $ ...$\\ 
\, 71803 & $ 161.6$ & $ 12.2 $ & $ 1$ & $ 0.338 $ & $ 0.044$ & $ 0.371 $ & $ 0.031 $& $ -4.954 $& $ 8.377 $\\
\, 76713 & $ 168.6$ & $ 15.4 $ & $ 1$ & $ -0.065 $ & $ 0.762$ & $ 0.083 $ & $ 0.707 $ & $ ...$ & $ ...$\\ 
\, 78395 & $ 301.1$ & $ 20.0 $ & $ 2$ & $ -0.446 $ & $ 0.015$ & $ 0.222 $ & $ 0.257 $ & $ ...$ & $ ...$\\ 
\, 81533 & $ 238.6$ & $ 39.7 $ & $ 1$ & $ -0.742 $ & $ 0.000$ & $ -0.743 $ & $ 0.000 $ & $ ...$ & $ ...$\\ 
\, 85158 & $ 225.2$ & $ 28.8 $ & $ 3$ & $ 0.156 $ & $ 0.409$ & $ 0.013 $ & $ 0.948 $& $ -4.831 $& $ 12.320$\\
\, 89354** & $ 132.3$ & $ 18.0 $ & $ 3$ & $ 0.190 $ & $ 0.165$ & $ 0.214 $ & $ 0.117 $ & $ ...$ & $ ...$\\ 
\, 95849* & $ 26.1$ & $ 25.7 $ & $ 2$ & $ -0.818 $ & $ 0.000$ & $ 0.828 $ & $ 0.000 $& $ -4.456 $& $ 15.736$\\
\, 99385 & $ 925.1$ & $ 69.5 $ & $ 2$ & $ -0.176 $ & $ 0.345$ & $ 0.303 $ & $ 0.141 $& $ -4.401 $& $ 9.333 $\\
108095* & $ 937.3$ & $ 74.7 $ & $ 2$ & $ -0.155 $ & $ 0.380$ & $ 0.489 $ & $ 0.003 $ & $ ...$ & $ ...$\\ 
109787 & $ 237.7$ & $ 15.0 $ & $ 2$ & $ 0.002 $ & $ 0.992$ & $ 0.174 $ & $ 0.284 $ & $ ...$ & $ ...$\\ 
110156 & $ 500.4$ & $ 31.5 $ & $ 2$ & $ -0.257 $ & $ 0.215$ & $ -0.103 $ & $ 0.640 $& $ -4.858 $& $ 8.139 $\\
110843 & $ 29.8$ & $ 11.7 $ & $ 1$ & $ 0.667 $ & $ 0.000$ & $ 0.228 $ & $ 0.283 $& $ -4.935 $& $ 8.573 $\\
112229 & $ 70.4$ & $ 15.1 $ & $ 1$ & $ 0.526 $ & $ 0.001$ & $ 0.197 $ & $ 0.241 $ & $ ...$ & $ ...$\\ 
115951* & $ 104.7$ & $ 10.6 $ & $ 1$ & $ -0.144 $ & $ 0.238$ & $ 0.333 $ & $ 0.005 $& $ -4.932 $& $ 8.606 $\\
116374 & $ 902.2$ & $ 69.1 $ & $ 3$ & $ 0.784 $ & $ 0.000$ & $ -0.095 $ & $ 0.682 $& $ -4.761 $& $ 8.377 $\\
6792-01746-1 & $ 262.2$ & $ 25.6 $ & $ 1$ & $ 0.396 $ & $ 0.014$ & $ -0.407 $ & $ 0.010 $ & $ ...$ & $ ...$\\ 
8681-00611-1 & $ 502.1$ & $ 31.6 $ & $ 1$ & $ -0.165 $ & $ 0.344$ & $ 0.107 $ & $ 0.541 $ & $ ...$ & $ ...$\\ 
\hline
\end{tabular}
\label{TABLE6}
\end{table*}

In the case of \textbf{\rm{HIP10278}}, \cite{rickman2020spectral} combined CORALIE and HARPS measurements to find $P = 123 \pm 41$ years. This period is too long to be detected with HARPS measurements, and accordingly we only find a first degree trend as the best fit to the data.

Similarly, we list in Table \ref{table:TrendActSophie} the stars observed with SOPHIE and ELODIE, having an acceleration greater than the threshold: $10 \,\rm{m\, s^{-1}yr^{-1}}$. Thirteen from SOPHIE exceed $25 \,\rm{m\, s^{-1}yr^{-1}}$.  
Among these stars, HIP26037, HIP29611, HIP71291, and  HIP76751 have an amplitude greater than the threshold for stability as defined in CS13. Among ELODIE stars, the following have an acceleration exceeding $25 \,\rm{m\, s^{-1}yr^{-1}}$: HIP24205 ($34.0 \,\rm{m\, s^{-1}yr^{-1}}$) and HIP61157 ($79.5 \,\rm{m\, s^{-1}yr^{-1}}$). The trend detected in ELODIE HIP24205 data is a portion of the full orbit of the known period \citep[2117.3 $\pm$ 0.8 d;][]{bean2007mass}.

\begin{table*}
%\tiny
    \caption{Same as Table \ref{tabe:TrendAct}, but for stars observed with SOPHIE and ELODIE.}
    \label{table:TrendActSophie}
    \begin{tabular}{|l|r|r|c|r|r|r|r|r|r|}
    \hline
     {HIP/TYC} & {$A$ [$\rm{m\, s^{-1}}$]} & {$\frac{A}{T} [\rm{m\, s^{-1}yr^{-1}}]$} & {Degree} & {$\rho_{\rm{BIS}}$} & {$\rm{pValue_{BIS}}$} & {$\rho_{S}$} & {$\rm{pValue_{S}}$}  & {log$R'_{\rm{HK}}$} & {RV jitter [$\rm{m\, s^{-1}}$]}\\
\hline
SOPHIE \\
\hline
\, 23209*& $      130.3 $ & $       13.0 $ & $  1 $ &   $ ...      $  &  $ ...$  &  $ -0.241 $  &  $ 0.217                         $ &  $ ...      $  &  $ ...$\\
\, 26037& $     1250.3 $ & $       90.0 $ & $  2 $ &   $ -0.134 $  &  $ 0.622$  &  $ -0.602 $  &  $ 0.006$  &  $ -4.909$  &  $ 8.864  $\\
\, 29611& $      402.3 $ & $       45.3 $ & $  1 $ &   $ -0.206 $  &  $ 0.227$  &  $ -0.073 $  &  $ 0.683                         $ &  $ ...      $  &  $ ...$\\
\, 31849& $      275.7 $ & $       46.6 $ & $  1 $ &   $ -0.895 $  &  $ 0.000$  &  $ 0.459 $  &  $ 0.014                          $ &  $ ...      $  &  $ ...$\\
\, 32906& $      167.3 $ & $       41.2 $ & $  2 $ &   $ 0.261 $  &  $ 0.367$  &  $ -0.699 $  &  $ 0.003                          $ &  $ ...      $  &  $ ...$\\
\, 35198& $       59.8 $ & $       63.7 $ & $  2 $ &   $ -0.052 $  &  $ 0.842$  &  $ -0.184 $  &  $ 0.494                         $ &  $ ...      $  &  $ ...$\\
\, 44324& $      216.7 $ & $       22.2 $ & $  1 $ &   $ ...      $  &  $ ...$  &  $ 0.133 $  &  $ 0.587                          $ &  $ ...      $  &  $ ...$\\
\, 61157& $      223.4 $ & $       27.6 $ & $  1 $ &   $ 0.114 $  &  $ 0.605$  &  $ -0.047 $  &  $ 0.832                          $ &  $ ...      $  &  $ ...$\\
\, 63346& $      192.4 $ & $       22.4 $ & $  2 $ &   $ -0.270 $  &  $ 0.058$  &  $ -0.282 $  &  $ 0.047                         $ &  $ ...      $  &  $ ...$\\
\, 68134& $      385.8 $ & $       31.5 $ & $  1 $ &   $ ...      $  &  $ ...$  &  $ 0.305 $  &  $ 0.219                          $ &  $ ...      $  &  $ ...$\\
\, 71291& $      779.7 $ & $       66.8 $ & $  1 $ &   $ 0.527 $  &  $ 0.064$  &  $ -0.398 $  &  $ 0.060                          $ &  $ ...      $  &  $ ...$\\
\, 74702& $       68.4 $ & $       39.3 $ & $  3 $ &   $ -0.846 $  &  $ 0.000$  &  $ 0.567 $  &  $ 0.011$  &  $ -4.457$  &  $ 9.177   $\\
\, 76751& $916.5$ & $81.2$ & $2$ & $-0.195$ & $0.544$ & $0.040$ & $0.908$ &  $ ...      $  &  $ ...$\\
\, 77718*& $      110.4 $ & $       22.1 $ & $  1 $ &   $ 0.043 $  &  $ 0.765$  &  $ -0.024 $  &  $ 0.881                          $ &  $ ...      $  &  $ ...$\\
\, 85653& $      130.2 $ & $       11.6 $ & $  1 $ &   $ 0.406 $  &  $ 0.026$  &  $ 0.206 $  &  $ 0.242$  &  $ -4.928$  &  $ 8.654    $\\
\, 91658& $       31.6 $ & $       31.8 $ & $  1 $ &   $ -0.312 $  &  $ 0.106$  &  $ -0.029 $  &  $ 0.884                         $ &  $ ...      $  &  $ ...$\\
\, 96184*& $      120.3 $ & $       11.6 $ & $  1 $ &   $ ...      $  &  $ ...$  &  $ -0.343 $  &  $ 0.017                         $ &  $ ...      $  &  $ ...$\\
\, 98828& $       37.0 $ & $       10.1 $ & $  2 $ &   $ -0.097 $  &  $ 0.720$  &  $ 0.827 $  &  $ 0.003$  &  $ -4.707$  &  $ 8.515   $\\
104587*& $      110.2 $ & $       12.0 $ & $  1 $ &   $ 0.843 $  &  $ 0.000$  &  $ 0.310 $  &  $ 0.004$  &  $ -4.994$  &  $ 7.814    $\\
111977& $       91.0 $ & $       10.1 $ & $  3 $ &   $ ...      $  &  $ ...$  &  $ -0.033 $  &  $ 0.899                         $ &  $ ...      $  &  $ ...$\\
113086& $      222.6 $ & $       19.9 $ & $  1 $ &   $ 0.261 $  &  $ 0.367$  &  $ 0.239 $  &  $ 0.325$  &  $ -4.815$  &  $ 12.665   $\\
114210& $       77.3 $ & $       12.6 $ & $  2 $ &   $ 0.190 $  &  $ 0.481$  &  $ 0.125 $  &  $ 0.589$  &  $ -4.735$  &  $ 14.5387  $\\
115714** & $    321.5$ & $     39.7$ & $ 1$ & $ 0.197 $ & $ 0.152$ & $ -0.213 $ & $ 0.033$& $ ...      $ & $ ...$\\
1991-01087-1*& $       27.8 $ & $       25.7 $ & $  1 $ &   $ -0.228 $  &  $ 0.075$  &  $ -0.313 $  &  $ 0.059                    $ &  $ ...      $  &  $ ...$\\
\hline
ELODIE\\
\hline
\, 19422&     $156.2$&$ 17.1$&$1$&  $ ...      $  &  $ ...$&  $ ...      $  &  $ ...$&  $ ...      $  &  $ ...$\\
\, 24205**&     $202.0$&$34.0$&$3$&  $ ...      $  &  $ ...$&  $ ...      $  &  $ ...$&  $ ...      $  &  $ ...$\\
\, 61157&     $191.8$&$79.5$&$1$&  $ ...      $  &  $ ...$&  $ ...      $  &  $ ...$&  $ ...      $  &  $ ...$\\
\, 77718&     $120.3$&$14.8$&$1$&  $ ...      $  &  $ ...$&  $ ...      $  &  $ ...$&  $ ...      $  &  $ ...$\\
\, 80264&     $ 96.3$&$14.0$&$3$&  $ ...      $  &  $ ...$&  $ ...      $  &  $ ...$&  $ ...      $  &  $ ...$\\
\, 80902**&     $110.6$&$12.4$& $2$&  $ ...      $  &  $ ...$&  $ ...      $  &  $ ...$&  $ ...      $  &  $ ...$\\ 
\, 85653&     $144.6$&$16.1$&$1$&  $ ...      $  &  $ ...$&  $ ...      $  &  $ ...$&  $ ...      $  &  $ ...$\\
108602&     $128.5$&$17.6$&$1$&  $ ...      $  &  $ ...$&  $ ...      $  &  $ ...$&  $ ...      $  &  $ ...$\\
\hline
\end{tabular}
\end{table*}
In order to be consistent with the $\sigma_{\rm{Model}}$ of stars with orbital solution, we also estimate the standard deviation of the trend model using Eq.\,\ref{Eq:trendSTD}. 

The considered stars should have a time span longer than 10 years. We list those with $\sigma_{\rm{Model}} \ge 100 \,\rm{m\, s^{-1}}$ over 10 and 12 years 
in Table\,\ref{tabe:sigmaModel}. We also list those with a scatter greater than $\ge 100 \,\rm{m\, s^{-1}}$ over a time span shorter than 10 years. 
\begin{table}
%\tiny
    \caption{Trend stars with model standard deviation exceeding  $100 \ \rm{m\, s^{-1}}$ in 10 years, 12 years, and during the time span. Asterisk symbols have the same meaning as in Table \ref{TABLE6}.}
    \label{tabe:sigmaModel}
    \begin{tabular}{|l|r|c|r|r|r|}
    \hline
       {HIP/TYC} & {$T$ [yr]} & {$\sigma_{10}$ $[\rm{m\, s^{-1}}]$} & {$\sigma_{12}$ $[\rm{m\, s^{-1}}]$} & {$\sigma_{{T}}$ $[\rm{m\, s^{-1}}]$} \\
\hline
HARPS \\
\hline
\, \, 5806** & $ 12.2$& ... & $ 113.5$ & $116.4$\\
\, 37233 & $ 16.1 $& $194.4 $ & $ 308.5$ & $637.4$\\ 
\, 64295** & $ 6.3$ & ... & ... & $ 103.2$ \\
\, 71001* & $14.8  $  & $ 1175.0 $ & $ 2358.4$ & $4950.0$\\ 
\, 99385 & $13.3  $ & $221.1 $ & $270.8$ &$304.2$\\
108095* & $ 12.6 $  & $ 144.8 $ & $ 156.1$ & $158.3$\\ 
116374 & $  13.0 $  & $ 191.3 $ & $ 192.6$ & $187.6$\\
8681-00611-1 & $ 15.9 $  & ... & $ 109.4$ &$144.9$\\ 
\hline
SOPHIE \\
\hline
\, 26037& $ 13.9     $  &   $809.4 $  &  $ 1056.5$ &$1316.7$\\
\, 29611& $   8.9    $  &   ... &  ...&$116.1$  \\
\, 32906& $   4.1    $  &  ...  & ... &$214.8$  \\
\, 68134& $  12.3    $  &   $ ...      $  &  $ 109.1$&$111.4$ \\
\, 71291& $  11.7 $  &   $ 192.9 $  &  $ 231.5$ &$225.1$ \\
\, 74702& $ 1.7  $  &   ...  & ... &$196.9$  \\
\, 76751& $ 11.3$  & $346.8$ & $434.1$ &$402.1$\\
114210& $ 6.1 $&   ...  & ... &$106.8$ \\
\hline
ELODIE \\
\hline
\, 80264 & $6.9$ &  ...  & ... & $108.1$ \\
\, 80902**& $8.9$ &  ...  & ... & $102.7$ \\
\hline
\end{tabular}
\end{table}

We combine the data for stars in common between HARPS, SOPHIE, and ELODIE, in addition to measurements from HIRES by \citet[][hereafter B17]{butler2017lces}. The trend amplitude, acceleration, and degree are listed in Table \ref{tabe:TrendCombine}. We also list the shift between instruments. The mean shifts with respect to B17 are much larger because their measurements are mean subtracted. After combination, HIP24205, HIP32906, and HIP74702 do not show a significant trend.
\begin{table*}
%\tiny
    \caption{Data for stars in common between HARPS, SOPHIE, ELODIE, and B17 showing a significant RV trend ($\frac{A}{T} > 10\,\rm{m\, s^{-1}yr^{-1}}$): RV amplitude, acceleration value, order of polynomial fit, shift between HARPS and B17, HARPS and SOPHIE, HARPS and ELODIE, and ELODIE and SOPHIE.}
    \label{tabe:TrendCombine}
    \begin{tabular}{|l|r|r|c|r|r|r|r|}
    \hline
     {HIP/TYC} & {$A$ [$\rm{m\, s^{-1}}$]} & {$\frac{A}{T} [\rm{m\, s^{-1}yr^{-1}}]$} & {Degree} & {$\rm{shift}_{\rm{HB17}}$ [$\rm{m\, s^{-1}}$]} & {$\rm{shift}_{\rm{HS}}$ [$\rm{m\, s^{-1}}$]} & {$\rm{shift}_{\rm{HE}}$ [$\rm{m\, s^{-1}}$]} & {$\rm{shift}_{\rm{ES}}$ [$\rm{m\, s^{-1}}$]}\\
     \hline
\, \, 1444&     $ 427.9$ & $      20.4$ & $ 2$ & $   28822.8 \pm 0.7$ &...&...&...\\
\, \, 5806** & $154.1$ & $12.6$ & $2$ & ... & $-25.1\pm 8.0$ & ...& ...\\
\, \, 7404& $     164.5$ & $ 12.8$ & 3 &... &$-108.8 \pm 7.7$ & ...& ...\\ 
\, 14614& $     246.9$ & $ 11.4$ & 3 & ...&$  18.2 \pm 2.2$ & $  -95 \pm 15$ &...\\
\, 26037& $    1250.5$ & $ 75.5$ & 2 & ...&... & ...& $ -326 \pm 16$\\
\, 29611& $     669.8$ & $ 45.3$ & 1 & ...&... & ...& $    2.3 \pm 16.1$\\
\, 42575& $     474.3$ & $150.0$ & 1 & ...& ...& ...& $288 \pm 29$ \\
\, 61157& $     578.8$ & $ 43.0$ & 3 & ...&... & ...& $-112 \pm 15$\\
\, 63346& $     267.2$ & $ 19.2$ & 3 & ...&... &...& $  -5.4\pm 11.7$\\
\, 68134& $     486.6$ & $ 31.5$ & 1 & ...&... & ...& $  71 \pm 23$\\
\, 71803&     $ 223.1$ & $      11.4$ & $ 2$ & $   12953 \pm 1$ &...&...&...\\
\, 85158&     $ 295.2$ & $      18.5$ & $ 3$ & $  -24065 \pm 0.8$  &...&...&...\\
\, 99385&     $ 925.2$ & $      69.5$ & $ 2$ & $   22239 \pm 2$  &...&...&...\\
104587& $     164.4$ & $ 13.4$ & 1 & ...&... & ...& $  75 \pm 11$\\
108602& $     307.5$ & $ 17.8$ & 1 & ...&... & ...& $ 277 \pm 23$\\
113086& $     209.5$ & $ 12.2$ & 2 & ...&... & ...& $  31 \pm 11$\\
115714**& $     448.4$& $  28.4$ & 1 & ...&... & ...& $ 140 \pm 10$\\
\hline
\end{tabular}
\end{table*}

In addition to stars with $\sigma_{\rm{Model}} \ge 100 \,\rm{m\, s^{-1}}$ for each instrument individually, and after combination of data from several instruments, there are four more stars listed in bold in Table \ref{table:sigmaModelCombined} having a high RV model scatter. We show their RV curves in Appendix \ref{append:Trend}.  
\begin{table}
%\tiny
    \caption{Similar to Table \ref{tabe:sigmaModel}, but for combined data sets. Stars in bold only exceed the Gaia threshold when data from different instruments are combined.}
    \label{table:sigmaModelCombined}
    \begin{tabular}{|l|r|c|r|r|r|}
    \hline
     {HIP} & {$T$ [yr]} & {$\sigma_{10}$ $[\rm{m\, s^{-1}}]$} & {$\sigma_{12}$ $[\rm{m\, s^{-1}}]$} & {$\sigma_{{T}}$ $[\rm{m\, s^{-1}}]$} \\
\hline
\, 5806** & $ 12.2$& ... & $ 113.9$ & $116.8$\\
\, {\bf 7404} & $12.9$ & ... & $160.2$ & $194.9$ \\
26037& $ 16.6$  &   $771.8 $  &  $ 1011.4$ &$1665.6$\\
29611& $14.8$  &   $130.8$ &  $157.0$&$193.4$  \\
{\bf 42575} & $3.16$ & ... & ... & $136.9$\\
{\bf 61157} & $13.5$ & ...& $138.3$ & $181.1$ \\
68134 & $  15.4$  &   ...  &  $109.0$&$139.9$ \\
{\bf 85158} & $15.9$ & $103.0$ & $142.3$ & $245.0$ \\
99385 & $13.3  $ & $219.8$ & $269.2$ &$302.4$\\
\hline
\end{tabular}
\end{table}

\subsection{Stars with companion detected by Adaptive Optics (AO)}
\label{sectAO}
The nature of the object causing the trend can be an unseen planetary or stellar companion. Under favourable circumstances in terms of magnitude difference and angular separation, imaging observations can be used to confirm the detection. Some of the trend stars have been observed using AO.
The archive data are provided by NIRC2, NACO \citep{rousset2003naos} and SPHERE \citep{beuzit2010sphere} instruments.

\begin{figure}
\begin{center}
\includegraphics[width=\columnwidth]{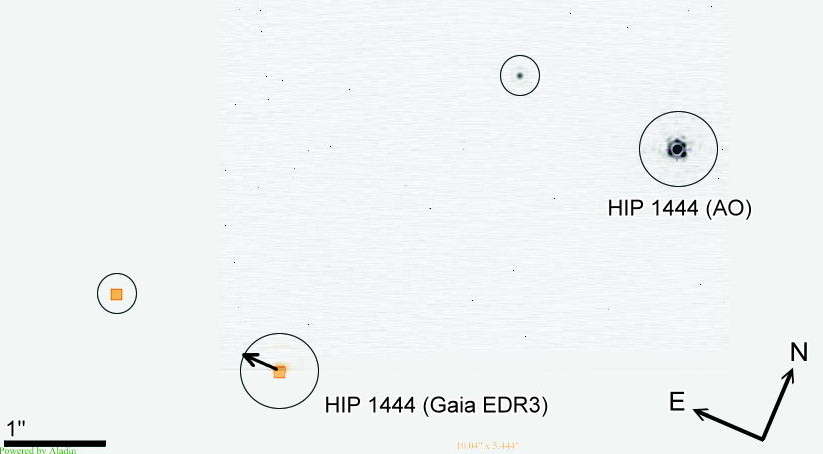}\end{center}
\caption{HIP 1444 NIRC2 adaptive optics image obtained on 09-08-2013 made available in Keck Observatory Archive (KOA) by the program PI, Barclay. We cropped the image and inverted the colours. The distance between the objects is $47.15 \pm 0.33$ AU and their flux ratio is 14.58 in the $K$ band. Gaia source positions at epoch 2016.0 (orange squares) were over-plotted using Aladin Sky ATLAS. 
The arrow shows the primary motion with the modulus corresponding to the proper motion.}
\label{hip1444Keck}
\end{figure}

An example of NIRC2 AO image is displayed in Fig.~\ref{hip1444Keck} for HIP1444 where the presence of an object at $1.75 \pm 0.01 ''$ from the target is revealed. For the twelve stars with available AO images, we calculated absolute magnitudes and projected physical separations using recent distance measurements from \cite{bailer2021estimating} based on Gaia EDR3. Supposing the system to be physical (i.e. that the companion is at the same distance as the primary), we obtained the absolute magnitudes in the $J$, $H$, and $K$ bands. The apparent magnitude of the companion, $m_2$, is given by $m_1 - m_2 = -2.5\,\mathrm{log}\frac{f_1}{f_2}$,  where $m_1$ is the apparent magnitude of the primary as given by the SIMBAD database. For some of these targets, there are objects with unknown parallaxes within a $5''$ radius from the sources according to Gaia EDR3 \citep{brown2021gaia}. There may be small differences between the angular separation, $\theta$, from Gaia and AO (compare column 3 and 4 of Table~\ref{tab:ao}) because these measurements were taken at different dates. 

The minimum mass of the companion, $M_{\rm{min}}$, which can cause a given acceleration,  $\gamma_r$, is given by Eq. \ref{Mmin} \citep{liu2002crossing}, 
\begin{equation}
\label{Mmin}
M_{\rm{min}} (M_{\odot}) = 5.34 \times 10^{-6} (d\, \theta)^2 {\lvert}\gamma_r{\rvert}\sqrt{27}/2,
\end{equation}
where $\theta$ is the projected angular separation between the components, $\gamma_r = \frac{dv}{dt}$ is the instantaneous radial acceleration (the derivative of the RV at the mean of dates), and $d$ is the distance to the system. The orbital function $F (i, e, \omega, \phi )$, which depends on the inclination, $i$, eccentricity, $e$, longitude of periastron, $\omega$, and orbital phase, $\phi$, has a minimum value of $\sqrt{27}/2$ \citep{torres1999substellar}.

Equation  \ref{Mmin} allows us to verify whether the observed trend can be caused by an orbiting body. We estimate the companion photometric mass using the colour-temperature conversions of \citet{pecaut2013intrinsic}\footnote{We used the  updated Version 2019.3.22 table available on E. Mamajek’s website: \url{http://www.pas.rochester.edu/\textasciitilde emamajek/EEM\textunderscore dwarf\textunderscore UBVIJHK\textunderscore colors\textunderscore Teff.txt}} by matching absolute magnitudes. We list the results of our photometric and astrometric analysis in Table~\ref{tab:ao}, along with the inferred measurements of physical separations, mass and period. The period is determined from Kepler's third law with the primary's mass extracted from literature.

\begin{figure}
\begin{center}
\includegraphics[width=\columnwidth]{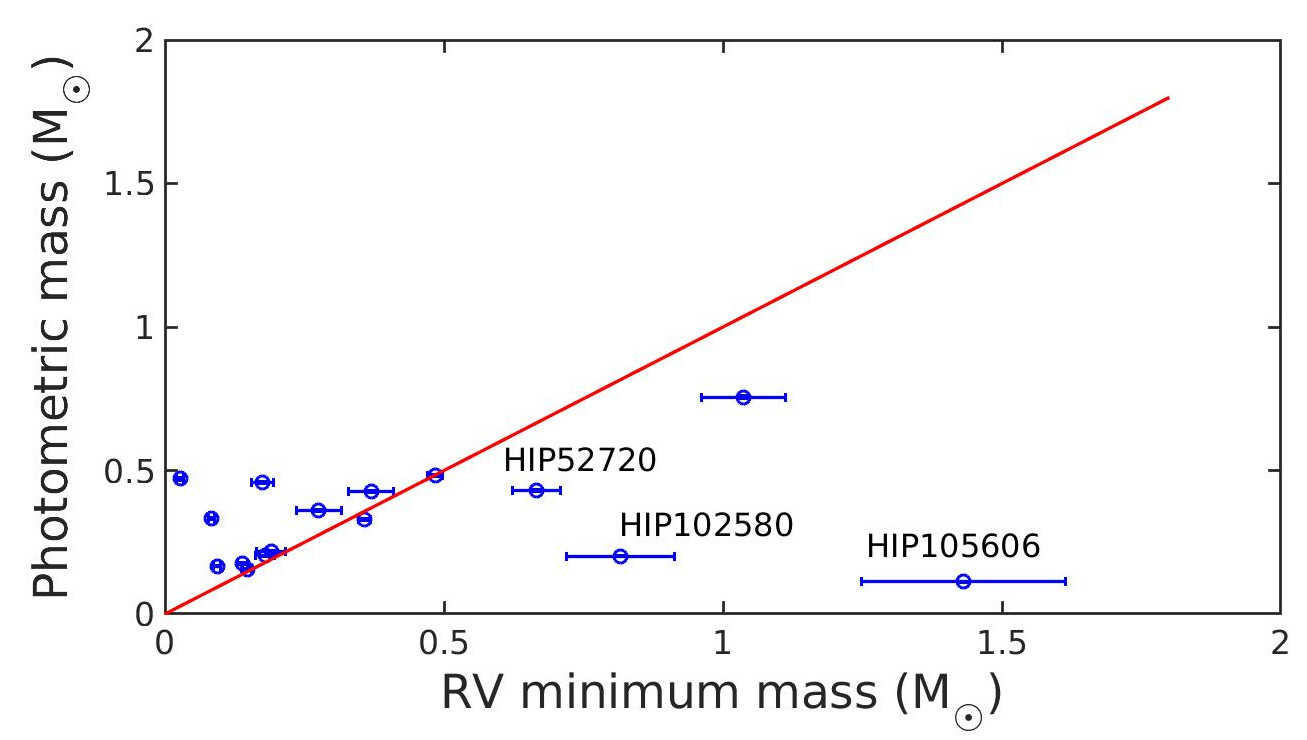}\end{center}
\caption{Comparison between photometric masses from AO and the minimum dynamic masses deduced from the trend acceleration. The 1:1 relation is represented by the red line. Stars above that line have companions that are likely responsible for the RV acceleration. Stars below it (HIP52720, HIP102580 and HIP105606) have visual companions not responsible for the RV acceleration at the corresponding measured separation.}
\label{massRatio}
\end{figure}

We compare the photometric masses, $M_{\rm{phot}}$, and the minimum dynamical masses, $M_{\rm{min}}$, in Fig.~\ref{massRatio}. If the mass determined from magnitude comparison is smaller than the minimum mass, another companion than the one identified in the AO image is responsible for the RV trend. There are three objects fulfilling this condition: HIP52720, HIP102580, and HIP105606. Therefore, the companion detected by AO is not responsible for the RV acceleration that might be caused instead by an unresolved tertiary companion. For HIP105606, Gaia EDR3 detects a fainter source at $0.85''$ from the target with the same parallax ($\pi = 28.4$ mas). This separation gives a minimum mass of 0.092 M$_\odot$, supporting the idea that this closer body is much more likely to be the source of the trend.

\begin{table*}
    \caption{Astrometric and photometric results for the stars observed with AO instruments allowing to compare Gaia and AO angular separations ($\theta_{\rm{Gaia}}$ and $\theta_{\rm{AO}}$), as well as minimum dynamical and photometric masses ($M_{\rm{min}}$ and $M_{\rm{phot}}$).}
    \label{tab:ao}
    \begin{tabular}{|r|r|r|r|r|r|r|r|r|r|}
    \hline
     {HIP} & {$\gamma_r [\rm{m\, s^{-1}yr^{-1}}]$} & {$\theta_{\rm{Gaia}} ['']$} & {$\theta_{\rm{AO}} ['']$ } & {$a$ [AU]} & {$K$ or $J$ [mag]} & {$M_{\rm{min}}$ [$\rm{M_{\odot}}$]}  & {$M_{\rm{phot}}$ [$\rm{M_{\odot}}$]}  & {$P$ [yr]} & {Instrument}\\%WDS
\hline
1444	& $18.574 \pm 0.052 $& $1.794$ &$1.574  \pm 0.020 $ & $42.3  \pm  0.6 $ & 5.744 (K) & $0.48 \pm 0.01$   & 	0.483 & $223.6 \pm 7.2$ & NACO
\\                                                                                                                    
7396    & $21.177  \pm 0.198 $& ... & $0.246  \pm 0.025 $ & $9.4  \pm  1.0$  & 6.054 (H)& $0.03 \pm 0.01$ &	0.471 & $26.0 \pm	4.2$ & NACO
\\                                                                                                                    
52720	& $-5.154 \pm 0.211 $& 1.677 &$1.691  \pm 0.023 $ &$ 94.2  \pm  1.4 $ & 6.093 (K)& $0.67 \pm 0.04$    & 	0.432 & $821 \pm 53$ & NACO
\\                                                                                                                    
58558	& $2.834 \pm 0.313$ &... & $ 1.617 \pm	0.019$ & $67.9 \pm 1.0$ & $8.793$ (J) &	$0.19 \pm	0.03$ &	$0.218$	& $520 \pm 17$ & NACO
\\
62039	&$ 7.118 \pm 0.052 $& 1.293 & $1.309 \pm	0.014$ & $58.8 \pm 0.7 $ & 7.667 (J) & $0.36 \pm 0.01$    & 	0.329 & $395 \pm 7$ & NIRC2
\\                                                                                                                    
71803	&$ -10.941 \pm 0.098$ & ... &$0.547  \pm 0.020 $ & $29.6  \pm  1.1 $ & 9.404 (J) & $0.14  \pm 0.01$   & 	0.176 & $143 \pm 11$ & NIRC2
\\                                                                                                                    
81533	&$ 39.727 \pm 0.374$ &... &$0.359	\pm 0.025$	& $25.3  \pm 	1.8 $ & 6.376 (H)& $0.37 \pm 0.04$  &	0.426 & $107 \pm 9$ & SPHERE
\\                                                                                                                    
85158	&$ -11.344 \pm 0.274$ &... &$0.680  \pm 0.019$  & $22.5  \pm  0.7 $ & 6.809 (K)& $0.08  \pm 0.01$ & 	0.332 & $90 \pm	4$ & NIRC2
\\                                                                                                                    
99385	& $-64.247 \pm 0.519$ &  ... &$0.876  \pm 0.048 $ & $13.9  \pm  0.6 $ & 8.119 (K)& $0.18 \pm 0.02$  &  0.202 & $54 \pm 4$ & NACO
\\                                                                                                                    
102580	&$ 9.2013  \pm 0.908 $&2.401 &$2.109  \pm 0.026 $& $78.0  \pm  1.1 $ & 8.123 (K)& $0.82 \pm 0.10$	 &  0.202 & $645 \pm 20$ & NACO
\\                                                                                                                    
105606	& $7.263 \pm	0.838$ &$0.849 $&	$3.318 \pm 0.019$ &	$116.4 \pm	0.8$ &	$10.010$ (K)&	$1.43 \pm	0.18$ &	$0.112$	& $1224 \pm 26$ &  NACO
\\                                                                                                                    
110843	& $-11.709 \pm 1.283 $&  ... &$1.089  \pm 0.025 $& $40.2  \pm  1.0 $ & 6.554 (K) & $0.28 \pm 0.04$  &	0.361 & $216 \pm 10$ & NACO
\\
\hline
    \end{tabular}
\end{table*}  

\begin{figure*}
\begin{center}
\begin{subfigure}{.48\textwidth}
\includegraphics[width=\linewidth]{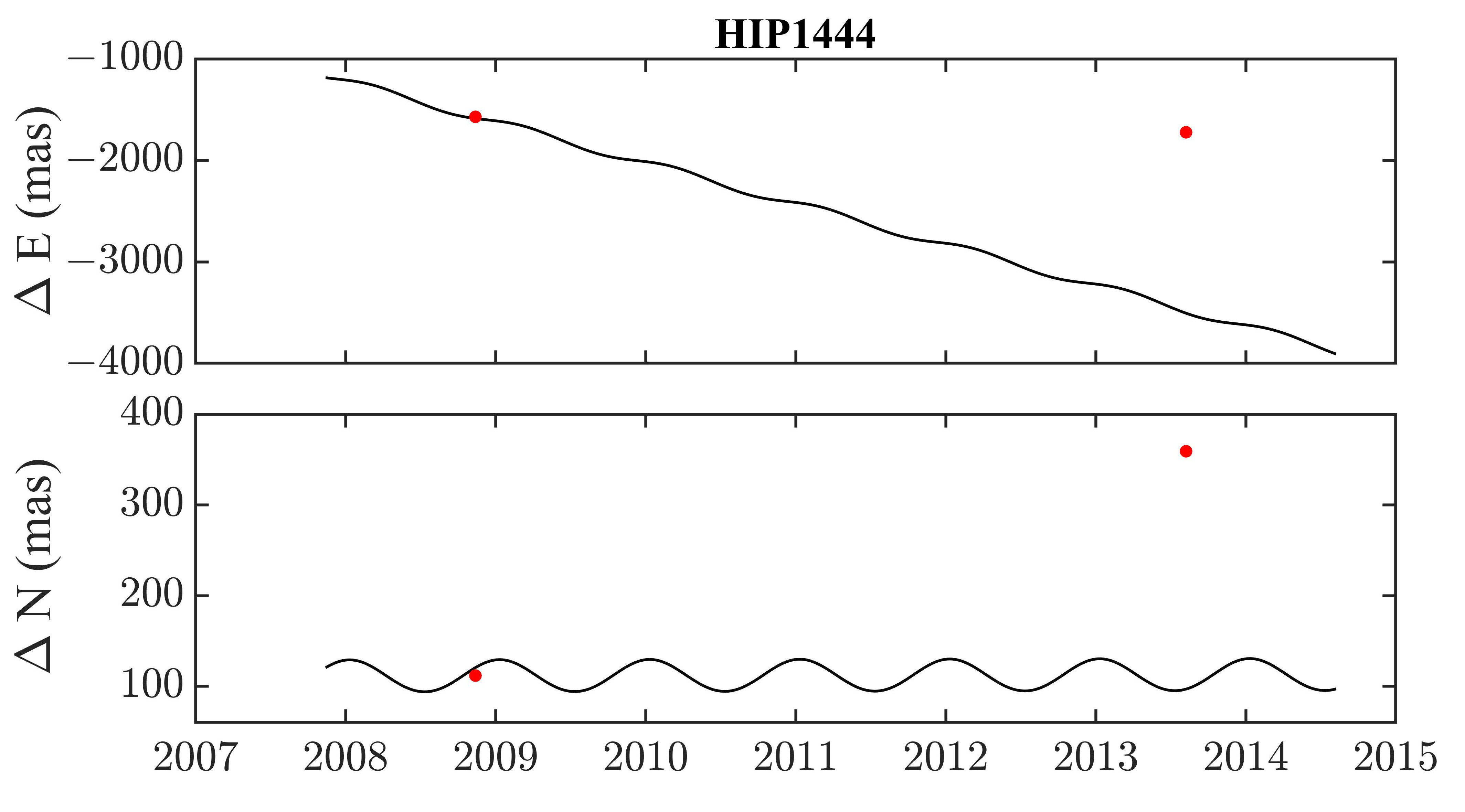}
\end{subfigure}
\begin{subfigure}{.48\textwidth}
\includegraphics[width=\linewidth]{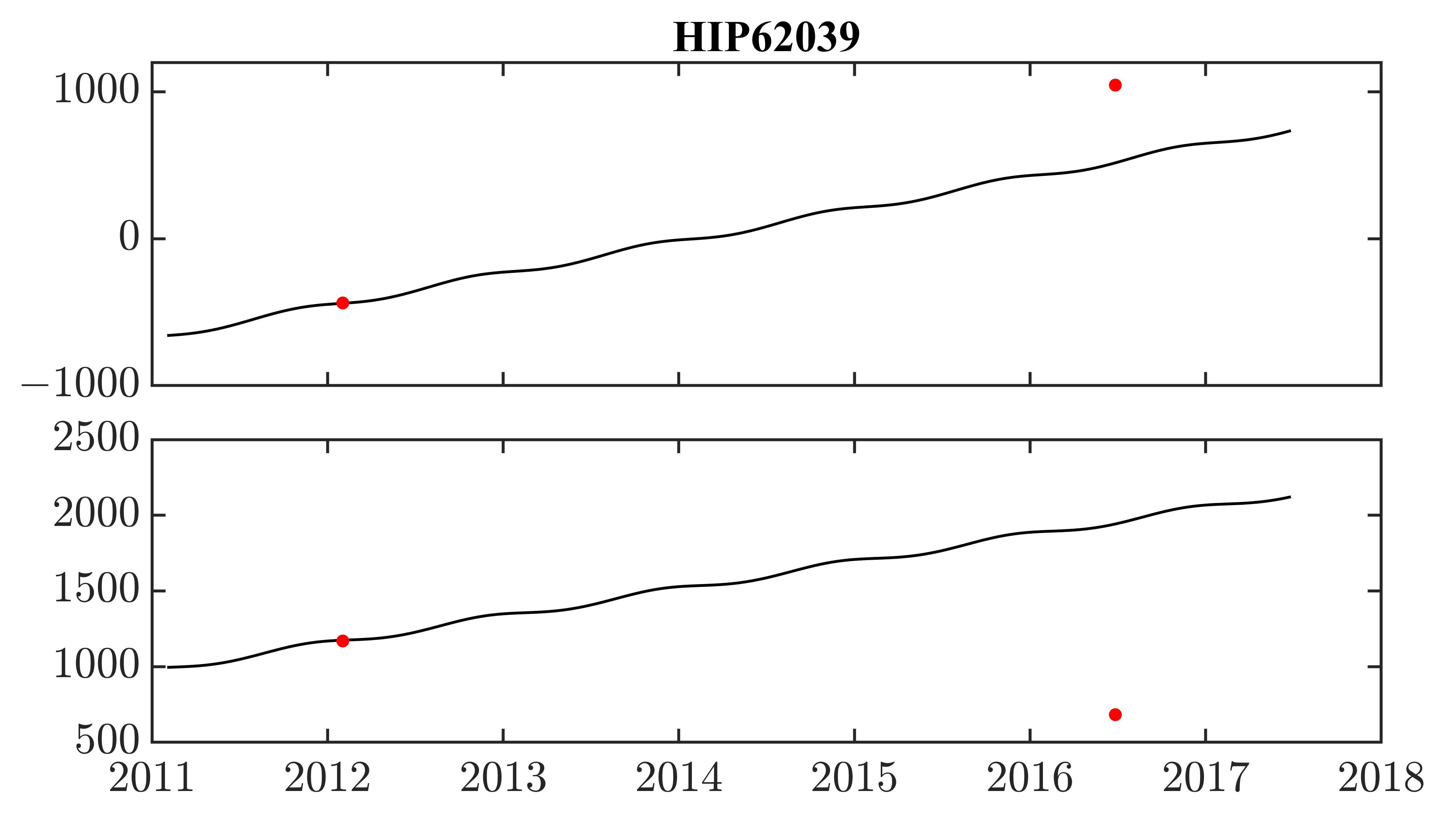}
\end{subfigure}
\begin{subfigure}{.48\textwidth}
\includegraphics[width=\linewidth]{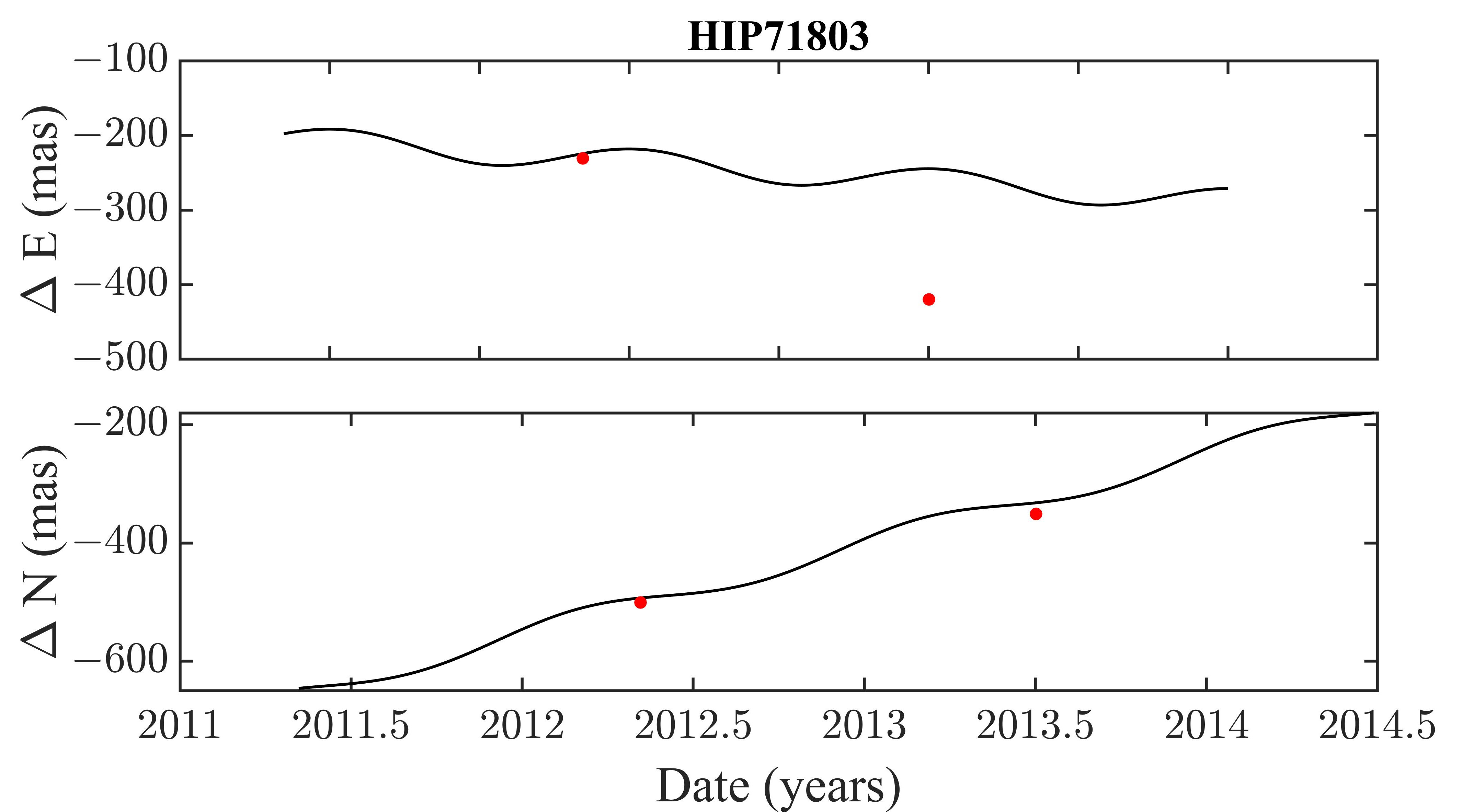}
\end{subfigure}
\begin{subfigure}{.48\textwidth}
\includegraphics[width=\linewidth]{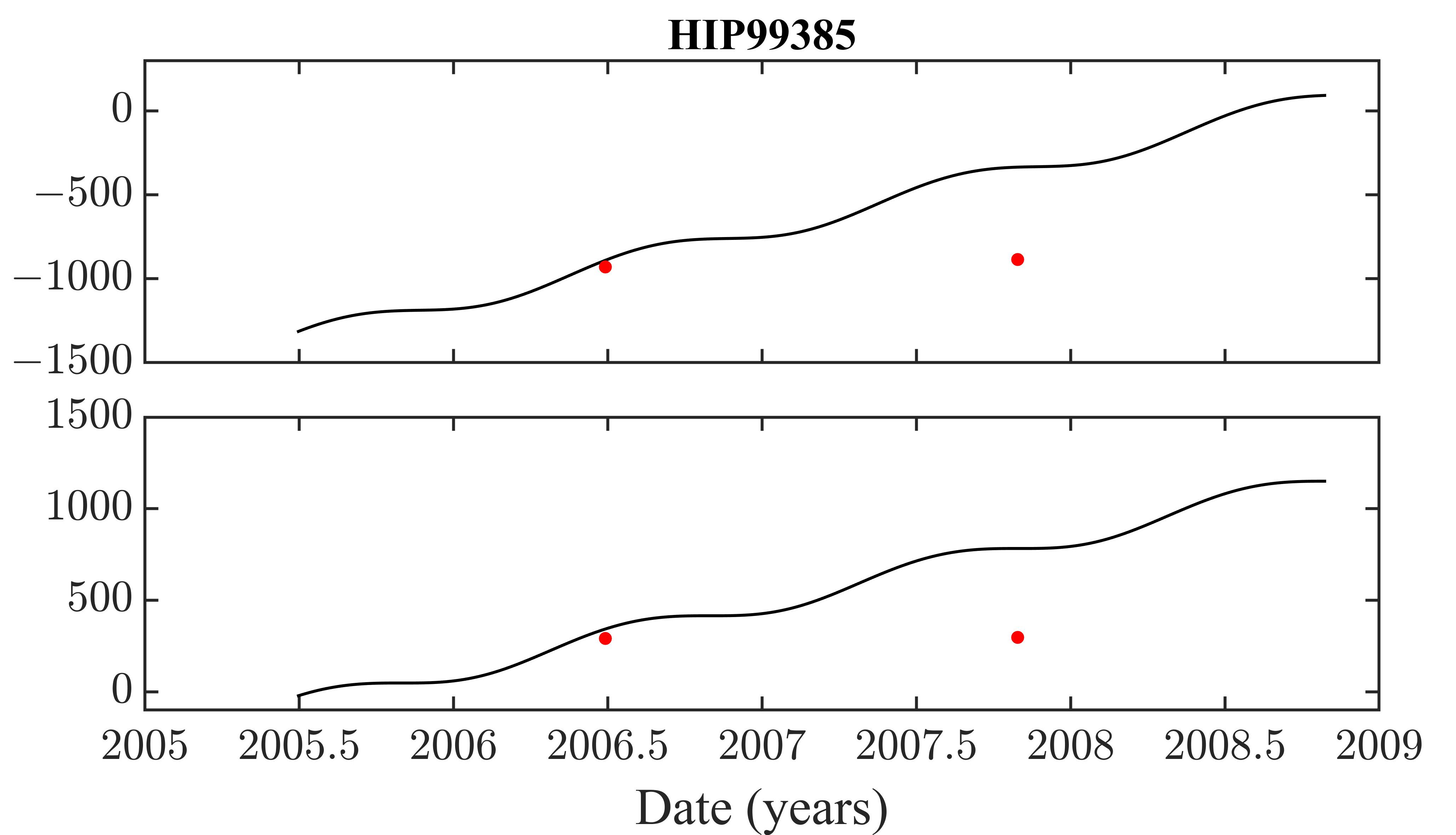}
\end{subfigure}
\end{center}
\caption{Astrometric data for the four stars observed in AO at different epochs. The black curves show the motion of a background object. The red points are the astrometric measurements of the visual companion. As the companion does not follow the motion of a background object, it is likely bound to the primary.}
\label{spaceM99385}
\end{figure*}

Among the twelve stars, there are four for which AO images are available for at least two different dates. It is therefore possible to check whether the observed companion is a background star lacking significant proper or parallactic motion. Following \cite{gonzales2020trends}, we calculated the difference in position (offset) between the primary and the putative companion ($\Delta E, \Delta N$) at two different epochs. $E$ and $N$ refer to East and North coordinates, respectively. If we consider a fixed background star, it would appear to shift relatively to the target by the amount $(-\Delta E_s, \Delta N_s)$ where $\Delta E_s = \Delta E_{PM} + \Delta E_{\pi}$ and $\Delta N_s = \Delta N_{PM} + \Delta N_{\pi}$; $\Delta E_{PM}/\Delta N_{PM}$ is the proper motion variation (to the East or North) between two different dates, and $\Delta E_{\pi}/\Delta N_{\pi}$ is the variation of the parallactic motion of the primary star. We show the space motion in Fig.~\ref{spaceM99385} where that of a background object is represented by the track plotted in black. The location of the companion is inconsistent with this track when it is comoving with the primary star.

\textbf{HIP1444}. The companion was observed with both NACO and NIRC2 instruments (see Fig.~\ref{hip1444Keck} in the latter case). In addition, Gaia EDR3 reports a source at nearly the same angular separation. Moreover, the proper and parallactic motions indicate that this object is bound to the target. B17 suggested that the primary could be a host to a candidate planet with an orbital period of $5082 \pm 99$ days which is much smaller than the value we found. Their adopted model consists of Keplerian signals and a linear acceleration.

\textbf{HIP62039}. Our measured acceleration of combined RVs is close to the $7.19 \pm  0.16 \,\rm{m\, s^{-1}yr^{-1}}$ value estimated by B17 who did not find a periodic signal. The lack of periodicity indicates the presence of a companion with a long period.

\textbf{HIP71803}. B17 found an acceleration of $-10.772 \pm  0.281\,\rm{m\, s^{-1}yr^{-1}}$ without a periodic signal. With nearly the same acceleration, $-10.592\,\rm{m\, s^{-1}yr^{-1}}$, \cite{hinkel2019recommendation} estimated a companion minimum mass of 6.8 M$_{\mathrm{Jup}}$. 

\textbf{HIP85158}. Even though the orbit is not fully covered, B17 considered it as a candidate to harbour a companion with a period of $5220 \pm 197$ days, a RV semi-amplitude of $29.20 \pm 2.24\,\rm{\; m\, s^{-1}}$ and an acceleration of $-16.372 \pm  0.432\,\rm{\; m\, s^{-1}yr^{-1}}$. The acceleration we estimated from combining RVs is much smaller.
\section{Activity indicators}
\label{activi}
The variation observed in the RV curves
can be related to chromospheric phenomena instead of being due to real dynamic changes in the star's centre of mass. We investigate this possibility below.
\subsection{RV jitter from chromospheric activity}
For the stars with putative companions (short periods or trends), we retrieved the well-known magnetic activity indicator inferred from \ion{Ca}{ii} H+K, log$R'_{\rm{HK}}$, from the catalogue of \citet{borosaikia2018chromospheric}. We then estimated the expected jitter from the calibration of \citet{santos2000coralie} relevant to F, G and K type stars, where $R_5 = 10^5 R'_{\mathrm{HK}}$:
\label{jitter}
\begin{equation}
\begin{cases}
\sigma_F (\rm{m\, s^{-1}}) = 9.2 \times \mathit{R}_5^{0.75}\\
\sigma_G (\rm{m\, s^{-1}}) = 7.9 \times \mathit{R}_5^{0.55}\\
\sigma_K (\rm{m\, s^{-1}}) = 7.8 \times \mathit{R}_5^{0.13}
\, \, \, \, . \\ 
\end{cases}
\end{equation}

The log$R'_{\rm{HK}}$ and RV jitter values can be found in Table~\ref{tabe:TrendAct} (HARPS) and Table~\ref{table:TrendActSophie} (SOPHIE+ELODIE) for stars with a trend (the existence of an additional Keplerian signal and a known period are noted by a single and double asterisk, respectively). 
 
A RV variability caused by a companion rather than by stellar activity is more likely for the stars in Table~\ref{tabe:TrendAct} that have a trend amplitude larger than five times their estimated RV jitter. Conversely, HIP45749, HIP95849, HIP98828, and HIP110843 have a RV amplitude below that limit and could have their RV variation arising from stellar activity.

\subsection{Line-profile indicators}
\label{Line profile indicators}
Another way to relate the RV variation to chromospheric activity is to examine the correlation between the former and some activity (line-profile) indicators, as done below. 
\subsubsection{Bisector velocity span}
\label{BVS}
To this aim, we first use the bisector velocity span ($BIS$), which is the difference between the velocity of the bisector at the top and bottom of the CCF. Defined by \citet{figueira2013line} and \citet{queloz2001no}, as the average of the mid-points between 60 and 90$\%$ for the top and 10 and 40$\%$ for the bottom. The $BIS$ are extracted from the instrument pipeline.

If the $BIS$ and RV data are independent, the variability signal may have its origin in an unseen companion. 
However, a negative correlation indicates that the RV variations arise from stellar activity, as it was concluded by \citet{queloz2001no} and \citet{fiorenzano2005line} for 
%their target 
HD 166435. On the other hand, a positive correlation may be the signature of a blended system or a substellar companion, as in HD 41004 where \citet{santos2002coralie} attributed the variation to a brown dwarf companion to the secondary. In the same vein, \citet{fiorenzano2005line} showed that the positive correlation between RVs and bisector spans in HD $8071$ is due to contamination by a companion with a nonlinear dependence. However, \cite{lovis2011harps} noted that a positive correlation could also indicate that a magnetic cycle is the source of the RV variation.
\citet{zechmeister2013planet} pointed out that the physical correlation cannot be confirmed on firm statistical grounds unless both variables vary with an identical period. In this case, the (sub)stellar companion option becomes invalid. 

We computed  Pearson correlation coefficients (hereafter $\rho$) between the RV and $BIS$ data, along with the associated confidence level, $p$. The $\rho$ and $p$ values are summarised in 
the fifth and sixth columns of 
Table~\ref{tabe:TrendAct} (HARPS) and Table~\ref{table:TrendActSophie} (SOPHIE+ELODIE) for stars with trends. 

Among the stars with a trend having an acceleration greater or equal to $10 \rm{\; m\, s^{-1}yr^{-1}}$, four have their RVs significantly anti-correlated to $BIS$, which indicates that the RV variation can be related to chromospheric phenomena. It concerns: HIP5806, HIP78395, HIP81533, and HIP95849.

However, as mentioned above, a significant correlation does not provide a firm identification of the source of the variation, and doubt remains for:

A companion is clearly detected in HIP81533 through AO SPHERE imaging (Sect. \ref{sectAO}), but we are unsure whether it is causing the trend. However, as the trend's acceleration is high, we doubt it would be caused by activity. 

There are six trend stars with RVs positively correlated to $BIS$ and without any companion detected previously (RV or imaging). For those, the variation can be caused by a blended system or a magnetic activity cycle. Among these stars, two have a high acceleration, which is more likely induced by a companion: HIP116374 and TYC6792-01746-1. 

As for SOPHIE targets, a significant, positive correlation is present for HIP85653 and HIP104587, which could indicate the cause of the trend to be a magnetic cycle. HIP31849 and HIP74702 have significant correlation between RV and $BIS$, however they have high acceleration.

\subsubsection{The S-index} 
The S-index is dependent on the strength of the magnetic field and compares the relative flux of the Fraunhofer emission lines (H \& K) of \ion{Ca}{ii} to the continuum value. It is used to determine stellar activity cycles and rotation periods.
\\We computed this index using  the method described by \citet{lovis2011harps} where $H$ and $K$ are the total fluxes in the bands centred at $3933.664$ \AA\,  ($K$) and $3968.470$ \AA\, ($H$), $R$ and $V$ are the values measured in adjacent continuum passbands, and $\alpha$ is a calibration constant:
\begin{equation}
S = \alpha \frac{H+K}{R+V}  \, \, \,  .
\end{equation}

As the activity level increases, active regions in which convection is frozen cover a larger fraction of the stellar surface. It implies that convective blueshift is globally reduced and the lines are slightly shifted to the red. As a net result, a  positive correlation between RVs and S-index is expected.

The $\rho$ and p-values diagnosing a possible correlation between these two quantities are listed in columns 7 and 8 of Table~\ref{tabe:TrendAct} (HARPS) and Table~\ref{table:TrendActSophie} (SOPHIE+ELODIE) for the stars with a trend. 

There are eight trend stars with RV positively correlated to S-index for which we assume that the long-term RV variation may be attributed to a magnetic cycle. HIP45749 and HIP71803 have a positive correlation with $BIS$ confirming that their RV variation is indeed caused by magnetic cycle. 

The RV variations exhibited by HIP1444 are confirmed to be related to a companion by AO imaging, although it shows a significant correlation with the S-index. Even though we are not sure if the companion in AO imaging causes the trend in the case of HIP81533, it is very unlikely to be activity related.   

The SOPHIE stars with positive correlation between RVs and S-index and low acceleration value are HIP98828 and HIP104587 (for which the $BIS$ is also positively correlated to RVs). Those with high acceleration value are HIP31849 and HIP74702.

\section{Conclusion}
\label{conc}
We present a work aiming at confirming the status of a sample of 2351 stars as RV standards for the Gaia mission. 
We estimated the offset of our RVs relative to other instruments and placed the zero point by adopting HARPS measurements as the reference. 
We find a mean absolute difference of $240.4 \rm{\; m\, s^{-1}}$ between our measurements and those determined by the instrument pipelines. 

We looked for RV variability due to the presence of a companion or due to the existence of stellar activity/magnetism. Furthermore, we find stars having a linear or polynomial trend variation for which the period cannot be determined with the actual data because it is longer than the observational baseline. Adaptive optics help in a few cases confirming the presence of distant companions potentially inducing a long-term trend, although the physical association must be proven. Five stars with AO data have a visual companion lying at an angular separation consistent with the observed RV amplitude, but  
it is not the case for the three others.
Among the first category, the analysis of the proper motions supports the bound status of the secondaries in  HIP1444, HIP62039, HIP71803, and HIP99385.  
As the stars exhibiting a RV trend have incomplete orbital coverage, they should be monitored on a long-term basis. The trends with high acceleration, on the other hand, could indicate that the companion orbiting these stars is a brown/red dwarf. 
However, as the RV variation can also be caused by chromospheric activity, we checked for possible correlations between the RV and activity data to ascertain that the observed RV variations have their origin in a low-mass companion.

\begin{figure}
\begin{center}
\includegraphics[width=\columnwidth]{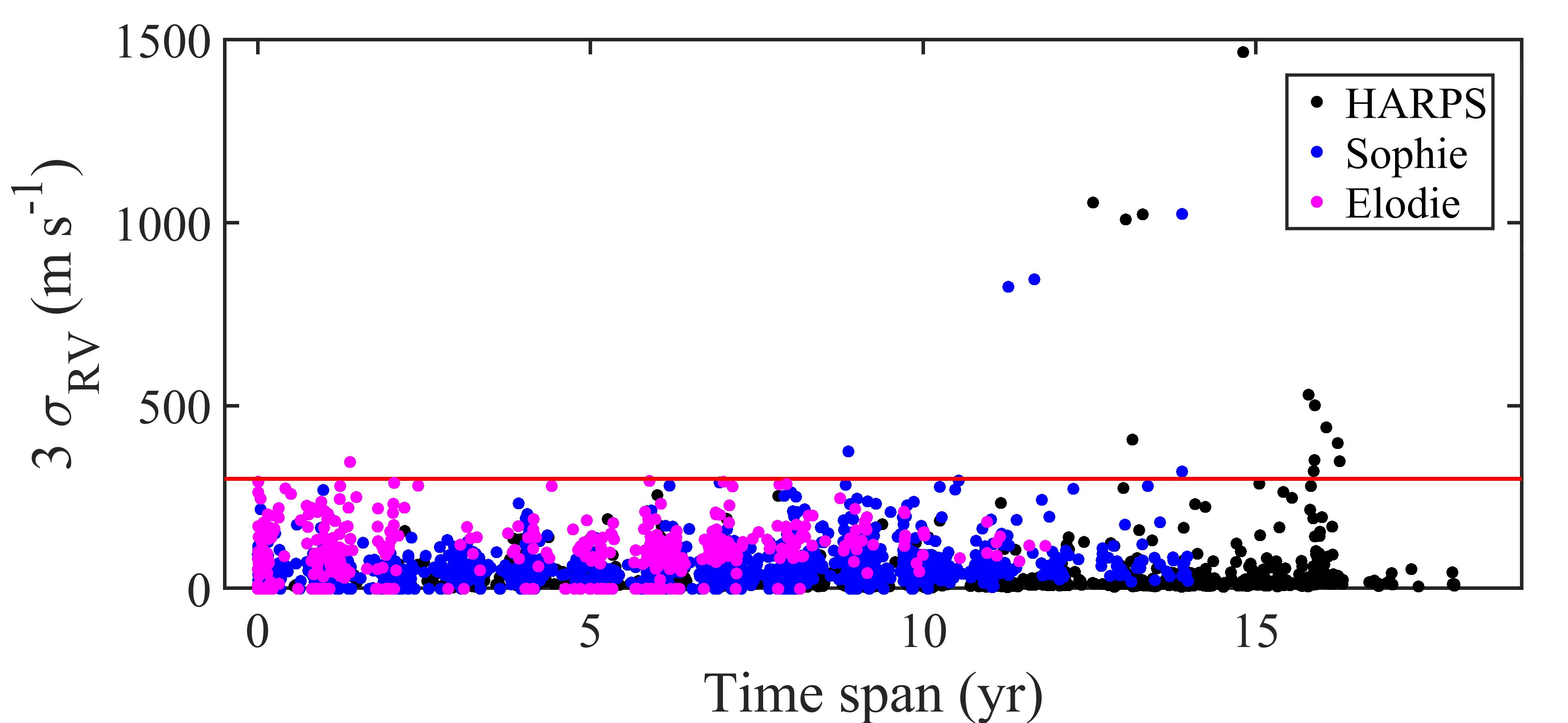}\end{center}
\caption{RV variation as a function of the time span with the Gaia RV variability threshold ($300\, \rm{m\, s^{-1}}$) depicted by the red horizontal line.}
\label{AvsT}
\end{figure}

Stars with a RV model scatter (orbital and/or trend) exceeding the threshold required for the Gaia mission, $300\, \rm{m\, s^{-1}}$, should be excluded from the calibrations for Gaia DR4. From the HARPS sample, it concerns five stars with a trend only (HIP37233, HIP64295, HIP99385, HIP116374, TYC8681-00611-1), three stars with a Keplerian solution only (HIP26394, HIP62534, and HIP113834), and three stars with both trend and Keplerian solution (HIP5806, HIP71001, and HIP108095). Additional variable stars are identified through their SOPHIE observations: HIP26037, HIP29611, HIP32906, HIP68134, HIP71291, HIP74702, HIP76751, HIP114210, HIP115714, and TYC3239-00992-1. The only ELODIE stars which show RV variability are HIP80264 and HIP80902. After combining data from several instruments four more stars can be considered variable: HIP7404, HIP42575, HIP61157, and HIP85158. All the calibration stars not fulfilling the {\it Gaia} criterion for a constant RV are listed in Table~\ref{table:removeDR4}. Other less obvious candidates are discussed in Sect. \ref{trend}. Stars for which the weighted standard deviation applied on data, $\sigma_{RV}$, exceeds the variability threshold set by CS13 are shown in Fig.~\ref{AvsT}. 
Because this subset constitutes a tiny fraction of our sample, it supports the validity of CS18 catalogue as a pool of RV calibrators for the RVS onboard Gaia.

\begin{table}
\caption{Targets with RV scatter greater than Gaia threshold, that should be removed from Gaia DR4.}
\label{table:removeDR4}
\begin{tabular}{|l|l|} 
\hline
{HIP/TYC} & {Instrument} \\
\hline
\, \, 7404 & H-S \\
\, \, 5806 & H \\
\, 26037 & S \\
\, 26394 & H\\
\, 29611 & S \\ 
\, 32906 & S \\
\, 37233 & H\\
\, 42575 & S-E \\
\, 61157 & S-E \\
\, 62534 & H\\
\, 64295 & H \\
\, 68134 & S \\
\, 71001 & H \\
\, 71291 & S \\
\, 74702 & S \\
\, 76751 & S \\
\, 80264 & E \\
\, 80902 & E \\
\, 85158 & H-B17 \\
\, 99385 & H \\
108095 & H \\
113834 & H \\
114210 & S \\
115714 & S \\
116374 & H \\
3239-00992-1 & S \\
8681-00611-1 & H \\
\hline
\end{tabular}
\end{table}

\section*{Acknowledgements}

YD, TM, YF and EG are greatly indebted to Belspo for a long-term support through a PRODEX contract related to the Gaia Data Processing and Analysis Consortium. This research was achieved using the POLLUX database ( \url{http://pollux.oreme.org} ) operated at LUPM  (Université Montpellier - CNRS, France) with the support of the PNPS and INSU. It also has made use of the Keck Observatory Archive (KOA), which is operated by the W. M. Keck Observatory and the NASA Exoplanet Science Institute (NExScI), under contract with the National Aeronautics and Space Administration. It also has made use of "Aladin sky atlas" developed at CDS, Strasbourg Observatory, France. We thank Dr N. Seghouani, Dr R. Mecheri and Dr K. Daifallah for providing us with calculus means. 

%%%%%%%%%%%%%%%%%%%%%%%%%%%%%%%%%%%%%%%%%%%%%%%%%%
\section*{Data Availability}
%
% 
%The inclusion of a Data Availability Statement is a requirement for articles published in MNRAS. Data Availability Statements provide a standardised format for readers to understand the availability of data underlying the research results described in the article. The statement may refer to original data generated in the course of the study or to third-party data analysed in the article. The statement should describe and provide means of access, where possible, by linking to the data or providing the required accession numbers for the relevant databases or DOIs.
%We provide RV and Vbroad measurements of the targets in online form, in addition to the co-added spectra of different epochs in the analyzed domains.
The data underlying this article are available in the article and in its online supplementary material.



%\bibliographystyle{mnras}
%\bibliography{biblio.bib} % if your bibtex file is called example.bib

%\begin{thebibliography}{99}
%\end{thebibliography}

%%%%%%%%%%%%%%%%%%%%%%%%%%%%%%%%%%%%%%%%%%%%%%%%%%

%%%%%%%%%%%%%%%%% APPENDICES %%%%%%%%%%%%%%%%%%%%%

\appendix
\section{Orbital solutions}
\begin{figure}
\begin{subfigure}{.5\textwidth}
\includegraphics[width=\linewidth]{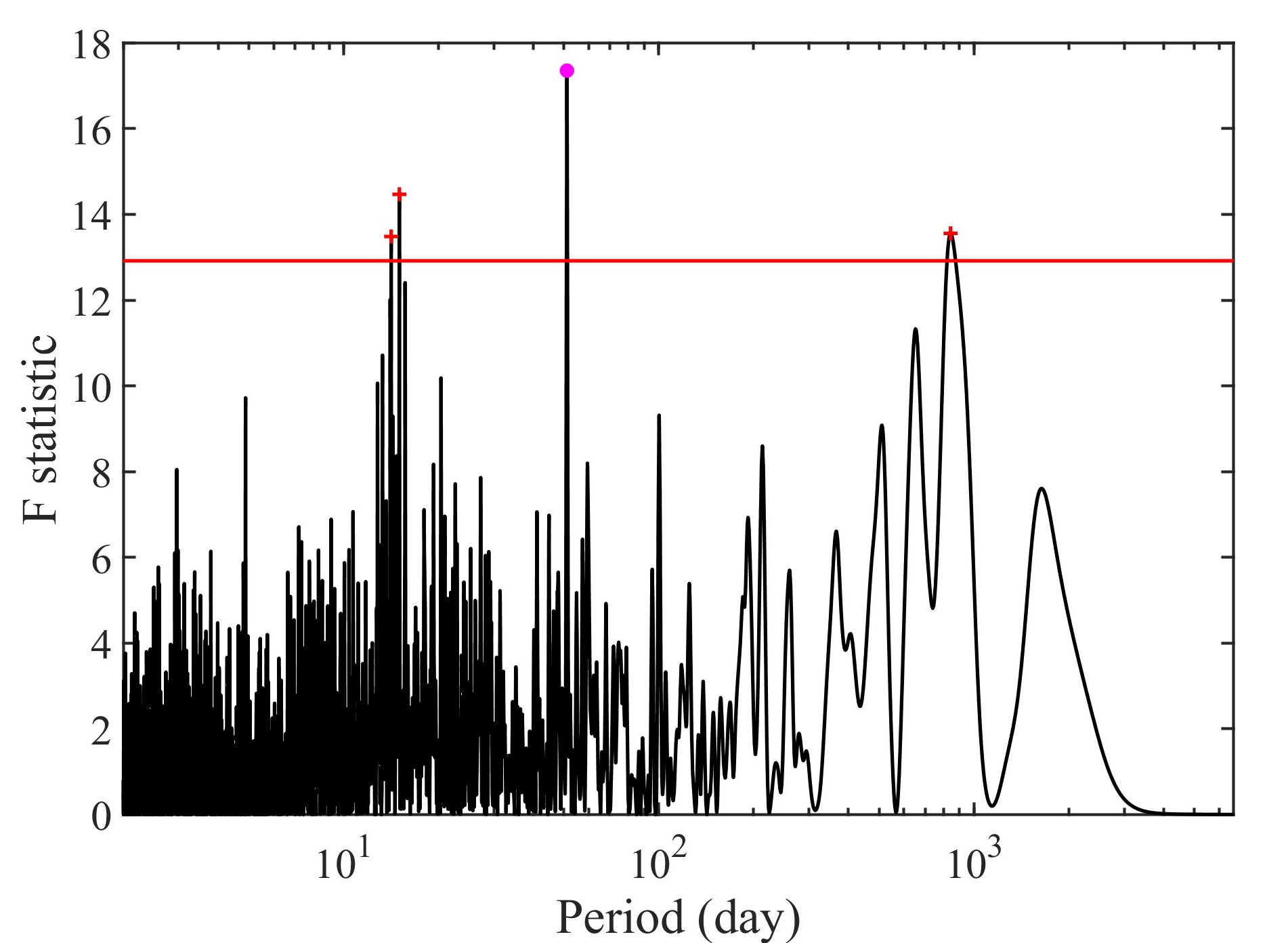}    
\end{subfigure}
\begin{subfigure}{.5\textwidth}
\includegraphics[width=\linewidth]{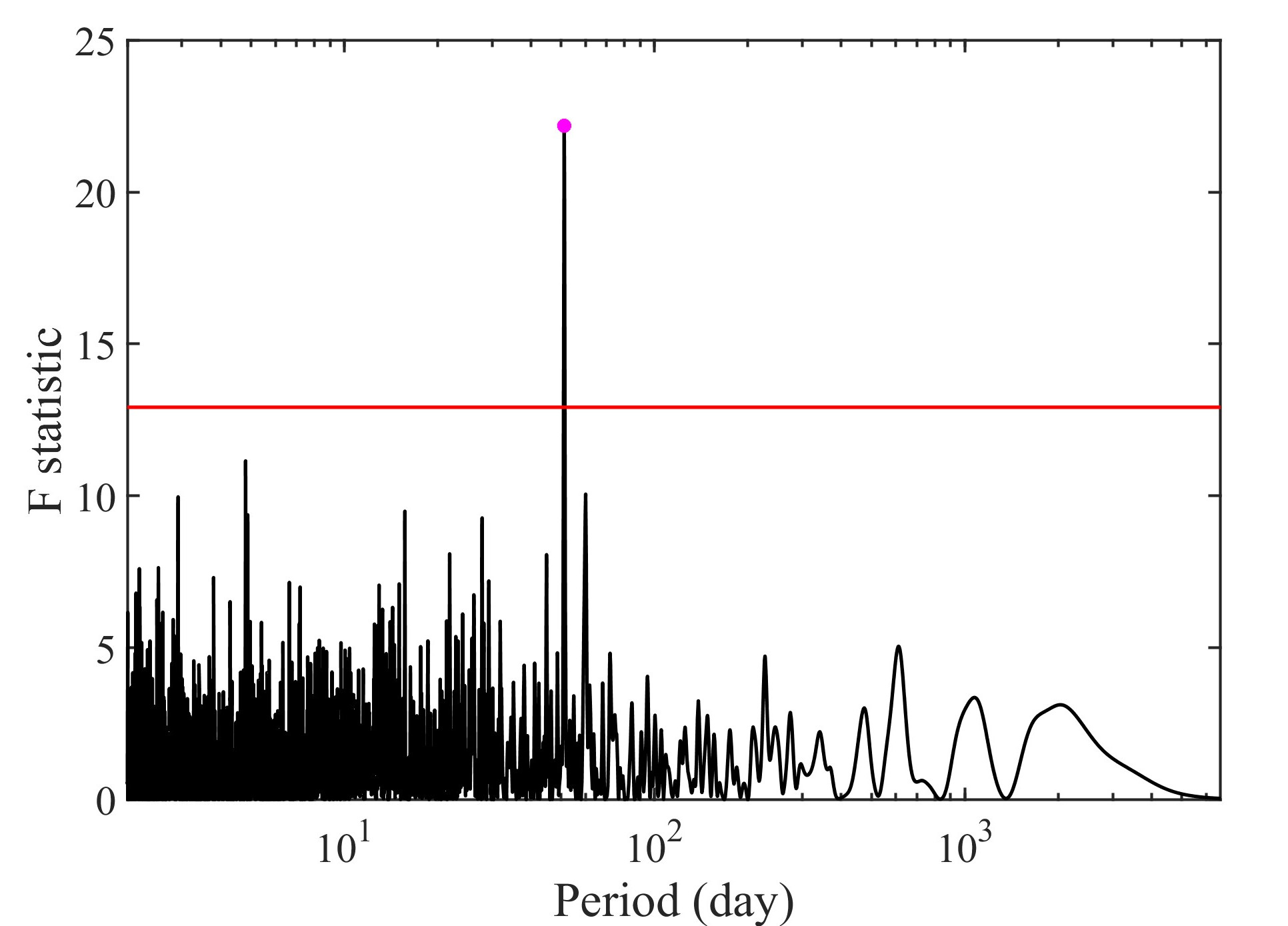}    
\end{subfigure}
\begin{subfigure}{.5\textwidth}
\includegraphics[width=\linewidth]{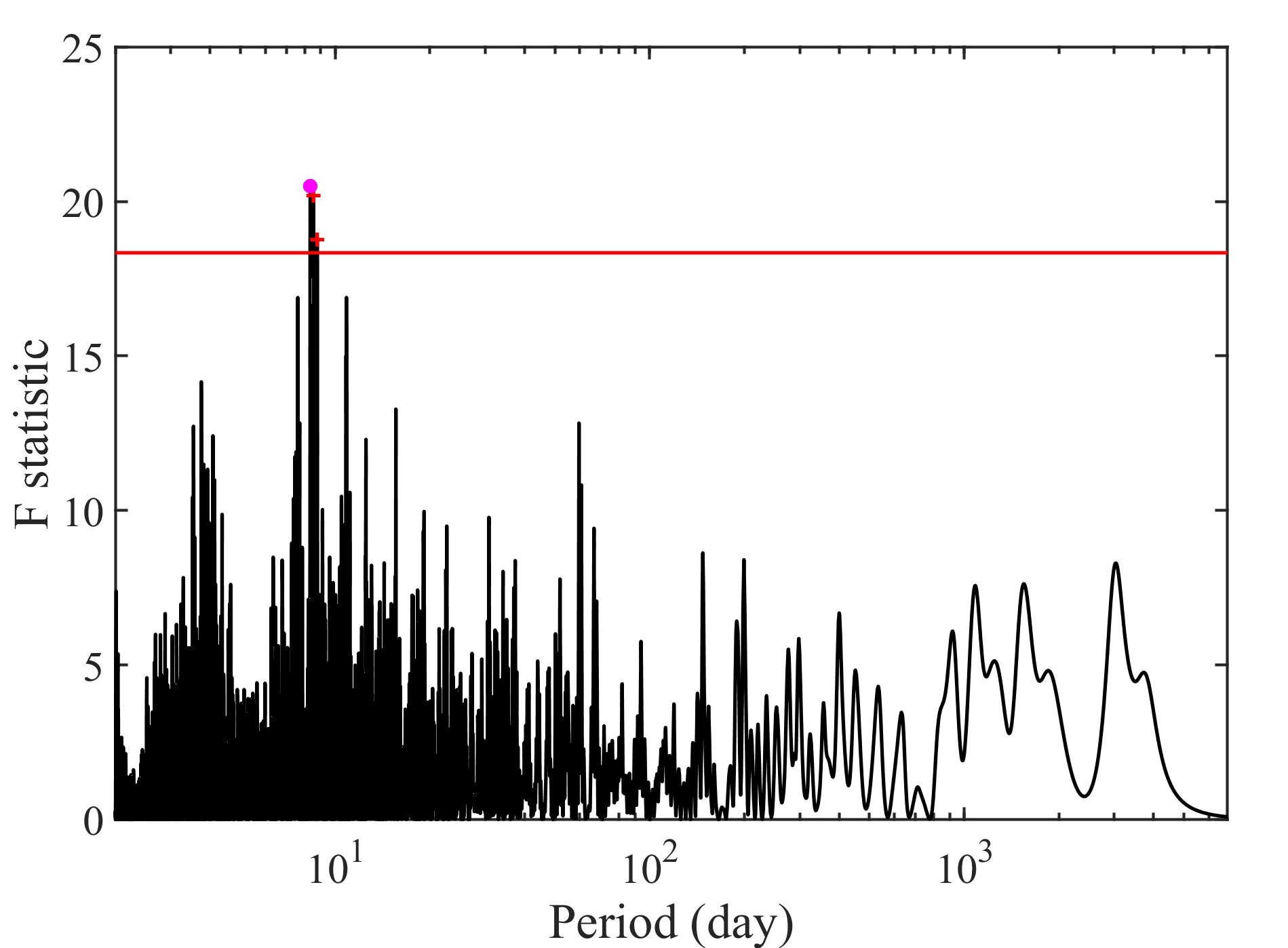}    
\end{subfigure}
\caption{Top panel: periodogram of HIP5806 with the orbital period of $806$ d. This period does not correspond to the highest peak, but it gives the best solution. Middle panel: periodogram showing HIP5806's third period, $51.154$ d. Bottom panel: periodogram of HIP108095 with the highest peak period of $8.53127$ d. Same symbols as in Fig. \ref{periodogram}.}
\label{fig:periodogramNew}
\end{figure}
\begin{figure}
\begin{subfigure}{.5\textwidth}
\includegraphics[width=\linewidth]{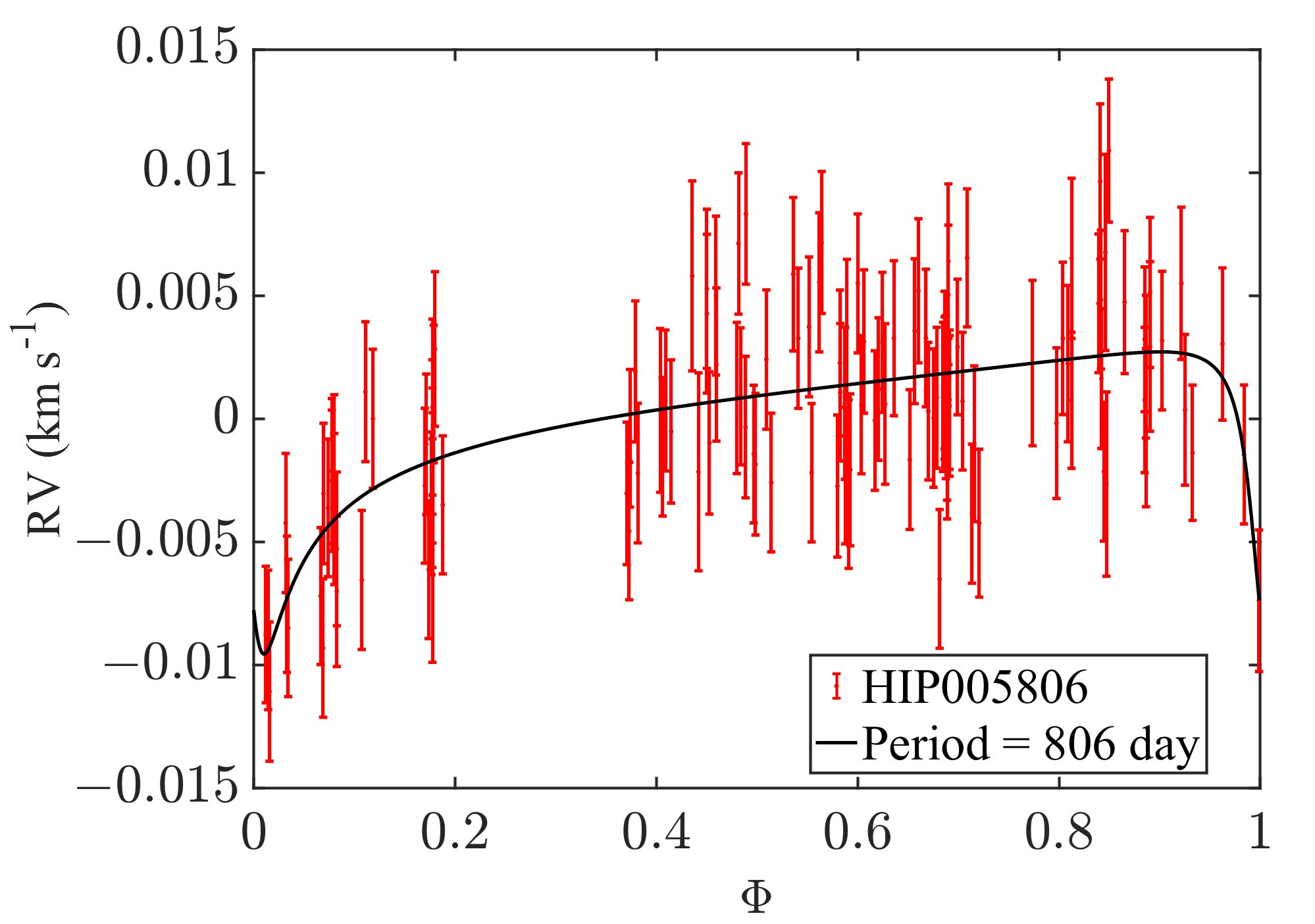}    
\end{subfigure}
\begin{subfigure}{.5\textwidth}
\includegraphics[width=\linewidth]{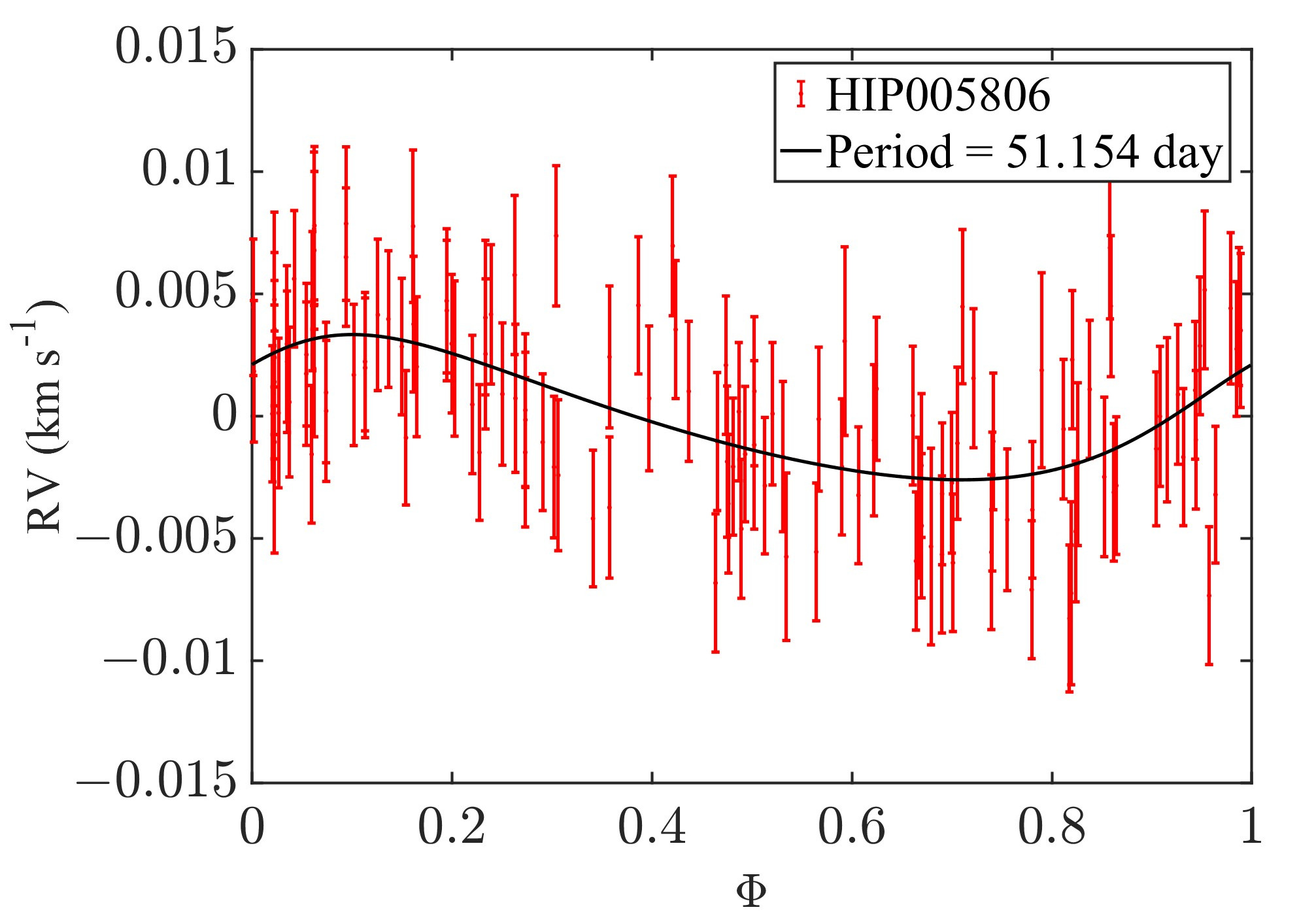}    
\end{subfigure}
\begin{subfigure}{.5\textwidth}
\includegraphics[width=\linewidth]{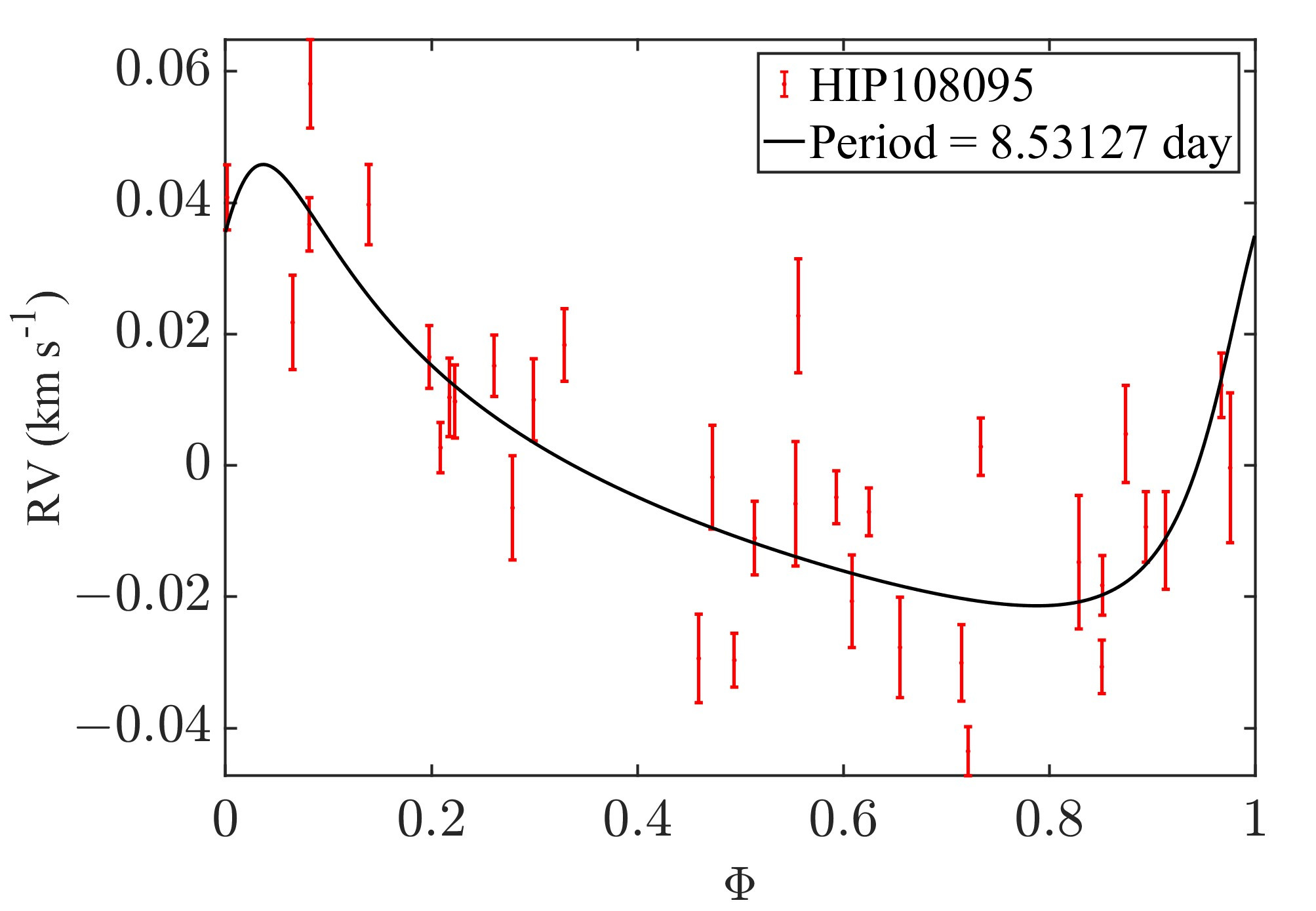}    
\end{subfigure}
\caption{Phase-folded RV curve for HIP5806 with the orbital periods of $806$ d (top panel), and of $51.154$ d (middle panel), and for HIP108095 with the orbital period of $8.53127$ d (bottom panel).}
\label{fig:orbitNew}
\end{figure}
\section{Computation of the standard deviation for trend solution}
\begin{equation}
\label{Eq:trendSTD}
	\sigma^2(P_n(RV)) = \sum_{i = 1}^{i = n} {\left[\frac{i^2\,a_i^2\,T^{2i}}{(2i+1)(i+1)^2} + 2 \sum_{j = i + 1}^{j = n} {\frac{i\,j\,a_i\,a_j\,T^{i + j}}{(i+j+1)(i+1)(j+1)}}\right]} 
\end{equation}

where :
\begin{equation}
	P_n(RV(t_k)) = \sum_{i = 0}^{i = n} {a_i\, t_k^i } 
\end{equation}
and $t_k$ is uniformly distributed in $[0, T]$.
\onecolumn
\section{RV trends}
\label{append:Trend}
\begin{figure}
\caption{RV variations of trend stars with $\sigma_{\rm{Model}} \ge 100 \rm{\; m\, s^{-1}}$ (the acceleration is indicated in each panel). Top row: HIP5806, HIP7404, and HIP026037. Middle row: HIP29611, HIP42575, and HIP61157. Bottom row: HIP68134, HIP85158, and HIP99385.}
\label{highAccTrend}
\begin{subfigure}{.36\textwidth}
\includegraphics[width=\linewidth]{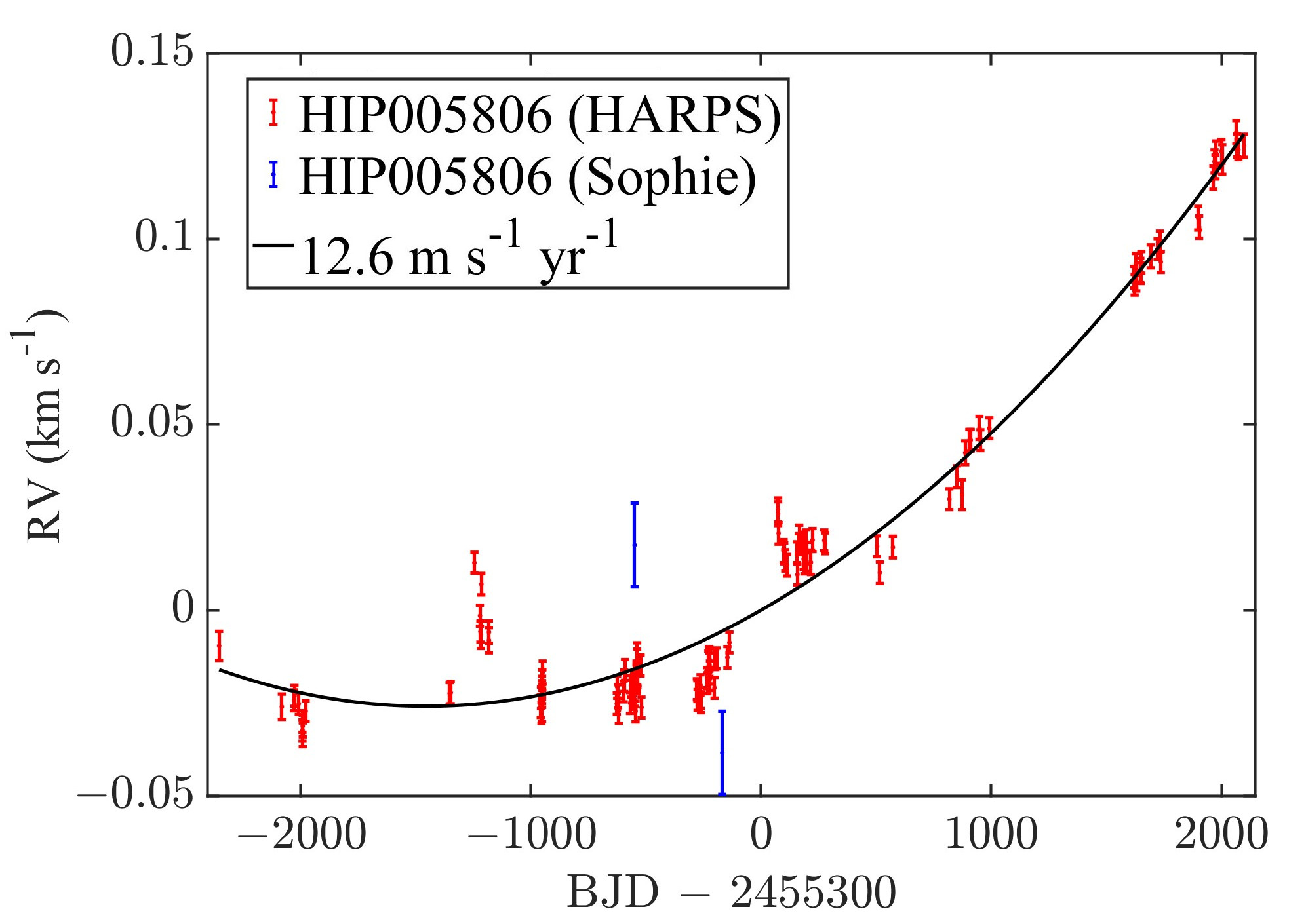}
\end{subfigure}
\begin{subfigure}{.35\textwidth}
\includegraphics[width=\linewidth]{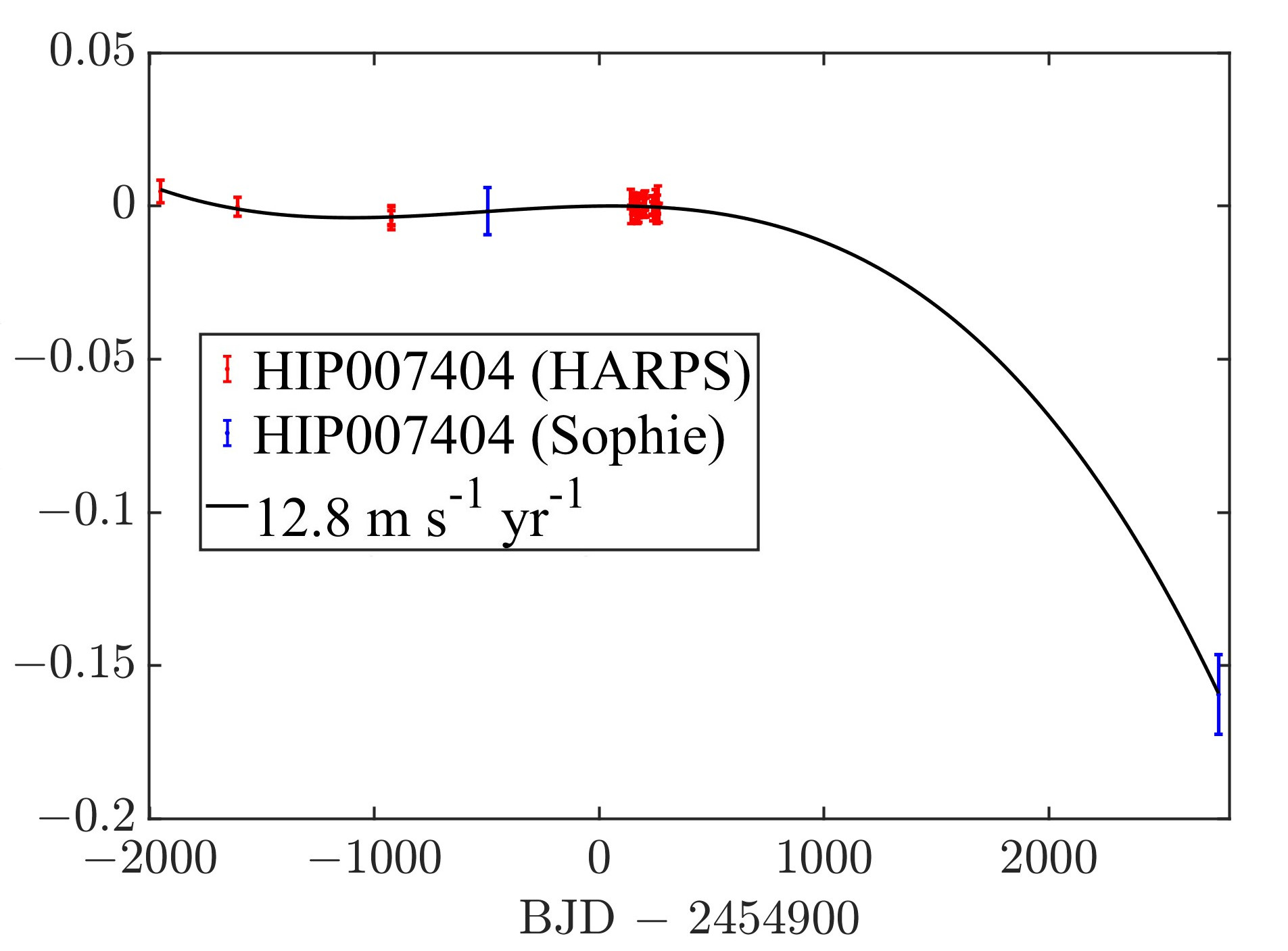}
\end{subfigure}
\begin{subfigure}{.35\textwidth}
\includegraphics[width=\linewidth]{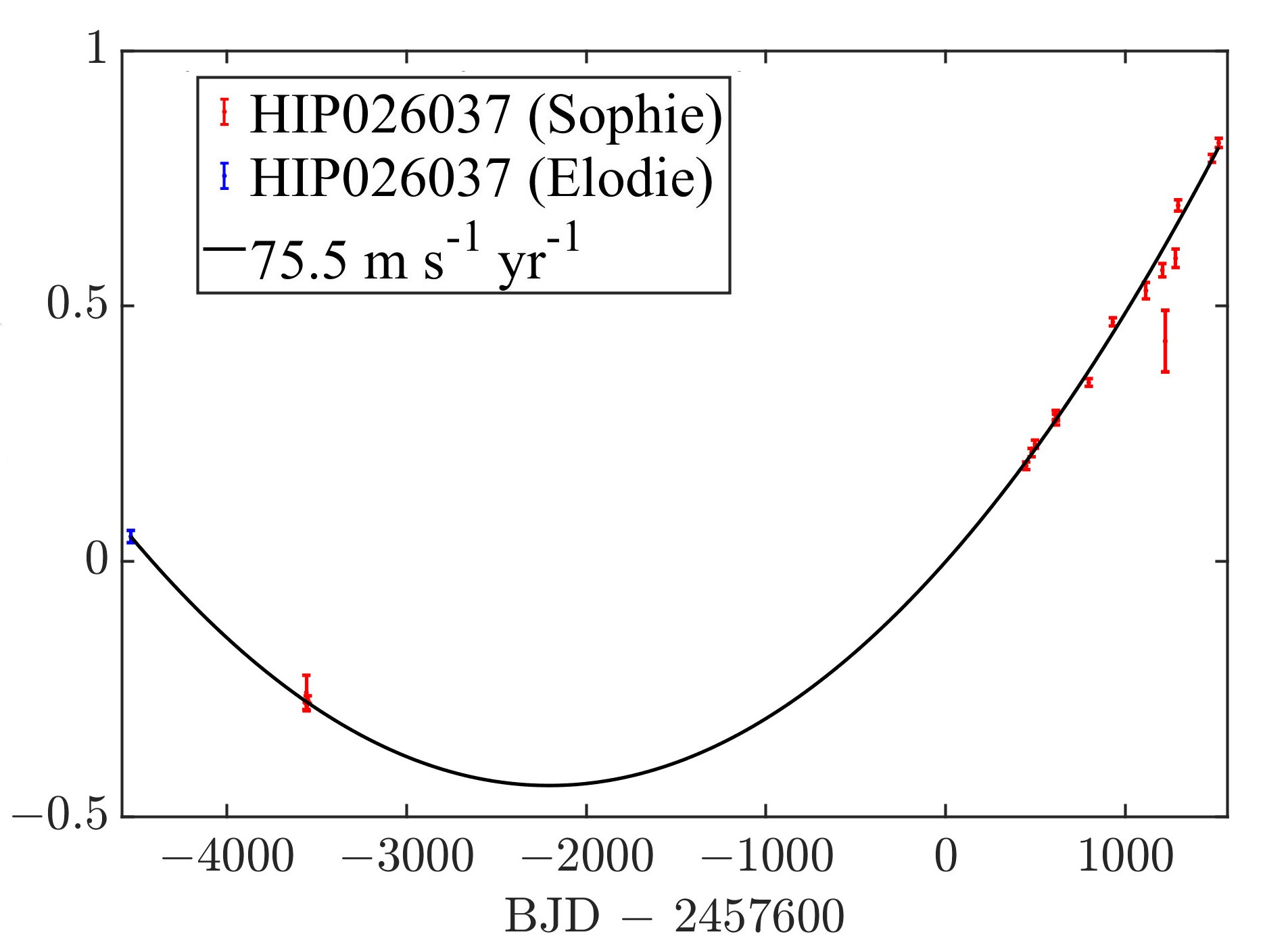}
\end{subfigure}
\begin{subfigure}{.36\textwidth}
\includegraphics[width=\linewidth]{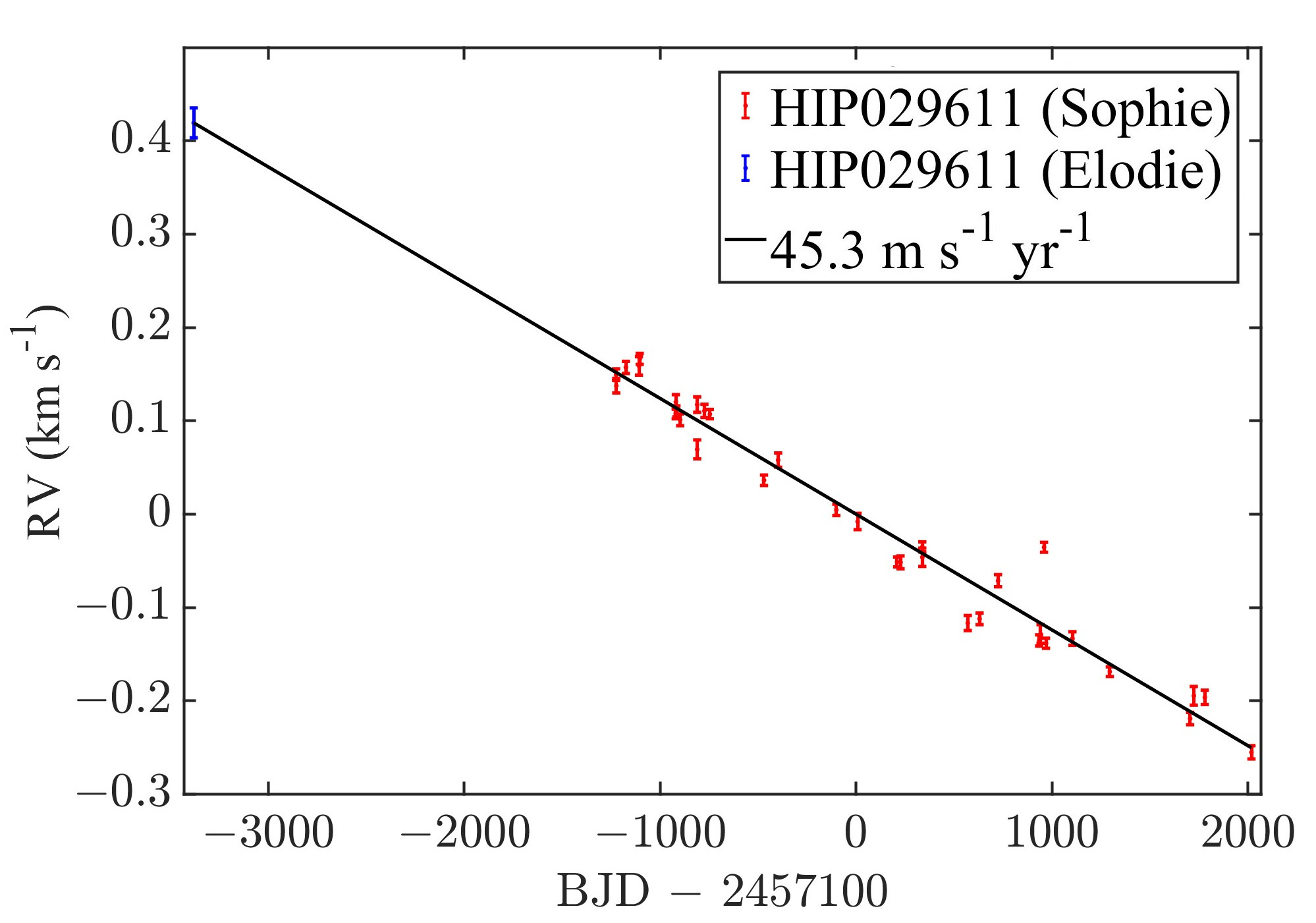}
\end{subfigure}
\begin{subfigure}{.35\textwidth}
\includegraphics[width=\linewidth]{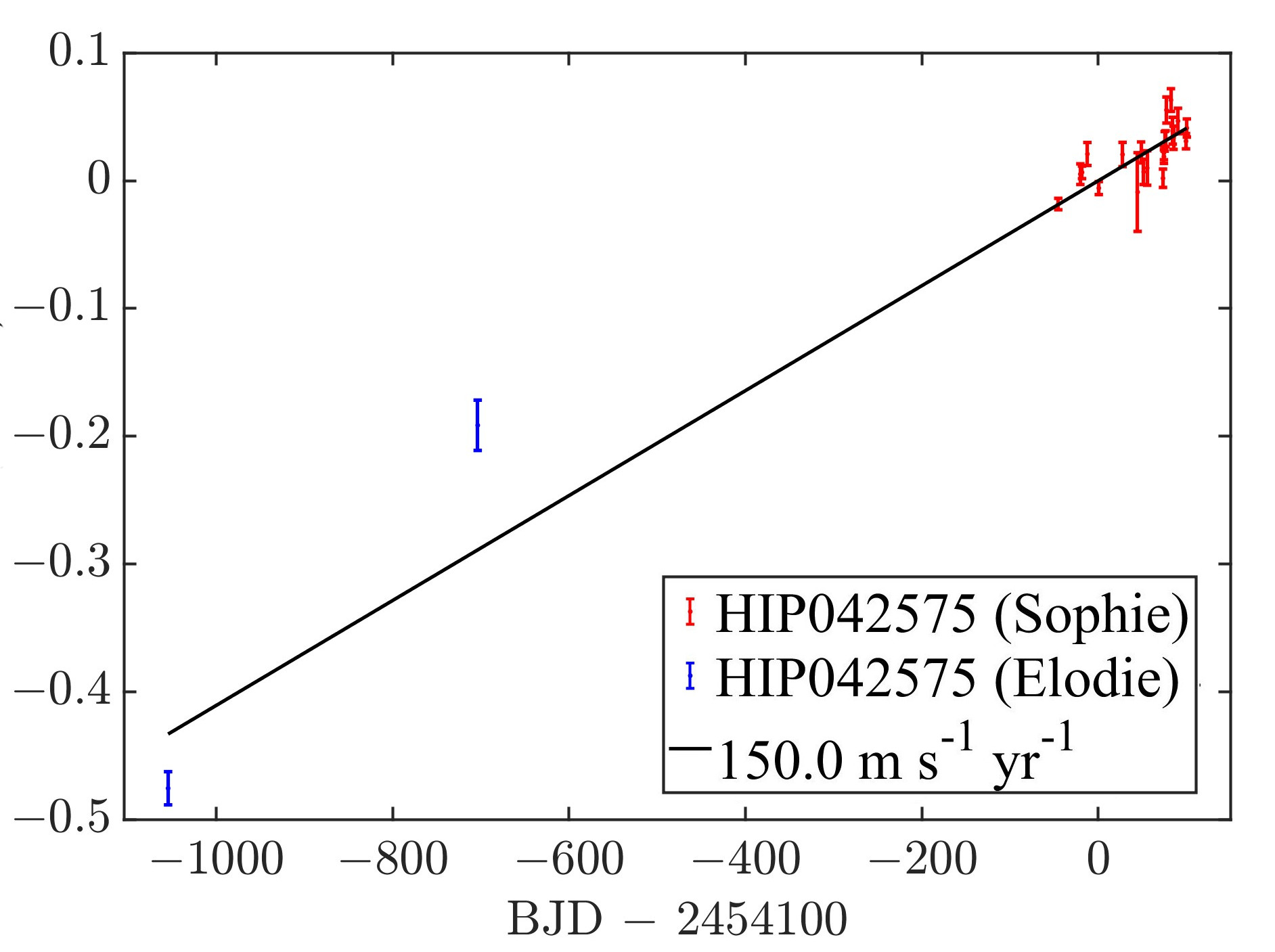}
\end{subfigure}
\begin{subfigure}{.35\textwidth}
\includegraphics[width=\linewidth]{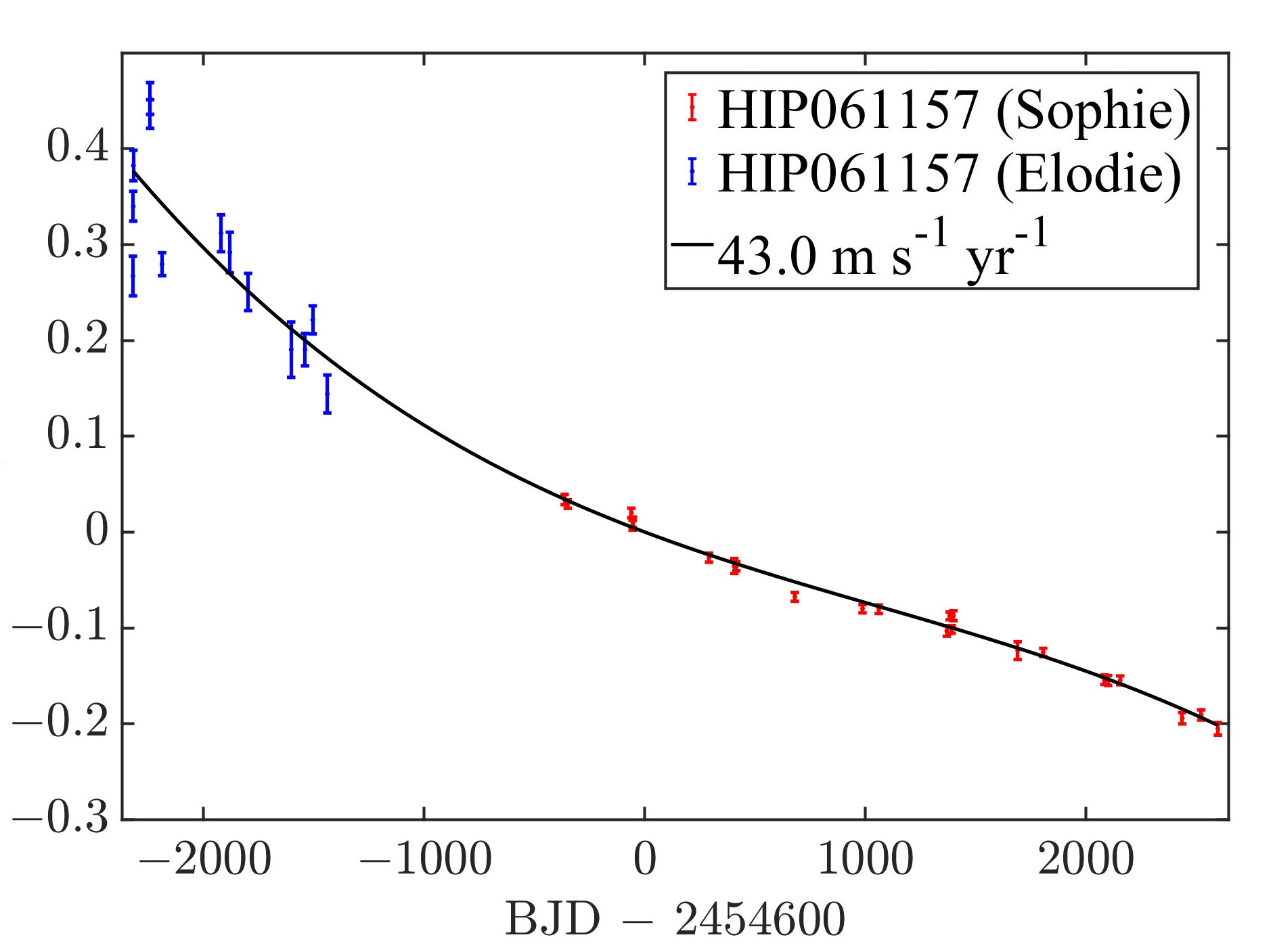}
\end{subfigure}
\begin{subfigure}{.36\textwidth}
\includegraphics[width=\linewidth]{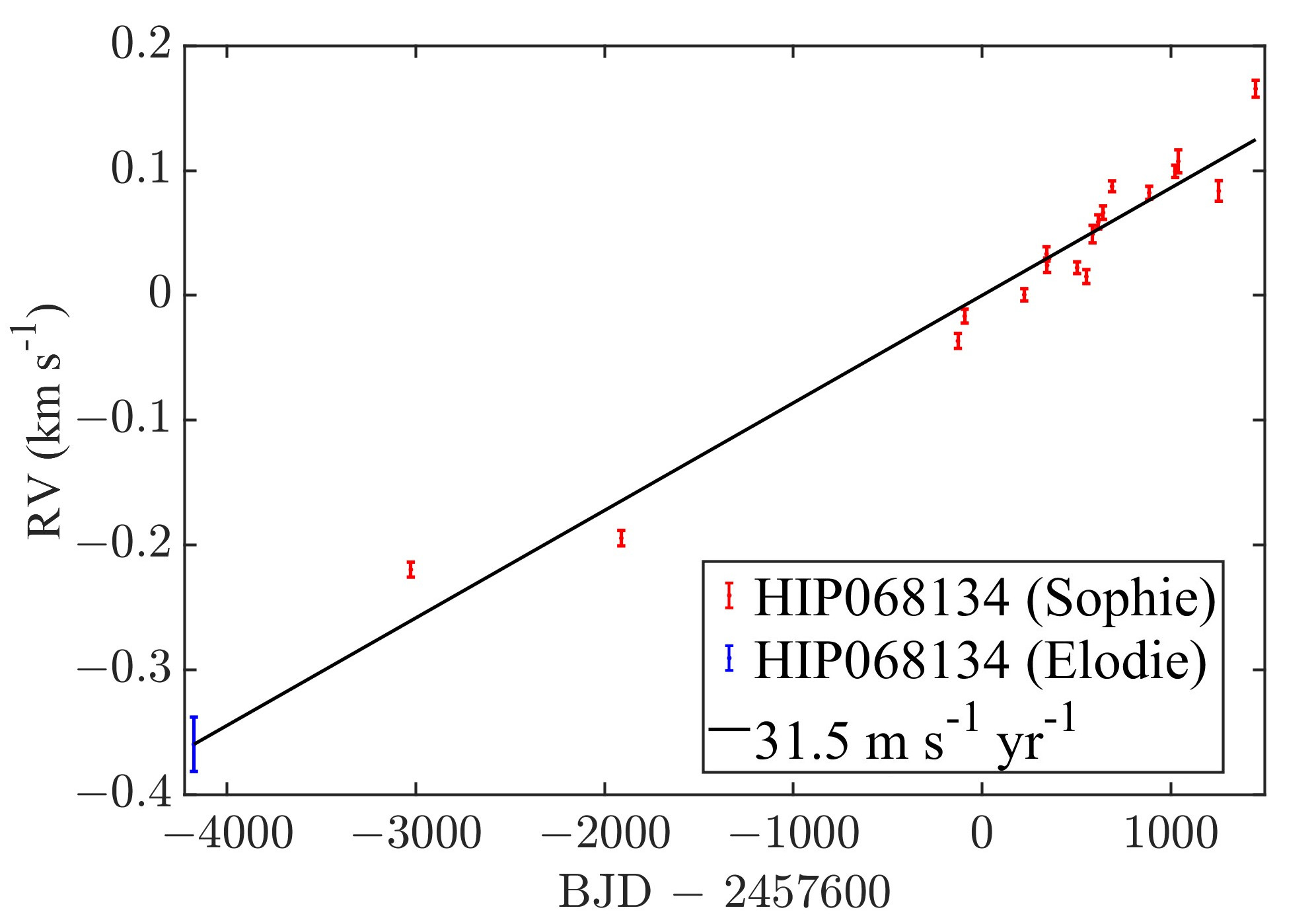}
\end{subfigure}
\begin{subfigure}{.35\textwidth}
\includegraphics[width=\linewidth]{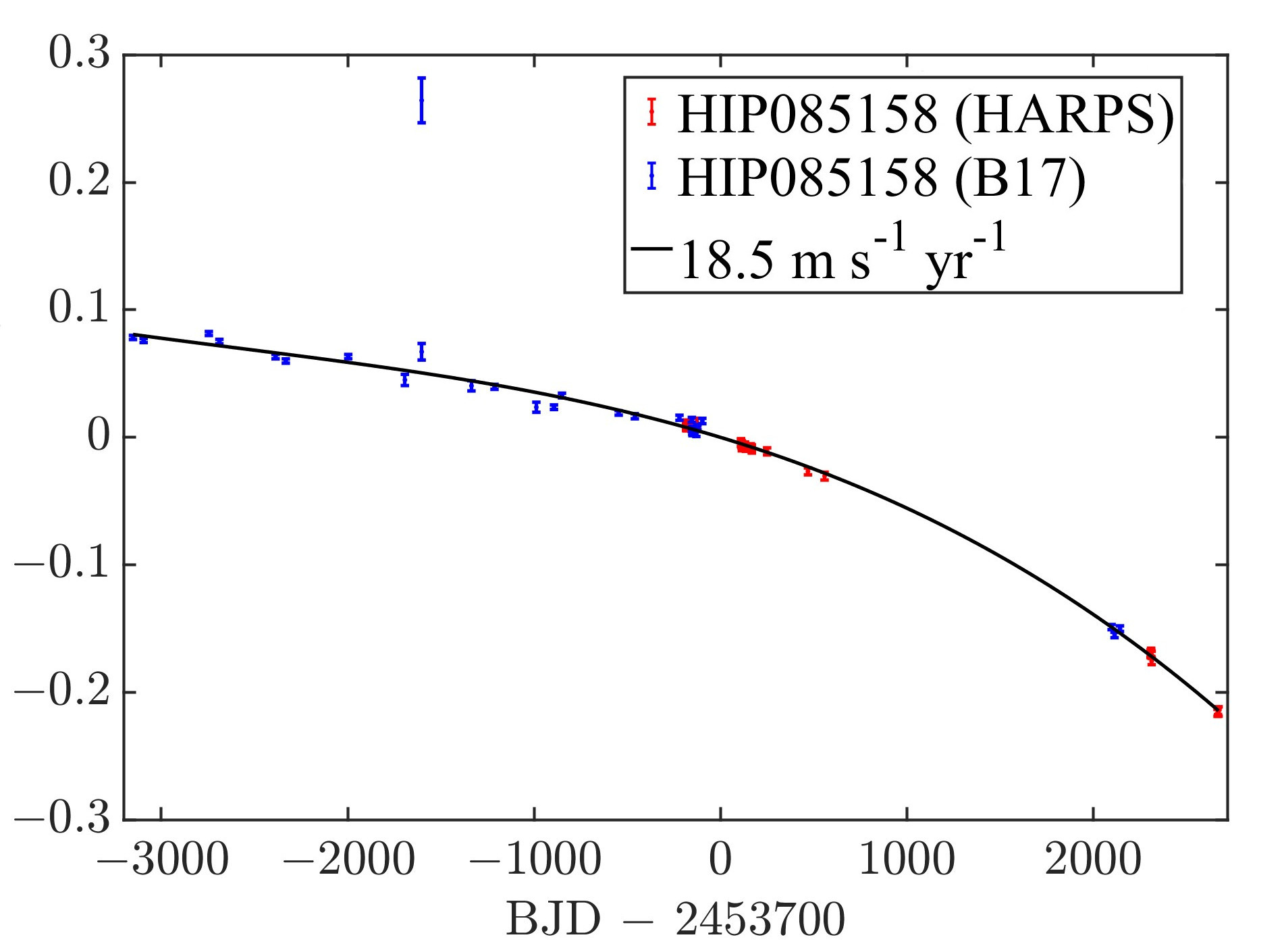}
\end{subfigure}
\begin{subfigure}{.35\textwidth}
\includegraphics[width=\linewidth]{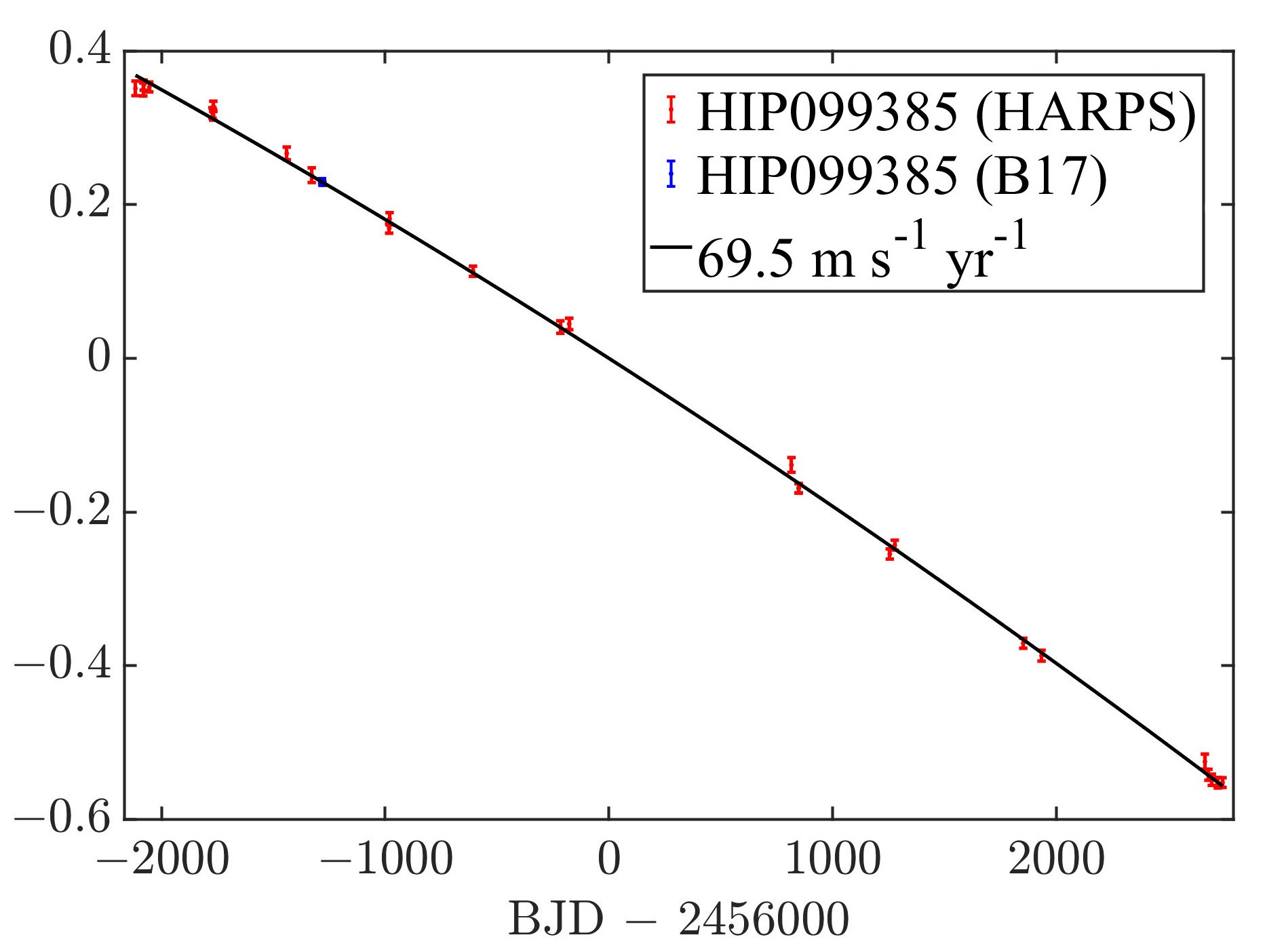}
\end{subfigure}
\end{figure}
\end{document}